\documentclass[phd,12pt,noupper,appendixpart,oneside]{ubcthesis}
% ...
\usepackage[numbers,sort&compress]{natbib}

\newcommand\href[2]{\texttt{#2}}

\usepackage{graphicx}

\usepackage{lscape}

\usepackage{psfrag}

\usepackage{amsfonts}

\usepackage{verbatim}

\institution{The University Of British Columbia}
\institutionaddress{Vancouver, Canada}
\department{Department of Physics and Astronomy}
\numberofsignatures{4}          

\previousdegree{B.Sc., The University of Victoria, 1998}
\previousdegree{M.Sc., The University of British Columbia, 2000}

\title{Tachyons, Boundary Interactions, and the Genus Expansion in String Theory}
\author{Mark Colin Andrew Laidlaw}
\copyrightyear{2003}
\submitdate{28 July, 2003}
\renewcommand\thepart         {\Roman{part}}
\renewcommand\thechapter      {\arabic{chapter}}
\renewcommand\thesection      {\thechapter.\arabic{section}}
\renewcommand\thesubsection   {\thesection.\arabic{subsection}}
\renewcommand\thesubsubsection{\thesubsection.\arabic{subsubsection}}
\renewcommand\theparagraph    {\thesubsubsection.\arabic{paragraph}}

\begin{document}

\newcommand{\ddf}{\bar \partial}
\newcommand{\df}{\partial}
\newcommand{\beq}{\begin{equation}}
\newcommand{\eeq}{\end{equation}}
\newcommand{\beqn}{\begin{eqnarray}}
\newcommand{\eeqn}{\end{eqnarray}}

 \linespread{1.3}
\frontmatter

\maketitle
\authorizationform

\begin{abstract}




This thesis examines the interaction of both bosonic- and superstrings with various 
backgrounds with a view to understanding the interplay between tachyon condensation
and world-sheet conformal invariance, and to understanding the d-branes that overlap with
closed string modes.  We briefly review the development of both 
background independent string field theory and cubic string field theory, 
as these provide insight into the problem of tachyon condensation.  We then develop the boundary state 
and show that in backgrounds of interest to tachyon condensation the 
conformal invariance of the string world-sheet is broken, which suggests
a generalized boundary state obtained by integrating over the conformal group of the disk. 
We find that this prescription reproduces particle emission amplitudes calculated
from the string sigma model for both on- and off-shell boundary interactions. 
The boundary state appears as a coherent superposition of closed string states, and 
using this a method for calculating amplitudes beyond tree level is developed.  The interaction of
closed strings with other backgrounds is also discussed.
An extension of the boundary state to
encode fields other than a gauge or tachyon field is described.
A modification of the  boundary state 
which encodes the time dependence of tachyon
condensation is reviewed, and an examination of spherically symmetric tachyon condensation
in the 1/D expansion is
presented.

\end{abstract}

\tableofcontents
\listoftables
\listoffigures

\preface
This work investigates a number of aspects of the interplay between 
the conformal invariance in string theory and 
interaction terms confined to the boundaries of the string world-sheet.  
A brief synopsis of some of the theoretical basis for the work is 
presented in
chapter \ref{ch:theoreticintro}, while chapters \ref{ch:boundaries} and \ref{ch:generalize}
 have sections of  extensive overlap
with, respectively, \cite{ Akhmedov:2001yh,Laidlaw:2001jt,Laidlaw:2002qu} and  
\cite{Grignani:2002rx}, works
on which the author collaborated.


\acknowledgements
This thesis would not have been possible without the advice and tutelage 
of my supervisor, Professor Semenoff,
and the members of my PhD. committee, Professors McKenna, Rozali, 
Schleich, and Zhitnitsky.  I am very grateful to them, and also my fellow 
graduate
students, who have answered more than their share of questions, both conceptual and technical,
while I was preparing this.  Finally, I am exceptionally grateful  for the  unwavering
support and encouragement of my wife, Susan, to whom I dedicate this work.


\newpage

\mainmatter


\chapter{Introduction}

Max Born was attributed, in 1928, with the statement that `Physics, 
as we know it,
will be over in six months'. \cite{Hawking:1988}  This confidence was 
reportedly based on the recent discovery of the Dirac equation describing the
electron, and the assumption that a similar equation 
could be found for the proton.
Indeed, the spectacular success in the development of quantum theory to 
describe the emission spectrum of hydrogen, the work functions of 
metals, and the
radiation of black bodies, as well as previous triumphs 
such as Maxwell's theory 
of electricity and magnetism can be seen as justifying 
that optimism.  It may, without
much exaggeration, be asserted that most of the progress 
in the discipline over the
past century has been related to the quantum effects that govern exactly those
particles of which Born was speaking, and that quantum 
mechanics forms the cornerstone
of our current understanding of the physics of the small.
Few physicists today would be willing to suggest that their 
discipline will be solved 
in short order, and many of the more pessimistic will suggest that the best
we can ever hope to do is achieve some effective field theory description 
of the world.  They would point to the difficulty quantizing gravity, and 
the success of the Standard Model in predicting and describing the results
of most particle scattering experiments, and might perhaps suggest that 
the vein of fundamental discoveries accessible to us is played out, nearly
exhausted.
We take a more optimistic attitude, and so  we ask the 
indulgence of the reader
as we briefly touch on some of the major developments in the field of physics over the past 
hundred years and allude to the current state of knowledge. It is our hope that this will serve 
to put the work contained within this thesis into perspective,
both as to its interest and its applicability
to future developments within the field.

The early years of the century were marked by two developments that forced radical changes in the way
the world was perceived.  The first was the exposition of the  theory of 
relativity which, for the 
first time, put the concepts of time and space on an equal footing and predicted apparently
counterintuitive effects such as length contraction and time dilation for objects moving close to the speed 
of light.  It 
was vindicated in many tests such as the precession of the perihelion of Mercury 
and the
aberration of stars' light by the sun.  The second was the discovery of the quantum nature of 
atoms, which 
facilitated an explanation of the spectra of the elements and compounds.  

Coming close upon the heels of the initial understanding of the quantum nature of atoms was the discovery 
of the constituents of the nuclei, protons and neutrons, and the tantalizing hint of more particles through 
clues such as $\beta$-decay and the observation in cosmic ray experiments
 of particles intermediate in weight between 
the nucleons
and the electron.  The 
promise of more `fundamental' particles was realized in a number of accelerator and reactor
experiments in the early 1950s, with the discovery of strange particles
and neutrinos, and the identification of the muon
as a lepton with similar properties to the electron (see \cite{Griffiths:1987tj} for a review).  
As the number of particles known to physicists increased, so did the ability of physicists to make sense of 
their interactions.  Between experiments which revealed the internal structure of the hadrons and 
mesons and others which sought to understand weak decay processes, a picture emerged of 
three 
families of particles, interacting with a spontaneously broken $SU(3) \times SU(2) \times U(1)$ 
gauge group.  This picture which was greatly strengthened by 
the discovery in the early 1980s of the $W$ and $Z$
bosons \cite{Arnison:1983rp,Arnison:1983zy,Arnison:1983mk}, and the $t$ quark in the 1990s 
\cite{Abe:1994st}.  

Simultaneously, the understanding of the large scale structure of the universe has undergone a
revolution in the past hundred years.  It was originally observed in the
1920s that distant galaxies recede from 
us faster than the nearby galaxies.  This observation admits the interpretation
 that we live in an expanding 
universe.  The
discovery of the cosmic microwave background in the 1960s was a window into an epoch when the 
universe was
both hotter and denser than it is now.  The powerful modern telescopes give a window into the past by 
allowing us to understand the formation of galaxies and the evolution of the universe.
The history and evolution of this universe  is 
also
explored by calculations like big bang nucleosynthesis, which predicts the abundances of the light
elements to great accuracy.  In addition, recent precision measurements of the cosmic
microwave background \cite{Bennett:2003bz,Jaffe:2000tx} give insight into the small fluctuations in 
density that were the seeds for 
the large scale structure of the universe.  

It appears that there exists a consistent and complete understanding of the world we live in.  
Many of the masses, couplings, and mixings of the Standard Model are known or measured, and the 
observed 
scattering processes are by and large calculated to better that 1\% 
accuracy.  We have a model
of the early universe that makes use of our knowledge of nuclear processes, predicts the abundances
of elements, and offers an explanation of the observed spectrum and
describes the fluctuations in the cosmic 
microwave
background.  The dynamics of large objects are described very well by classical general relativity 
which has also been tested in an astrophysical setting by watching the decay of rotation time for 
binary 
pulsars.  In short, a large variety of physical processes on many scales are well known
and well described by current understanding.

However, there are, just as there were one hundred years ago, a number of gaps in our understanding 
that may well provide windows into new and exciting regimes and effects.  One particle not yet observed
to complete the description of the Standard Model is the Higgs boson, and its absence raises a
question, is it truly a fundamental particle, distinguished as the only such scalar in nature, or does 
its mass-generating 
effect come from a more complicated mechanism such as 
technicolor?  Recent observations have discovered masses and mixings
between the species of neutrinos, which are not predicted in the Standard Model \cite{Ahmad:2002jz,
Ahmad:2002ka,Fukuda:2001nk}.  
Recent 
cosmological observations have shown two facts that are very interesting, that the matter
content of the universe accounts for roughly 30\% of its observed energy density, with the other
70\% coming from so-called vacuum energy \cite{Riess:1998cb}, and 
further that the familiar particles from the Standard Model represent only a small 
fraction of the
matter content of the universe, 
a fact 
previously suggested by data on galactic rotation curves \cite{Ellis:2003ug}.  
In addition to this there are serious suggestions that gravity might be testably modified, both at
the sub-millimeter level, and at length scales much greater than the size of our galaxy 
\cite{Dvali:2002vf}.  

While this list is far from a comprehensive exposition of all the current areas of research, it
suffices to give the impression that there are a number of very interesting and currently unresolved issues 
in the field.  In a very real sense
the discipline of Physics is currently at an exciting crossroads where it is possible to get 
precision experimental information about the parameters in a number of theories spanning 
orders of magnitude in size and energy.  However, as many of the fundamental questions about
the nature of the universe are laid open to inspection and resolution by diligent work, other 
questions arise to which the answers are not currently known.

A concrete question that is often asked is whether the current known particles exhaust the 
spectrum of the theory describing the world, and there are many currently popular suggestions.  It may
be that the world exhibits broken supersymmetry, in which case for each known particle there will 
exist a superpartner with identical charges and couplings, and many theorists expect that the lightest
of these superpartners is a viable candidate for the dark matter that affects galaxy rotation curves
\cite{Falk:1995fk}.  
Another possibility is that there may exist a larger gauge group which unifies the existing particles and
couplings, but is broken at some higher scale.  The simplest forms of such a grand unification 
which sees $SU(5)$ break to $SU(3)\times SU(2) \times U(1)$ have been experimentally
ruled out, and the related supersymmetric models have been strongly constrained
 \cite{Ellis:1992ri,Anselmo:1991uu}, but larger groups have not.
The breaking of a large 
enough group could result in 
matter
in a `hidden sector', which is to say light matter that is uncharged with respect to the matter we are made
of, but which may couple at higher energies through interactions mediated by 
massive particles much like the 
leptoquarks in standard GUTs.

Other intriguing scenarios have been proposed as well 
\cite{Randall:1999ee,Antoniadis:1998ig,Barbieri:2000vh}. A
 prominent recent theme being the 
existence 
of
extra dimensions in addition to the three spatial and one time dimension so familiar from everyday experience.
This idea has a number of interesting consequences, the first being that for extra dimensions with a 
very small spatial extent, wrapped up on themselves (compactified), there could very well be 
an infinite number 
of new particles which are massive partners of the known particles coming from the Fourier modes of the
known particles around these compact dimensions.  Equally well, larger, but still small, compact extra dimensions
have been proposed \cite{Antoniadis:1990ew} which give a natural way to interpret the relative 
weakness of gravitational interactions
as compared with the other forces of nature.  These and similar ideas have given rise to a number of
scenarios in which large, or even non-compact extra dimensions are invoked with the assertion that
our universe resides on some topological feature which describes a subspace of the extended space.
In addition to all of these things, a large body of work describes the attempts to quantize gravity
(for example, 
\cite{Ashtekar:1986yd,Polyakov:1987zb,Gibbons:1977ue,Dewitt:1967ub,Arkani-Hamed:1998nn}), which
is currently a classical theory.

There are thus a large number of directions in which physical research can progress, adding more particles
to the theory of the universe with interactions described by larger groups, adding extra dimensions to space
and observing their effect, examining dynamics on a topological defect in this larger space, quantizing
gravity.  To try to do any one of these is a non-trivial task, and it would appear that to attempt to do
many simultaneously would be much more difficult, but there has emerged over the past decades a physical
theory that can apparently address 
all of these called `String Theory'.  String theory can naturally 
accommodate many of these directions, it can describe gravitons, it can have particles with 
complicated gauge interactions, it must describe a world with more dimensions than our familiar 
space and time.  For all this, there is a challenge, that many of the descriptions of string theory 
have a tachyon, a particle that travels faster than light.  We neither see nor expect such a 
particle, and to explain why it is not present is a challenge that many have undertaken.  
In this thesis, we discuss a possible mechanism that explains the absence of the tachyon, and also 
can explain why we observe less dimensions than the number one might expect from string theory.
This mechanism is tachyon condensation, and the details presented later show how it can 
force particles to inhabit a small subspace within a higher dimensional volume.

We will now present a brief overview of string theory,
both with a view to narrative exposition, and with a view to fixing some 
conventions (for the most part following \cite{Green:1987sp,Green:1987mn,Polchinski:1998rq,
Polchinski:1998rr}) that will 
be used later in this work.

\begin{figure}[tp]
  \begin{center}
\psfrag{Psi1}{$\psi_1$}
\psfrag{Psi2}{$\psi_2$}
\psfrag{P1P2}{$\psi_1*\psi_2$}
\includegraphics[width=0.8\textwidth]{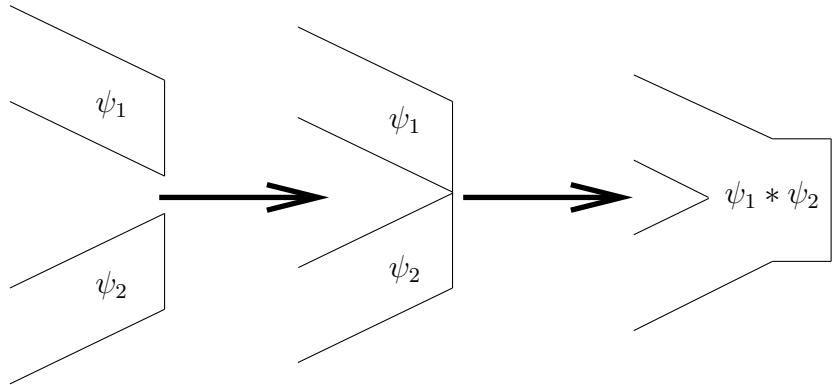}
\caption[Open string theory interaction]{A representation of 
world-sheet interactions
between open strings.  On the left two open strings, $\psi_1$ and $\psi_2$, are
 propagating.
(The previous positions of the two strings are indicated by the regions diagonally above and
below $\psi_1$ and $\psi_2$ respectively.)
In the center they 
interact by connecting at one end, and on the right they propagate 
as a single string, $\psi_1*\psi_2$ which encodes the particle information
in both of the original strings.  This can also be thought of as a series of incomplete pictures
of the string world sheet, the first showing only the portion with $t<t_0- \epsilon$, the second
showing the portion $t<t_0$, and the third $t<t_0+ \epsilon$, where $t_0$ is some measure of the 
time coordinate where they appear to merge, and $\epsilon$ is some small constant, and target space
time increasing on the horizontal axis.}
\label{fig:cubicsft}
  \end{center}
\end{figure}

Field theories are naturally concerned with point-like 
quanta and so a natural generalization 
is to ask 
how to quantize extended objects.  These would have a generalization of
a world-line with  more than one dimension.  
For a point-like object the 
action is the proper length of the world-line swept out by the propagation 
in space and time, and for a one dimensional extended object the natural 
action is the surface area swept out
by its propagation in time.  
This area can be calculated as an integral over the 
world sheet of the positions of each point along the sheet in space and 
time, which will be regarded as fields in what follows.  The action is 
given as
\beqn
S = \int d^2\sigma \sqrt{h} h^{\alpha\beta}(\sigma) g^{\mu\nu}\left(X_\gamma \right)
\df_\alpha X_\mu(\sigma) \df_\beta X_\nu(\sigma)
\eeqn
where $X_\mu$ is the position of some point of the world-sheet
in space and time,  the string world-sheet has a metric 
$h^{\alpha\beta}$ and $h$ is defined as the determinant of 
$h^{\alpha\beta}$.  The pair of coordinates denoted 
$\sigma$ parameterize the world-sheet, 
and $g_{\mu\nu}$ is the space-time metric, which is generically a function 
of the position.  This action has both Weyl and reparameterization 
invariance and these can be used to eliminate  the world-sheet metric 
from this equation \cite{Green:1987sp}.
  It is also possible to add a term proportional to
the two dimensional Ricci scalar $R$, 
\beqn
S = \int d^2 \sigma \sqrt{h} R(h)
\eeqn
but this is purely a total derivative and, while not important in 
determining the spectrum of this theory, and it is possible to see that this
term  is 
responsible for the coupling 
constant that governs the string loop expansion because up to a constant 
this term evaluates nothing but the Euler number of the string world-sheet.
As mentioned the reparameterization can eliminate the metric and this gives 
a sigma model action for the $X$s, which is free when expanding around 
Minkowskian space.  It can also be shown that not specializing to 
$g_{\mu\nu} \rightarrow \eta_{\mu\nu}$ will give the spacetime gravity 
action and stringy corrections that vanish in the limit of large string tension 
\cite{Green:1987sp}.  

In the free case, which is of interest for perturbative calculations, 
it is possible 
to make a Laurent expansion of the modes of $X_\mu$, observing that the right 
and left movers decouple in the bulk of the string world sheet.  
The conformal invariance of the string world sheet can be used to fix a flat metric 
and then it is possible to Wick rotate from a Minkowskian signature to a Euclidean signature 
through the transformation $\sigma^0 \rightarrow i \sigma^2$ \cite{Green:1987sp}.  The
left and right movers can 
be expressed in terms of the holomorphic and antiholomorphic 
coordinates ($z$ and $\bar z$) on the
Euclideanized world sheet using the expansion 
\cite{Polchinski:1998rq}
\beqn
X^\mu(z,\bar z) = x^\mu + p^\mu \ln |z^2| + \sum_{m\neq 0} \frac{1}{m}
\left( \frac{\alpha^\mu_m }{z^m} + \frac{\tilde \alpha^\mu_m}{\bar z^m} 
\right).
\label{xexpansion}
\eeqn
In the following
the terms holomorphic and antiholomorphic will be used
interchangeably with left and right mover.  When 
quantized the commutation relation between the 
Fourier modes of 
$X$ is
\beqn
\left[ \alpha^\mu_a , \alpha^\mu_b\right] = a \eta^{\mu\nu} \delta_{a+b,0}
~~~
\left[ \tilde \alpha^\mu_a , \tilde \alpha^\mu_b\right] = \eta^{\mu\nu} 
\delta_{a+b,0}
\eeqn
In the same way the Fourier coefficients of the two-dimensional energy 
momentum tensor can be written in terms of these $\alpha$s, and for the
holomorphic part we find
\beqn
L_m = \frac{1}{2} \sum_{n= -\infty}^{\infty} : \alpha_{m-n} 
\cdot \alpha_n :
\label{virasoro1}
\eeqn
where $:\left\{~~\right\}:$ denotes the  normal ordering of
any expression within  $\left\{~~\right\}$, which is moving the creation 
(negatively moded) operators to the right, and the dot represents 
contraction with respect to the Lorentz indices and $\alpha_0$ is 
proportional to the momentum.  An identical expression 
for the antiholomorphic $\tilde L$s can also be written.  Since the energy 
momentum 
tensor is traceless it appears that all the $L_m$s should annihilate the
physical states, however this strong condition would eliminate 
the spectrum of the theory, so the condition is relaxed to be that 
positively moded $L$s will annihilate physical states. This coincides 
with the choice of positively moded $\alpha$s as the annihilation 
operators, and also imposes that physical states are eigenstates of $L_0$ with 
eigenvalue $a$.
Further, the $L$s obey the Virasoro algebra
\beqn
\left[ L_m, L_n \right] = \left( m-n\right) L_{m+n} + \delta_{m+n,0} A(m)
\label{virasoro}
\eeqn
where the $A(m)$ is the central charge which turns out to be proportional 
to the dimension of space-time, which is the number of world-sheet fields $X$s.  

Furthermore, it was briefly discussed above 
that the spacetime metric is expanded around 
the Minkowskian metric, which gives rise to a particular difficulty due to its negative signature, 
namely that there may be some excitations in the spectrum which have a 
negative norm.  It is not difficult to show, demonstrated in { \cite{Green:1987sp,
Polchinski:1998rq} } 
that the condition for eliminating these negative norm states is 
equivalent to a condition on the number of $X$s and on the value of $a$.
It turns out that the number of dimensions for the
bosonic string must be either 2 or 26, and the
value of $a$ in units of the string tension $\alpha'$ is fixed to $-1$ when 
the dimension is 26 \cite{Green:1987sp}.  The consequences of this
are interesting to investigate.   First the spectrum of this theory is built 
from a Fock space vacuum with the $\alpha$s and $\tilde \alpha$s, and it must satisfy level 
matching conditions, as well as conditions on the polarization tensors for 
the various states which are obtained by requiring that the positively 
moded $L$s do annihilate the state.  The ground state of this theory is 
tachyonic, as it has negative mass squared, and the massless state consists of  
a symmetric traceless tensor, an antisymmetric tensor, and a trace term. 
These may be identified as a graviton, some gauge field, and a dilaton.  
This appears to be both good and bad, because while a particle exists with the appropriate
quantum numbers for a graviton, 
the tachyon mode intimates an instability in the vacuum, which will be 
explored more later in this work.

Secondly, there are a number of extraneous degrees 
of freedom in the bosonic string, as witnessed by the restrictions on 
possible polarization tensors for the various excited states.  One way to 
accommodate this is to work in so called light cone gauge where two 
directions are singled out as distinct and only oscillations transverse to 
those are permitted to propagate {\cite{Goddard:1973qh}}.  While very effective at reducing 
the number of degrees of freedom and enforcing the no-ghost conditions, 
this has the price of eliminating the manifest Lorentz invariance of the
theory.  There 
is a more elegant way to compensate for these extra degrees of freedom, 
and that is to introduce ghost fields in the manner of Fadeev and Popov to 
the string action, as exemplified in \cite{Polyakov:1981rd,Polyakov:1981re}.  
These will be a pair of 
anticommuting fields, $b$ and 
$c$ with conformal weights $2$ and $-1$ respectively whose action term 
\beqn
S_{b,c} = \int d^2\sigma b \df c
\eeqn
These can also be broken into holomorphic and antiholomorphic degrees of 
freedom, and satisfy anticommutation relations
\beqn
\left\{ c_n, b_m \right\} &=& \delta_{m+n,0} \nonumber \\ 
\left\{ b_m, b_n \right\} &=& 
\left\{ c_m, c_n \right\} = 0.
\eeqn
Their fermionic nature gives a contribution to the determinant of the 
path integral which cancels the contributions of two of the $X$ fields.
Naively these ghosts appear to add to the number of possible Fock 
space excitations, but there is now an additional constraint, that the 
physical states must be annihilated by an operator composed of these 
ghosts, namely the BRST operator  $Q$ \cite{Becchi:1976nq}
\beqn
Q = \sum : \left( L^\alpha_{-m} + \frac{1}{2} L^{b,c}_{-m} -a \delta_m 
\right) c_m :
\label{qbrst1}
\eeqn
with
\beqn
L^{b,c}_m = \sum_n (m-n) b_{m+n} c_{-n}
\eeqn
The constraint is then that $Q + \tilde Q$ must annihilate a physical 
state.  This can be thought of as analogous to a gauge condition, that 
just as in the case of an Abelian gauge theory the transformation
$A^\mu \rightarrow A^\mu + \df^\mu \lambda$ for some scalar function 
$\lambda$ leaves the field strength $F$ invariant, the BRST transformation
$| \phi \rangle \rightarrow | \phi \rangle + (Q + \tilde Q) |\psi \rangle$
will result in a state that is still annihilated by the BRST operator 
even if $|\psi\rangle$ is not, because $Q$ is nilpotent.

This is an attractive picture so far, but to mimic nature there is still a 
need for fermions charged under gauge groups and the corresponding 
gauge bosons in the spectrum.
The simplest way to add fermions to the action is to generalize to a 
supersymmetric theory on the world-sheet \cite{Green:1987sp}.  
The result of this is that the
action changes
\beqn
S \rightarrow S_{bosonic} + \int d^2\sigma 
i \bar \psi^\mu \rho^\alpha \df_\alpha \psi_\mu
\eeqn
where $\rho$ is the world-sheet $\gamma$ matrix, and $\psi$ is a two
dimensional Majorana spinor which can be decomposed into holomorphic and 
antiholomorphic parts which decouple.  The convention
which will be used for the Laurent expansion of the $\psi$ field into 
modes (following {\cite{Polchinski:1998rr} }) 
is 
\beqn
\psi^\mu = \sum_n \frac{\psi^\mu_n}{ z^{n+\frac{1}{2} } }
\eeqn
where $n$ is either an integer or a rational number of the form 
$n=\frac{2m+1}{2}$ for integer $m$.  This condition occurs the 
boundary conditions on the fermions impose that fermion bilinears are
single valued.  
The anticommutation relations which
arise are
\beqn
\left\{ \psi^\mu_n , \psi^\nu_m \right\} = \eta^{\mu\nu} \delta_{m+n,0}
\label{fermcomrel}
\eeqn 
The choice between integral and half integral modes for the fermions 
arises in the following way in the case of open strings. 
We require that the boundary variation vanishes which in turn implies 
 the equality of certain holomorphic and antiholomorphic fermion 
bilinears, and this means that on the boundary the fermions are equal up 
to a sign, and the relative sign between the boundaries determines whether 
the fermions can admit a zero mode.  In the case of closed strings the same 
considerations apply, only the choice of periodic and antiperiodic is 
independent for $\psi$ and $\tilde \psi$, with the periodic sector known 
as the Ramond (R) sector and the antiperiodic as the Neveu-Schwarz (NS) sector
\cite{Polchinski:1998rr}.  These 
periodicities have the consequence of producing space-time fermions 
because the 
zero modes in the Ramond sector carry a representation of a Clifford 
algebra.  This can be seen by examining equation \ref{fermcomrel}
and noting that $\psi^\mu_0$ acting on any state in the Fock space will
not change the eigenvalue of that state under action by $L_0$.  

The addition of the new fermionic term to the string world sheet action 
has the following consequence, that there are additional parts in the 
two dimensional energy momentum tensor, coming from the fermions, and also 
a sort of superpartner for this, the Noether current for non-constant 
supersymmetry transformations over the world-sheet.  In terms of modes 
these are \cite{Green:1987sp}
\beqn
L_m &\rightarrow& L_m^\alpha + \frac{1}{2} \sum_r \left( r + \frac{m}{2} 
\right) : \psi_{-r} \cdot \psi_{m+r} : \\
G_m &=& \sum_r \alpha_{-r} \cdot \psi_{m+r}
\eeqn
where the sum is implicitly over integers or half integers as appropriate 
for the sector of the theory.  The Virasoro algebra remains unchanged, but
there are additional terms that must be calculated
\beqn
\left[ L_m, G_n \right] &=& \left(\frac{m}{2} -n \right) G_{m+n} \\
\left\{ G_a, G_b \right\} &=& 2 L_{a+b} + B(a) \delta_{a+b,0}
\eeqn
where as in (\ref{virasoro}) $B(a)$ is a
central charge.
The physical states are annihilated by the positive modes of 
these currents (and in the R sector, the zero mode of $G$).

Again, it is possible to reduce constraints such as the imposition of the 
light cone gauge by the introduction of commuting ghost fields with 
conformal weights $\frac{3}{2}$ and $-\frac{1}{2}$, $\beta$ and $\gamma$.
These will be integrally or half integrally moded as appropriate from the 
fermionic sector, 
and contribute to the operators $L$, $G$, and $Q$ in a well known way.

A similar exercise to that performed in the case of purely bosonic string reveals 
that the critical dimension is $10$, and that depending upon whether the 
string is in the NS or R sector (for left and right movers) there is a 
different ground state energy.   Between this and level matching it is 
possible to determine the spectrum in each of the sectors: for the case of 
both left and right moving NS sectors, there is a tachyon, massless modes 
with quantum numbers matching those of the graviton, Kalb-Ramond field, and the dilaton,
in addition to  the
spectrum of massive modes \cite{Green:1987sp}.  When both left and right 
movers are in the R sector there is no tachyon but a massless field that 
transforms with two spinorial indices under Lorentz transformation. In the sectors where
the left and right moving fermions obey different boundary conditions level 
matching makes 
the lowest state massless and it has both vector and spinor indices, 
making it a combination of spin $\frac{1}{2}$ and spin $\frac{3}{2}$.
The number of fields described here is apparently too many 
to fill a supergravity multiplet, but more sophisticated analysis 
reveals that there is a condition which reduces the spectrum, the GSO 
projection \cite{Gliozzi:1977qd}, which gives supersymmetry by projecting out particles of a 
given chirality.  At the level of interacting 
theories it is necessary to have a number of different combinations of 
left and right moving boundary conditions and GSO projections.  This 
construction reveals essentially two types of spacetime supersymmetric 
theories, those that are chiral and non-chiral (respectively IIA and IIB 
theories).

\begin{figure}[tp]
  \begin{center}
\psfrag{E8}{{\Huge $E_8 \times E_8$ }}
\includegraphics[width=0.7\textwidth]{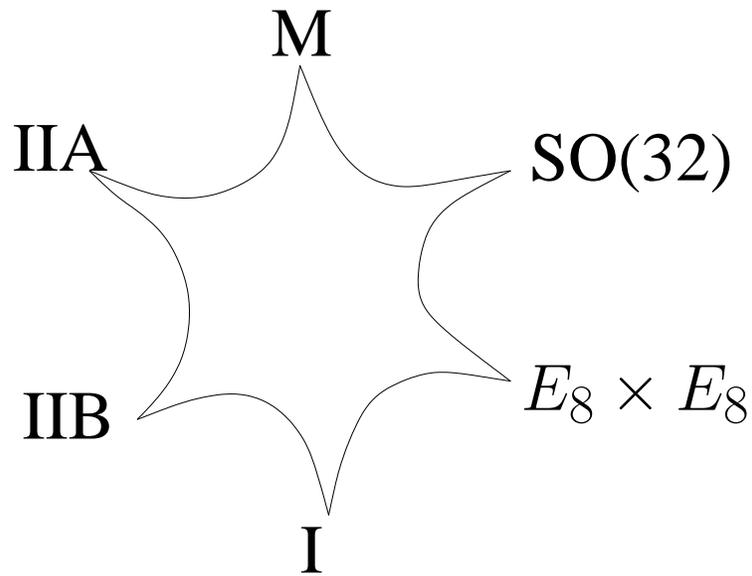}
\caption[String Theory Modulus Space]{The modulus space of string theory, which
consists of the five known perturbative string theories, and $M$ theory.  The 
edges on the figure represent duality transformations which will map one theory,
usually at strong coupling to another at weak coupling, allowing perturbative 
calculations in one to probe non-perturbative effects in the other.}
\label{fig:stmodspace}
  \end{center}
\end{figure}

Any orientation in string theory would be incomplete without mention of 
another class of theories, the heterotic string theories.  The fact that 
the left and right movers decouple makes it possible to write different 
theories for the left and right movers.  In the heterotic theories the 
left moving degrees of freedom are written as fermions which obey some 
internal symmetry \cite{Gross:1985dd,Gross:1985fr}.  Consistency
requires that there be 32 such fermions, 
and further that they either all obey the same boundary conditions (in 
that case there is an $SO(32)$ symmetry), or half obey one set of 
conditions and half obey the other, and when properly projected this 
theory has gauge group $E_8 \times E_8$.  For these theories, the index 
from a left moving fermion is a gauge index, and the theories both have a 
symmetry group large enough to be broken to the Standard Model.

There remains one small difficulty: the world neither 
exhibits ten dimensions nor supersymmetry, nor these large gauge groups, and 
it is another matter to further constrain the theories to give a good 
simulation of the particles seen now.  This is obtained in many ways, all 
essentially similar in that they force several dimensions to have another 
topology to that of the real axis, they are either periodically 
identified, or identified under reflection, or both, and this has the 
effect of breaking the large number of symmetries of the system.
It is these techniques that allow the various string theories outlined to 
be related to one another.  There is a famous `web' of dualities that 
allow one theory compactified in a certain way at strong coupling to be 
related to another at weak coupling  with a different compactification.

From this overarching framework, in this thesis we concentrate on the problem 
of tachyon condensation.
In a number of the string theories described there is a 
tachyon in the spectrum:  The state which is annihilated by all positively 
moded oscillators.  Since 
tachyons are not observed in nature this indicates that the 
naive Fock space vacuum we 
have chosen to expand around is not the true ground state of the
theory.  The naive Fock space vacuum is the one, identified above, which is annihilated by all
positively moded Virasoro generators $L$, and by all positively moded $\alpha$s and $\phi$s.  By
contrast, the operational definition we choose for the `true ground state' is one where all
the excitations have positive semi-definite mass squared, eliminating tachyonic modes.  We 
must therefore attempt to describe the ground state of the theory.
This problem is commonly addressed  in string field theory, of which there are
two principal types.  
The first is the open string field theory \cite{Witten:1992qy,Witten:1993cr,Shatashvili:1993kk,
Shatashvili:1993ps} which concerns itself solely with interactions at the string's boundaries.
The second is the cubic, Chern-Simons like, string field theory
\cite{Witten:1986cc} which consists solely of a kinetic term for the string field
and a cubic interaction vertex.
For each of these, the problem is the same, to describe what happens to the
open string degrees of freedom as the tachyon condenses.  From the world-sheet point
of view it is possible to consider the following picture of what is 
happening:  The volume in which open strings
can end describes a d-brane.  As the tachyon condenses the number of dimensions of this
brane decreases, and the final stage of the condensation is a state in which
the brane has been reduced to a point and the two ends of the open string coincide.
This gives rise to closed strings, so at the endpoint of tachyon condensation only
closed string degrees of freedom remain, and there are also predictions for the height of the 
tachyon potential \cite{Sen:1999mh}.

As noted in \cite{Kutasov:2000qp} it is difficult to reproduce the properties 
of tachyon condensation in the cubic string theory because the
calculations involve interactions of an infinite number of fields whose mass can be
arbitrarily high.  It is possible in some cases to investigate the structure of
cubic string theory using techniques such as level truncation \cite{Moeller:2000xv,Taylor:2002uv}.
This method yields results which tend to agree with those expected for some quantities,
such as the vacuum energy of the condensate, but it remains unclear 
why the procedure works and whether level truncation is generally applicable, considering that 
there is no natural small parameter being expanded in.

In this thesis we study tachyon condensation within the framework of the open string field
theory better known as `boundary string field theory'.  The 
idea is to consider backgrounds that interact with the boundary of the string, and analysis
suggests that there is a set of coordinates on `the space of all string fields' \cite{Kutasov:2000qp}
that is better suited to the study of tachyon condensation.  Further from the point of
view of the world-sheet the tachyon condensation is described by a renormalization group flow from
the UV, where the open strings end on d-branes, to the IR where only closed string degrees of 
freedom persist.  

Here we formulate a boundary state in order to reproduce string sigma model amplitudes.
This is constructed by modifying the definition of the boundary state \cite{Callan:1987px} to 
include an integral over conformal reparameterizations.  This boundary state then encodes
the effect of these conformal reparameterizations, and is useful for many circumstances
such as computing d-brane tensions, cylinder amplitudes, and looking for the gravity counterparts
of d-branes \cite{DiVecchia:1999fx,DiVecchia:1999rh}.  In the operator approach to 
string perturbation theory the boundary state contains the couplings of closed strings
to a d-brane.  This method gives an algebraic approach for calculations, and it
suggests a method to generalize to higher loops, which reproduces the known 
results for the annulus. 

The plan of the thesis is as follows.  In Chapter \ref{ch:theoreticintro}
we introduce both boundary string field theory and cubic string theory, and we also 
describe in the language of boundary string field theory how to obtain
actions for the fields which parameterize the boundary interactions.  The discussion of 
cubic string field theory is intended to illustrate another approach to the problem of tachyon
condensation which has offered some concrete evidence of the dynamics and end point of this process,
and to stand in contrast with the methods used in subsequent chapters.
In Chapter \ref{ch:boundaries}
we develop the `boundary state' describing the boundary interactions that
parameterize tachyon condensation.  This state is a generalization of that discussed in
\cite{Callan:1987px}, and it correctly reproduces the sigma model amplitudes for 
emission of  closed string states.  It is also a description of a state which
is at neither fixed point of renormalization group flow and so interpolates between
Neumann and Dirichlet boundary conditions.  We also show that this
boundary state  can be used to reproduce known partition functions for boundary fields
at the closed string tree level in the case of conformal invariance,
compare with recent work on the definition of boundary string field theory at 
open string one loop level, and speculate on the applicability to more 
complicated surfaces.  We also briefly develop boundary states for the world-sheet fermions
of the superstring and the (super)conformal ghosts.  In Chapter \ref{ch:generalize}
we present some other calculations related to the question of tachyon condensation.  We
write a boundary state describing non-local boundary interactions \cite{Li:1993za}.
We also summarize some recent work on time dependent tachyon condensation
to show the general applicability of the boundary state method, and investigate the issue of
spherically symmetric tachyon condensation.  Chapter \ref{ch:conclusions} we conclude 
and mention some ideas for future directions of research, and 
certain calculational details have been relegated to Appendices \ref{app:conftrans}
and \ref{app:gf}.

\chapter{String Field Theory}

\label{ch:theoreticintro}

A classic problem in string theory is to understand how the background
space-time on which the string propagates arises in a self-consistent
way.  For open strings, there are two main approaches to this problem, discussed
below,
cubic string field theory  \cite{Witten:1986cc} and background
independent string field theory \cite{Witten:1992qy,Shatashvili:1993kk}.

The latter approach is defined as a problem in boundary conformal
field theory, and the analysis begins with the partition function of open-string
theory where the world-sheet is a disc.  The strings in the bulk are
considered to be on-shell and a boundary interaction with arbitrary
operators is added.  The configuration space of open string field
theory is then taken to be the space of all possible boundary
operators modulo gauge symmetries and the possibility of field
redefinition.  Renormalization fixed points, which correspond to
conformal field theories, are solutions of classical equations of
motion and should be viewed as the solutions of classical string field
theory.

Despite many problems which are both technical and matters of
principle, background independent string field theory has been useful
for finding the classical tachyon potential energy functional and the
leading derivative terms in the tachyon effective action
\cite{Gerasimov:2000zp,Kutasov:2000qp,Tseytlin:2000mt}.  Boundary
field theories which can be used to study tachyons 
are the subject of the
a large portion of the presented work.

\begin{figure}[tp]
  \begin{center}
\includegraphics[width=0.6\textwidth]{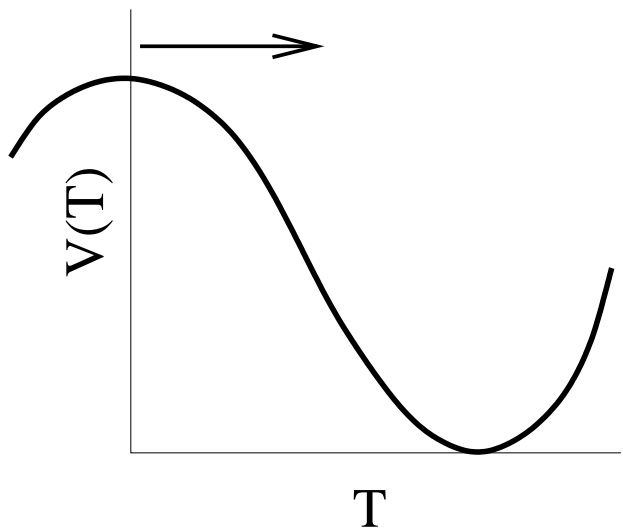}
\caption[Tachyon Condensation]{Tachyon Condensation:  This schematic representation
shows the idea behind tachyon condensation, that the tachyon is indicative of an
instability in the perturbative vacuum.  
The perturbative (Fock space) vacuum is defined at the 
maximum of $V(T)$, and as the (open string) tachyon $T$ rolls toward
the minimum of its potential $V(T)$ 
it represents a decay of the space filling brane.  At the 
minimum
of the tachyon potential only closed string degrees of freedom survive.}
  \end{center}
\end{figure}

The existence of a tachyon in the bosonic string theory indicates that
the 26-dimensional Minkowski space background about which the string  
is quantized is unstable. An unstable state is likely to decay
and the nature of both the decay process and the endpoint of the decay
are interesting questions \cite{Moeller:2000xv}.  Recently, some understanding of this
process has been achieved for the open bosonic string.  The picture
is that elaborated by
 Sen \cite{Sen:1999mh,Sen:1999xm}, that the open bosonic string tachyon 
reflects the instability of the d-25 brane.  This unstable d-brane  
should decay by condensation of the open string tachyon field.  The 
energy per unit volume released in the decay should be the d-25 brane
tension and the end-point of the decay is the closed string vacuum   
\cite{Sen:1999mh,Elitzur:1998va,Harvey:1999gq,Kutasov:2000qp}. There 
are also intermediate unstable states which are the d-branes of all  
dimensions between zero and 25.

These considerations are generally stated in the following way, that the
tachyon field has a potential of the form
\beqn
V(T) = M f(T)
\eeqn
where $f(T)$ is a function of the tachyon field which is both universal and independent
of the field theory describing the d-brane, and $M$ is the mass of the d-brane for vanishing
tachyon field.  Further, the potential is defined
with an additive constant such that at its minimum, $T_0$, it cancels the mass of the
d-brane \cite{Sen:1999xm}.  With these conventions the tachyon potential can be
written
\beqn
V(T) = M \left( 1+ f(T) \right) \mathrm{~~with~~} 1 + f(T_0) = 0.
\eeqn 
The conjectures about the dynamics of tachyon condensation also contend that 
at the minimum of the tachyon potential the corresponding brane system is
indistinguishable from that where there is no d-brane \cite{Sen:1998sm,Sen:1999xm}.
This is to say that the open string degrees of freedom have condensed 
to leave only closed string modes.

\section{Background Independent String Field Theory}

In this section we review the basic formalism of background independent off-shell string theory
because
 they will be of some use in the motivation of the subsequent work on the boundary state.
This review follows very closely the work presented in
 { \cite{Witten:1992qy,Witten:1993cr,Witten:1993ed,Shatashvili:1993kk} } 
and, for concreteness, 
focuses on 
the
tachyon field, although the results are much more general.  
It was first demonstrated in { \cite{Witten:1992qy} } that in an attempt
 to define a Lagrangian in the `space of all
open-string world-sheet theories' we discover that the boundary interaction on the string 
world-sheet
is constrained in a certain interesting way.  In particular we find that  the classical equations of
motion which are derived from $S$, the Lagrangian on the space of possible interactions, are equivalent
to BRST invariance of the theory on the string world-sheet.  Further 
we find that if it is possible to
decouple the matter and the ghosts in the world-sheet theory by a 
gauge condition on the boundary interaction, 
then the equations of motion in that particular gauge are equivalent to conformal invariance, and that
the infinitesimal generators of a gauge transformation are the BRST operators.  The result from 
all of this 
is that if matter and ghosts are decoupled then the on shell action $S$ for the particular 
interaction is equal
to the partition function of the world-sheet matter.  \cite{Witten:1993cr} 

\subsection{Bosonic String Case}

The starting point for this analysis is the string action with both bulk 
and boundary terms
\beq
S = S_0 + S_{bdy}.
\eeq
In this equation $S_0$ is the standard action on a closed string world sheet
\beq
S_0 = \int_M d^2 z (\df X_\mu \ddf X^\mu + b^{ij} \df_i c_j ),
\eeq
and we have already specialized to the case of flat metrics on both 
the world sheet and a Minkowskian (or Euclidean) metric on space-time. The
terms $b$ and $c$ are the standard anticommuting ghosts from the quantization of
bosonic string theory.  Similarly $S_{bdy}$ is a local boundary term made up of
both the bosonic fields and ghost terms on the string world sheet and
so can be written
\beq 
S_{bdy} = \int_{\df M} {\cal V}.
\eeq
In particular the boundary operator ${\cal V}$ satisfies 
\beq 
{\cal V} = b_{-1} {\cal O}
\eeq
where $b$ is a ghost field and ${\cal O}$ is some combination of fields of ghost number
one \cite{Witten:1992qy}.  The reason for this choice stems particularly from the Batalin-Vilkovisky formalism
\cite{Batalin:1981jr,Batalin:1983jr}, and 
can be summarized in the following way.
In this treatment we consider the string world-sheet as a super-manifold with a $U(1)$
symmetry, referred to as a ghost number symmetry in the literature \cite{Witten:1992qy}.  
For this 
kind of manifold the defining characteristic is a structure $\omega$ which is
a non-degenerate fermionic two-form that is closed; $\omega$ can be thought of, and has been
motivated in the literature as a fermionic symplectic form.  
With this symplectic form it is possible to define a Poisson bracket on the space
\beq
\left\{ A, B \right\} = \frac{ \df A}{\df u^K} \omega^{KL} \frac{ \df B }{ \df u^L },
\eeq
and in terms of this bracket the Master equation for the action is $\left\{ S,S \right\} =0$.
In this definition of the Poisson bracket the $u$s are local super-coordinates.  It is also possible
to define a vector field $V$ (notice that it is distinct from ${\cal V}$ the boundary interaction
term) which is a contraction on the symplectic form satisfying
\beq
V^K \omega_{KL} = \frac{ \df}{\df u^L} S
\eeq
and in this case $S$ need not satisfy the Master equation.  This condition is equivalent to (with 
indices suppressed)
\beq
V \omega = dS.
\eeq
Now, under a diffeomorphism such as 
\beq 
u^L \rightarrow u^L + \epsilon V^L
\eeq
the two-form $\omega$ transforms as
\beq
\omega_{KL} \rightarrow \omega_{KL} + \epsilon \left( V^M d_K \omega_{ML} + d_K V^M \omega_{ML} \right)
\eeq
but since $\omega$ is closed we have $d\omega=0$ and then $V$ generates a symmetry of $\omega$ if
$d (V^M \omega_{ML} )=0$.
This also implies that for a given vector $V$ it is possible to construct an $S$ that will give 
the required $V$ according to the definition above.  
It is also possible to write the two-form as
\beq
\omega = \oint d\theta_1 d\theta_2 \langle {\cal O} (\theta_1) {\cal O} (\theta_2) \rangle
\eeq
for some basis of operators ${\cal O}$ of appropriate ghost number, and 
with $\langle \ldots \rangle$ denoting an expectation value calculated through the
usual path integral weighted by  the string action.  It is possible to 
decompose a particular vector in terms of these bases and we find that 
\beqn
dS &=& \oint d\theta_1 d\theta_2 \langle V {\cal O} (\theta_1) {\cal O} (\theta_2) \rangle.
\eeqn
Thus for the special case of choosing the vector $V$ as the BRST operator
\beqn
dS &=& \oint d\theta_1 d\theta_2 \langle d {\cal O} (\theta_1)
\left\{ Q_{BRST} , {\cal O} (\theta_2) \right\} \rangle.
\label{dSeqn}
\eeqn

Now we can specialize to a particular theory that contains the germs of generality.  In particular
we note that the matter part of ${\cal V}$ can be Taylor expanded in the bosonic field $X$ as 
\beq
{\cal V} = T(X) + A_\mu (X) \df_\theta X^\mu + B_{\mu\nu} \df_\theta X^\mu 
\df_\theta X^\nu + \ldots
\eeq
and we now restrict our
attention to the tachyon field term, $T(X)$.  We also note in passing 
that later in this
work the $A_\mu (X) \df_\theta X^\mu$ term, which gives rise to a background gauge field, will also be
important.  A particularly simple solvable model is that of the quadratic tachyon
\beq
T(X) = \frac{a}{2\pi} + \sum_i \frac{u_i}{8\pi} X^2_i.
\label{eq:wittensquadtachyon}
\eeq
This model was originally considered \cite{Witten:1993cr}
because it was a simple quadratic model, and 
therefore
exactly solvable.  The addition of non-zero $u_i$s corresponds to a breaking of translational 
invariance, and because the term adds a potential energy to the zero mode of the string the 
strings oscillations are limited to  a finite volume, and in the limit of a particular $u 
\rightarrow \infty$ the string end is fixed to a particular point in the space in which it is 
embedded, and we will argue later that this provides an interesting model to describe a d-brane.  
(This model can easily 
be generalized to the more general quadratic term $U_{\mu\nu} X^\mu X^\nu$ and
this is often desirable if additional background fields are also being considered.)
For this interaction term we note that the ghost fields decouple, and so the world sheet action
 can be
written as
\beq
S = \int_M d^2z \df X^\mu \ddf X_\mu + \int_{\df M} d\theta \left( \frac{a}{2\pi} + \sum_i \frac{u_i}{8\pi}
X_i^2 \right).
\eeq
After some manipulation \cite{Witten:1993cr}  (presented explicitly in section \ref{bosonicsigmamodel}) 
we can find 
the partition function
for this world sheet models
\beq
Z = e^{-a} \prod_i \sqrt{ u_i} \Gamma (u_i).
\eeq
Note that this differs from the result found in { \cite{Kraus:2000nj} } 
by a factor relating to the 
normalization
of the zero modes.  It is also possible to note in passing that there are several good features of this 
function that are suggestive of it playing a role similar to the action for a space-time theory.  First
note that for any individual $u$ the function goes as $\frac{1}{\sqrt{u}}$ for 
$u \rightarrow \epsilon > 0$. This arises
because $u$ plays the role of localizing the function near $X=0$, in fact it is remarked elsewhere in the 
literature that $u$ interpolates between Neumann and Dirichlet boundary conditions. 
(A convenient way to see this is in the bosonic component of the boundary state 
written in (\ref{eq:nmbs1}) and (\ref{eq:nmbs2}) go between the the boundary states for
Neumann and Dirichlet boundary conditions \cite{Callan:1987px,DiVecchia:1999rh} as $U$ goes from
$0$ to $\infty$.) 
This implies that
the divergence near $u=0$ can be interpreted as associated with the delocalization
of the string over the volume of spacetime.  
Also, 
the 
expression for the world sheet partition function has divergences when $u<0$, reflecting the fact 
that in that case the
world-sheet action is not bounded below.

Now, following the previous derivation in (\ref{dSeqn}) and explicitly writing the
dependence of the fields $X$ on the boundary coordinate $\theta$, we find that 
\beq
dS = \oint d\theta_1 d\theta_2 \langle T(X)(\theta_1) \left( 1 + \epsilon^i
\partial_i \right) T(X)(\theta_2) \rangle.
\eeq
Now, use the fact that (as in \cite{Witten:1992qy,Witten:1993cr}) 
the derivative with index $i$ 
refers to the parameters
within the tachyon field, and we have
\beq 
\left( 1 + \epsilon^i \df_i \right) T(X) = \left(1+ 2 \sum_i \frac{\df^2}{\df X_i^2} \right) T(X).
\eeq
Including the ghost contribution, and using both the explicit form of the $X$ two point function, 
and the relationships
\beqn
\oint d\theta \langle X^2_i(\theta) \rangle &=& -8\pi \frac{\df Z}{\df u_i} 
\label{eq:intwithghosts1} \\
\oint d\theta_1 d\theta_2 \langle X^2_i(\theta_1) X^2_j(\theta_2) \rangle &=& (8\pi)^2 \frac{ \df^2 Z}{
\df u_i \df u_j}, \label{eq:intwithghosts2}
\eeqn
it is easy to obtain
\beq
dS = d\left( \sum_i u_i Z - \sum_j u_j \frac{\df}{\df u_j} Z + (1+a) Z \right).
\eeq
This is equivalent to obtaining the action for the boundary fields
\beq
S = \left( \beta_i \frac{\df}{\df \lambda^i} +1 \right) Z
\label{bosonicbfaction}
\eeq
where $\beta_i$ is the $\beta$-function for the $i$th coupling $\lambda^i$, which is a parameter of the 
boundary interaction terms.
This way of looking at the effect of terms on the string world sheet boundary will be particularly 
useful while considering the boundary state in subsequent sections.

\subsection{Superstring Case}

There have been several attempts to generalize the method 
given above to the superstring \cite{Kutasov:2000qp,Marino:2001qc,
Kutasov:2000aq}, and
we give an account of one of them here \cite{Marino:2001qc}, 
which has the consequence of proposing a modification
to the boundary field action (\ref{bosonicbfaction}).  The proposal for the action is
\beq
S = S_0 + S_\Gamma + \int_{\df M} G_{-1/2} {\cal O}
\label{sustractionequation}
\eeq
where $S_0$ is the usual bulk action of the RNS superstring, including both the bosonic and fermionic 
ghosts,
\beqn
S &=& \int d^2 z \left( \df X^\mu \ddf X_\mu + \psi^\mu \ddf \psi_\mu + \tilde \psi^\mu \df
\tilde \psi_\mu + b \ddf c + \beta \ddf \gamma \right)
\eeqn
with $b,c$ and $\beta,\gamma$ anticommuting and commuting ghosts respectively. 
The second and third terms of (\ref{sustractionequation}) can be thought 
of as the perturbation due to 
the addition of the field on the 
boundary.  Explicitly they are given as
\beq
 S_\Gamma  = \int_{\df M} \int d\theta \Gamma D \Gamma,
\eeq
with the defining relation for $\Gamma$,
\beq
\Gamma = \mu + \theta F
\eeq
where $\mu$ is a fermionic component of the boundary action, and $F$ is a bosonic component, and 
$\theta$ is a supercoordinate on the string world sheet.  Finally, the derivative
operator $D$ is defined as 
\beq
D = \df_\theta + \theta \df_\parallel
\eeq
on the boundary of the string world sheet.  The subscript $\parallel$ refers to the tangential 
orientation
with respect to the boundary while $\df_\theta$ is a Grassmanian derivative.
${\cal O}$ is also identified as the lowest component of a world-sheet superfield, $\Psi$ satisfying
\beq
\Psi = {\cal O} + \theta G_{-1/2} {\cal O}.
\eeq 

The proposal for the string field action, in analogy with the development leading up to
(\ref{dSeqn})  is 
to write the two form $\omega$ as 
\beq
\omega( {\cal O_1}, {\cal O_2} ) = 
\frac{1}{8} \oint \frac{d\tau d\tau'}{ 4 \pi^2} \langle {\cal O_1}(\tau) 
{\cal O_2}(\tau') \rangle
\eeq
where the contribution from the conformal ghosts has been suppressed for clarity 
(as in  (\ref{eq:intwithghosts1}) and (\ref{eq:intwithghosts2}))
as well as the factors appropriate to the inclusion of bosonized fermions $e^{-\phi}$ which 
are appropriate for the (-1) picture.
Similarly we can write 
\beq
dS = \frac{1}{8} \oint \frac{d\tau d\tau'}{ 4 \pi^2} \langle d{\cal V}(\tau) 
\left\{ Q_{BRST}, {\cal V}(\tau') \right\} \rangle
\eeq
which completes the analogy with equation (\ref{dSeqn}).

As discussed in {\cite{Harvey:2000na,Kutasov:2000aq}} the superfield which describes a 
tachyon profile is
\beq
\Psi = \Gamma T(X),
\eeq
so following their argument we get
\beq 
{\cal O} = \mu T(X)
\eeq
so it is easy to show that
\beq
G_{-1/2} {\cal O} = F T(X) + \psi^a \mu \df_a T(X).
\eeq
Inserting these expressions and integrating we obtain 
\beqn
S_\Gamma + \int_{\df M} G_{-1/2} {\cal O}
&=& \frac{1}{2\pi} \int d\tau d\theta
\left( \mu + \theta F \right) \left(
\df_\theta + \theta \df_\parallel \right)
\left( \mu + \theta F \right) 
\nonumber \\
&~&  + 
\frac{1}{2\pi} \int d\tau \left( F T(X) + \psi^a \mu \df_a T(X) \right)
\nonumber \\
&=& \int  \frac{ d\tau}{2\pi} \left( F^2 + \mu \df_\parallel \mu + F T(X) + \psi^a \mu \df_a T(X) \right).
\nonumber \\
\label{superboundint}
\eeqn
It is immediately apparent that integrating out the auxiliary field $F$ will give a term like $e^{-T^2}$
 in the partition function.  This is appropriate because the tachyon profile $T$ 
used in the superstring case is analogous to the 
square root of that used in the previous section.

Two cases are of special note in the literature, the first is the case of the constant tachyon, in
which, again up to ghost contributions we find that 
\beq
\left\{ Q_{BRST}, \mu T(X) \right\} \propto T(X)
\eeq
and upon integration of the various modes we have
\beqn
dS &=& -\frac{1}{2} T dT e^{-\frac{1}{4} T^2} \nonumber \\
&=& d\left(  e^{-\frac{1}{4} T^2} \right)
\eeqn
and since $Z \propto e^{-\frac{1}{4} T^2}$ we find upon integration
$S=Z$.
For the case of a linear tachyon 
\beq
T(X) = u_\mu X^\mu
\eeq
the calculation is somewhat more involved, but the result is known
\cite{Marino:2001qc}, and can be summarized in the following
way.  The world sheet action (\ref{superboundint}) includes a term linear in $\mu$ now and
with a field redefinition to account for this 
\beq 
\mu \rightarrow \mu + \frac{1}{2} \frac{1}{\df_\parallel} \psi^\nu \df_\nu T
\eeq
the integral becomes simpler in terms of the redefined field, but there is now a term of the form
$\psi \frac{1}{\df_\parallel} \psi$ in the action.  It is well known how to show 
that the 
expression for $dS$ becomes \cite{Marino:2001qc}
\beq
dS = -\frac{1}{4} \langle X^2 + \psi \frac{1}{\df_\parallel} \psi \rangle Z(y) dy
\eeq
where $y=u^2$.
However, since 
\beq
\frac{ d \ln Z}{dy} =  -\frac{1}{4} \langle X^2 + \psi \frac{1}{\df_\parallel} \psi \rangle
\eeq
we have that the world-sheet partition function is equal to the action for the boundary field.

\section{Cubic String Field Theory}

In addition to the discussion of background independent string field theory
above, another important motivation for the discussion of tachyons and other string fields comes
from understanding cubic string field theory, and the recent conjectures of Sen  
\cite{Sen:1999mh,Berkovits:2000hf,Sen:1999nx}  about
the condensation of open string tachyons.  This brief review draws heavily on lectures on cubic string
field theory delivered at TASI-2001 by Taylor and Zwiebach {\cite{TASI1,TASI2,Taylor:2002uv}.  This 
method is interesting in the context of this thesis for two reasons.  The first
reason is that it provides
a distinct and independent check on the ideas describing the condensation of tachyons, and the final
state of this decay.  The second is that the coherent states that can be used to describe the product 
of string states resemble those which we will detail in constructing boundary states, and that 
similar
manipulations can be performed on both.

Cubic string theory is an attempt to
treat string theory as a field theory, with 
a kinetic term and a cubic, Chern-Simons like,
interaction term.  The string fields are constructed 
by operating creation operators on a vacuum, and since
there are an infinite number of such possible interactions
it is possible to think of string field theory as an interacting 
field theory with an infinite number of massive particles
in its spectrum.  The recent interest in cubic string field theory 
can, 
in 
large part, 
be 
traced to 
work 
\cite{Recknagel:1998ih}
 on 
the 
conjectures recently raised by Sen {\cite{Sen:1998sm,Sen:1998sp}}, which address 
the question of the bosonic open
string tachyon in the following way.  
\begin{itemize}
\item{} In analogy with the Higgs mechanism familiar from the study of the Standard Model,
the bosonic open string tachyon can be thought of as an instability of a space filling D-brane.
\item{} There exists a locally stable minimum of the tachyon potential, and around that minimum
there exist no open string excitations.
\item{} That the height of this potential is given by
$\frac{1}{2 \pi^2 g^2} = \frac{\Delta E}{V}$, where $g$ is the string coupling constant, 
and $V$ is the volume of space time.
\end{itemize}
These conjectures appeal to our physical intuition and are under active investigation.

We start with the proposal by Witten {\cite{Witten:1986cc}} for a cubic string field theory action, 
particularly
\beq
S = -\frac{1}{2} \int \psi * Q \psi - \frac{g}{3} \int  \psi *  \psi *  \psi
\label{comeback}
\eeq
where $\psi$ is an open string field, $g$ is the string coupling constant, 
$*$ is a product on the space of string fields, and $Q$ is an 
operator which is roughly analogous to a derivative operator (in fact it will be the BRST operator).
A number of properties that are important in the
realization of the theory.  
First, with respect to ghost number, the $*$ product is additive, which is to say that if
\beq
\psi_1*\psi_2 = \psi_3 \rightarrow G_{\psi_3} = G_{\psi_1} + G_{\psi_2}
\eeq
where $G_\psi$ is the ghost number of the string field $\psi$.  In a similar way to this, the operator $Q$ adds
one to the ghost number of the field:
\beq
Q \psi = \psi' \rightarrow G_{\psi'} = 1+ G_\psi,
\eeq
and the integration picks out the components of the integrand which have ghost number 3,
\beq
\int \psi = 0 ~~~\forall ~~~\psi:G_\psi \neq 3.
\eeq
To avoid difficulty with boundary terms, it 
is desirable for the integration to vanish for total derivatives 
\beq
\int Q \psi = 0 ~~~ \forall ~~\psi.
\eeq
There are further properties motivated
by thinking of $Q$ as an exterior derivative. The first is that $Q$ is nilpotent 
(satisfied for the BRST operator
in the critical dimension), secondly a Leibnitz rule for the $Q$ operator
\beqn
Q \left( \psi_1 * \psi_2  \right) &=& \left( Q \psi_1 \right) * \psi_2 + (-1)^{G_{\psi_1}} \psi_1 * Q \psi_2, 
\eeqn
and also a commutativity condition
\beq
\int \psi_1 * \psi_2 = (-1)^{G_{\psi_1} + G_{\psi_2} } \int\psi_2 * \psi_1.
\eeq
It is also interesting to note that under the analog of a non-Abelian gauge transformation
$\psi \rightarrow \psi + \delta \psi$ subject to 
\beq
\delta \psi = Q \Lambda + g \left( \psi * \Lambda - \Lambda * \psi \right)
\eeq 
and $G_\Lambda =0$, the cubic string field theory action is invariant.  

The Fock space which the bosonic string fields inhabit is defined by the general state
\beq
\prod \alpha_{-m}^\mu \ldots c_{-n} \ldots b_{-k} \ldots |0\rangle
\nonumber
\eeq
where the $\ldots$ refer to any arbitrary insertion of oscillators similar to the preceding $\alpha,~c$,
or $b$.  The 
$\alpha$ 
oscillators come from the quantizing of the bosonic $X$ field
and the $b$, $c$ are the ghost fields.  These have the property that on the Fock space 
vacuum {\cite{Green:1987sp}}
\beqn
b_{n} | 0 \rangle &=& 0, ~ n\geq -1 \nonumber \\
c_{n} | 0 \rangle &=& 0, ~ n\geq 2 \nonumber \\
\alpha_n | 0 \rangle &=& 0, ~ n\geq 1
\eeqn
and $\alpha_0$ is the momentum operator.
It is well known that
it is possible to write both the BRST operator and the Virasoro generators in terms of the 
raising and lowering operators $\alpha$, $b$, and $c$, 
\beq
Q_B = \sum_n c_n L_{-n}^m + \sum_{mn} \frac{m-n}{2} : c_m c_n b_{-m-n} : -c_0
\eeq
where the matter part of the Virasoro generator is given by 
\beq
L_k^m = \frac{1}{2} \sum_n :\alpha^\mu_{k-n} \alpha_{\mu n} : + a \delta_{k0}
\eeq
with $a$ a normal ordering constant associated with the mass-shell condition for the strings as 
in (\ref{virasoro1}) and (\ref{qbrst1})
Given this information it is relatively easy to explicitly construct Fock space states which
have inner products with the string fields which satisfy all the enumerated
requirements for the integral.  
For the `kinetic' term of the string field theory action 
we need to construct a state which is the tensor product of
two such Fock spaces
because the string fields each carry
a raising operator Fock space.  In particular it can be shown that the state
\beqn
\langle I_2 | &=& \int dp \left( \langle 0,p |_1 \otimes \langle 0,p |_2 \right) 
\exp \left[ - \sum_{n=1}^\infty \frac{ (-1)^n }{n} \alpha^{(1)}_n \alpha^{(2)}_n - \right.
\nonumber \\ &~& ~~~~~~~~~~~~~ \left.
\sum_{n=0}^\infty (-1)^n \left( c_n^{(1)} b_n^{(2)} + c_n^{(2)} b_n^{(1)} \right) \right]
\eeqn
is a representation of the integral.  In the above, note that the operators numbered
$1$ and $2$ operate on their respective ground states only.  We note in passing that
this is tantalizingly similar to the boundary states to be discussed in chapter \ref{ch:boundaries},
 in 
that it is a 
coherent superposition of independent Fock spaces.
In a similar manner it is possible to construct a state which is  a 
tensor product of three Fock spaces to represent the integral for the interaction term, schematically
given by
\beq
\langle I_3| = \int dp \left( \langle 0,p |_1 \otimes \langle 0,p |_2  \otimes \langle 0,p |_3
\right) \exp \left[ - \alpha^{(i)} N_{ij} \alpha^{(j)} - c^{(i)} X_{ij} b^{(j)} \right]
\eeq
where the oscillator indices have been suppressed, and the terms $N_{ij}$ and $X_{ij}$ are known
\cite{Taylor:2002uv} and of similar form to the terms in $\langle I_2 |$.  At this 
point, in
principle it is now possible to calculate all terms and contributions to the cubic string field
theory action.  This approach has not been applied for arbitrary 
excitations because there 
are an infinite number of terms in the expansion of the string field and 
the number of terms at each level increase very quickly.

This problem has been approached with some success \cite{Moeller:2000xv} 
using level 
truncation.  The general 
idea of this method is to include only states up to some finite level in 
both oscillators and the sum of the level numbers for those oscillators.  A complete exposition of this
interesting field of study is well beyond the scope of this discussion, however we pause to note that 
this line of research has yielded some compelling `experimental' evidence in favor of 
the famous conjectures
about the minimum energy of the open string tachyon.

\begin{table}
\begin{center}
\begin{tabular}{ c | c | c }
Level & \# Fields & $V/T_{25}$ \\
\hline
(0,0) & 1 & -0.685 \\
(2,4) & 3 & -0.949 \\
(4,8) & 10 & -0.986 \\
(6,12) & 31 & -0.995 \\
(8,16) & 91 & -0.998 \\
(10,20) & 252 & -0.999 \\
\label{table:levtrunc}
\end{tabular} 
\end{center}
\caption[Level truncation in string field theory]{Level truncation in string field theory.  
The term $V/T_{25}$ would be $-1$ to confirm Sen's conjectures. \cite{Taylor:2002uv}}
\end{table}

To illustrate the method of level truncation we perform explicitly the calculation at
the lowest level.  The potential can be written as
\beq
V = \sum d_{ij} \phi^i \phi^j + g \kappa \sum t_{ijk} \phi^i \phi^j \phi^k
\eeq
where $d_{ij}$ and $t_{ijk}$ can be calculated, $g$ is the string coupling constant
of equation \ref{comeback}, and $\kappa$ is chosen to
be $\kappa= \sqrt{3^7}/2^6$ so that $t_{111}=1$.  {\cite{TASI1} }
With this convention we find that for the ansatz
\beqn
\phi = \frac{t}{g} |0\rangle
\eeqn
we obtain
\beq
V = -\frac{1}{2} \frac{t^2}{g^2} + \kappa g \frac{t^3}{g^3}
\label{tachvacenergy}
\eeq
and inserting the value for $\kappa$ we get the equation that must be satisfied for $t$ 
to find an extremum of the string field theory action is
\beq
-t + \frac{3^{9/2}}{2^6} t^2 =0
\eeq
and solving for $t$ and substituting into (\ref{tachvacenergy}) we obtain
\beq
V = - \frac{2^{11}}{3^{10}} \frac{1}{g^2}
\eeq
and since the tension of the d-25 brane is $\frac{1}{2\pi^2 g^2}$ their ratio is 
$-0.685$ as mentioned in Table \ref{table:levtrunc}.

There is another way to examine the cubic string field theory, which is  equivalent
to that described above, but offers a more geometric picture of the construction.  It is to represent
the star product as a functional integral over the fields of the string world sheet.
Explicitly, because of the decoupling between the matter and ghost sectors of the string, it can be 
written for an open string world sheet with width $\pi$ as
\beqn
\left(\psi_1 * \psi_2 \right) \left[z(\sigma)\right]
&=& \int \prod_{0\leq \tau \leq \pi/2} dy(\tau) dx(\pi-\tau) \times
\nonumber \\
&~& \prod_{\pi/2 \leq \tilde \tau \leq \pi} \delta \left[ x(\tilde \tau) - y(\pi - 
\tilde \tau) \right] \psi_1 \left[ x(\tau) \right] \psi_2 \left[y(\tau) \right]
\label{sftproduct}
\eeqn
subject to the identification that
\beqn
x(\sigma) &=& z(\sigma),~ 0 \leq \sigma \leq \frac{\pi}{2} \nonumber \\
y(\sigma) &=& z(\sigma),~ \frac{\pi}{2} \leq \sigma \leq \pi.
\eeqn
This can be thought of as joining the right half of one string to the left half 
of another to make a single string.
\begin{figure}[tp]
  \begin{center}
\includegraphics[width=0.8\textwidth]{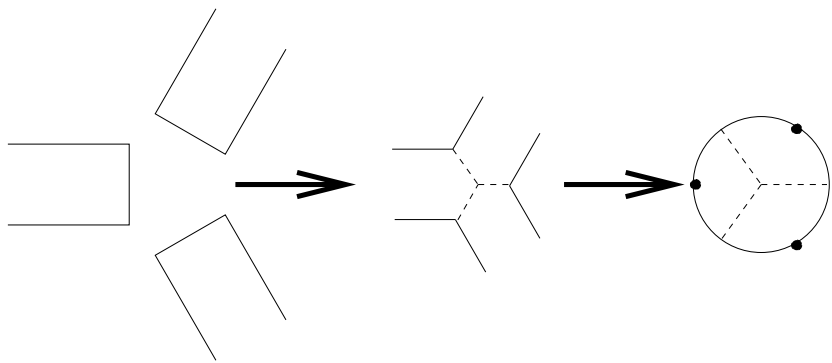}
\caption[Cubic String Field theory integration]{Cubic String Field theory integration:
this schematic diagram outlines the three string field interaction.  The string world-sheets
are conformally mapped onto the unit disk, with the boundaries forming triangular wedges as 
indicated, with identification along the boundaries.  The arrows indicate 
the two steps involved in the process, first identifying boundaries of 
the open string using the $*$ product, and then mapping the resulting world sheet to a disk
using residual conformal invariance.} 
\label{fig:cubicint}
  \end{center}
\end{figure}
A similar expression to (\ref{sftproduct}) can be found for the integral over string fields,
and again because of the decoupling the matter integral is
\beq
\int \psi = \int \prod_{0 \leq \sigma \leq \pi} dx(\sigma) \prod_{0 \leq \tilde \sigma
\leq \pi/2} \delta \left[ x(\tilde \sigma) - x(\pi - \tilde \sigma) \right] \psi\left[ x(
\tilde \sigma) \right].
\label{sftintegral}
\eeq
Similar to (\ref{sftproduct}) being thought of as gluing the left and right halves
of their respective strings together the integral (\ref{sftintegral}) 
can be thought of as gluing the left and right halves of the same string together 
with this $\delta$ function interaction.
This is illustrated for the cubic interaction term in Figure \ref{fig:cubicint},
and the residual conformal invariance of the string world sheet is used to map the 
semi-infinite string world sheets to a disk, with vertex operator insertions 
containing the asymptotic description of the open strings.   

}

\chapter{Boundary States}

\label{ch:boundaries}

In this chapter we use the boundary state formalism for both the bosonic string 
and the superstring to calculate the 
emission amplitude for closed string states from particular d-branes and show that the 
amplitudes 
are exactly those obtained from world-sheet sigma model calculations.  We find that
the construction of the 
boundary state automatically enforces the requirement for integrated vertex operators, even in the 
case of an off-shell boundary state.  Using the boundary state and a similar expansion for the 
cross-cap, we  produce higher order terms in the string loop expansion for the partition 
function of the backgrounds considered.

\section{Introduction}

The study of off-shell string theory has been addressed many times in the literature
 within the context 
of
background independent string
field theory
\cite{Witten:1992qy,Witten:1993cr,Witten:1993ed,Shatashvili:1993kk,Shatashvili:1993ps}
which 
has been the subject of a considerable amount of interest in that it can provide useful 
information about the properties of unstable d-branes \cite{Gerasimov:2000zp,
Gerasimov:2000ga,Kutasov:2000qp}.
  Despite this
there are several subtleties that have been examined, and in particular  
a great deal of effort has been 
expended
in determining an action for a tachyon field coupled to a bosonic string \cite{Gerasimov:2000zp,
Gerasimov:2000ga,Kutasov:2000qp,Akhmedov:2001jq,Gerasimov:2001pg,Kraus:2000nj,
Craps:2001jp,
Viswanathan:2001cs,Rashkov:2001pu,Alishahiha:2001tg,Andreev:2000yn,Arutyunov:2001nz,
deAlwis:2001hi}, and while 
great progress had been made the understanding of higher loop effects is incomplete at best. 

The boundary state for the superstring was first examined 
in \cite{Callan:1987px}  and the overarching idea
of the system is to produce a state that vanishes when the boundary conditions, acting as operators,
act on it.  This state is then supposed to reproduce the overlap with closed string states,
and if the surface upon which the strings end and have their boundary conditions is regarded
as a dynamical object with fields upon it, the amplitudes for the emission of various closed 
string states determine the stringy self-interactions, the brane coupling to bulk fields, and the
brane-brane interaction by string exchange.  A source for a great deal of the formalism is 
\cite{DiVecchia:1999fx,DiVecchia:1999rh}  
and this idea has been generalized to non-quadratic interactions in, among other places,
\cite{Sen:2002vv,Sen:2002in,Sen:2002nu,Rey:2003zj}.  The 
principal
focus of this chapter is to first develop the boundary state for the case of a background 
tachyon field and a background constant Abelian gauge field strength, then 
examine the effect of world sheet coordinate reparameterization
invariance upon these states, and finally to examine and explore a way in which the boundary state
could be used to generate amplitudes more complicated than simply tree level closed string exchange.
The general form of the boundary state is particularly simple for the case of quadratic boundary interactions,
precisely because they are an exactly solvable model, and while more general interactions are discussed
in Chapter \ref{ch:generalize}
 and in  \cite{Sen:2002vv,Recknagel:1998ih}, in this section the emphasis is on conformal properties, and 
for the moment
we restrict attention to the quadratic case.

Since the bosonic and fermionic world-sheet oscillators, as well as the 
conformal ghosts, do not have 
non-trivial (anti)commutation relationships it is possible to decompose the boundary state into the 
direct product of boundary states $|B\rangle$ for the bosons and fermions respectively,  
\beqn
|B\rangle = {\cal N} |B_X \rangle |B_\psi \rangle |B_{bc}\rangle |B_{\beta\gamma} \rangle
\eeqn
where ${\cal N}$ is a normalization constant which generically depends upon the various 
background fields which appear as coupling constants in the string $\sigma$ model.
In addition to this property, the $(b,c)$ and $(\beta,\gamma)$ ghosts do not interact with any of the
fields on the world-sheet boundary and so in a sense these contributions are trivial, and do not contribute
to the amplitudes other than as a multiplicative factor which reproduces the known, conformally invariant,
free case.
There are a number of constraints that the boundary state must satisfy in order to encode physical 
degrees of freedom  \cite{Callan:1987px,DiVecchia:1999rh}, specifically
\beqn
\left( Q_{BRST} + \tilde Q_{BRST} \right) |B\rangle &=& 0 
\eeqn
which is to say that it is BRST invariant.   
The strategy espoused here to determine the boundary state will be to 
examine in detail, 
in the next few sections, the bosonic string in particular backgrounds,
and then look at the fermions and ghosts in a similar manner.

\section{The Bosonic Boundary State}
\label{sec:bbs}

A tractable problem 
within this genre is the study of the off-shell theory in the background of a quadratic tachyon 
profile, a problem that is similar in spirit and detail to the examination of string theory in the 
background of a constant electromagnetic field \cite{Fradkin:1985qd}.
In the following we combine these naturally compatible studies using the 
boundary state formalism
 \cite{Bardakci:2001ck,
Lee:2001cs,Lee:2001ey,Fujii:2001qp,DiVecchia:1999fx,Akhmedov:2001yh,Laidlaw:2001jt, 
Billo:1998vr,DiVecchia:1997pr,DiVecchia:1999uf}.  
It 
allows us to calculate the probability for a topological defect which supports these quadratic 
fields to emit any number of closed string states into its bulk space-time.  The loss of conformal 
invariance introduced by the background tachyon field is naturally accommodated by a conformal 
transformation which induces a calculable change in the boundary state.  This new boundary state can 
be shown to reproduce the sigma model expectation values for the insertion of a vertex operator at 
an arbitrary point on the string world-sheet.  

Using the correspondence between the sigma model calculation and that in the operator formalism the 
question of higher genus surfaces with some number of boundaries interacting with the background 
fields is considered.  The insertion of both loops and boundaries is included naturally in this 
method, and the results obtained are compared with known results.

Throughout this work the bosonic action under consideration is
\beqn
S\left( g, F, T_0, U \right) &=&
\frac{1}{ 4 \pi \alpha'} \int_\Sigma d\rho d\phi ~ 
g_{\mu\nu}  \df^a X^\mu \df_a X_\mu 
\nonumber \\
&~& + \int_{\df \Sigma} d\phi \left(\frac{1}{2} 
F_{\mu\nu} X^\nu \df_\phi X^\mu + \frac{1}{2\pi}
T_0 + \frac{1}{8\pi} U_{\mu\nu} X^\mu X^\nu \right),
\nonumber \\
\label{action}
\eeqn
where $\alpha'$ is the inverse string tension, $\Sigma$ is the string world-sheet, $\df \Sigma$ is
the boundary of the string world-sheet, $d\rho d\phi$ 
is the integration measure of the string bulk, $d\phi$ is the integration measure of the string 
world-sheet boundary, and $\df_\phi$ is the derivative tangential to that boundary.  
This action is motivated in \cite{Witten:1993cr,Kraus:2000nj}.  The field 
content in this are a constant $U(1)$ gauge field strength $F_{\mu\nu}$ 
and the tachyon profile, 
\beqn 
T(X) = \frac{1}{2\pi}
T_0 + \frac{1}{8\pi} U_{\mu\nu} X^\mu X^\nu,
\label{tachyonprofile}
\eeqn
is characterized by a constant, $T_0$, and a constant 
symmetric matrix $U_{\mu\nu}$. 
This provides a simple generalization for the discussion given 
in 
{\cite{Craps:2001jp,Witten:1993cr}} and (\ref{eq:wittensquadtachyon}), and similarly to avoid
divergences we impose that it is positive semi-definite.

The virtue of the boundary state as a tool in the analysis of the action above is that it 
allows calculations that previously took careful integration to be reduced to algebraic 
manipulations.  We wish to carefully construct the boundary state and to show that it 
reproduces with ease the particle emission amplitudes that would be obtained from the string 
sigma model.  The starting point for this analysis is the action (\ref{action}).  By varying 
it, 
we obtain the equation
\beq
\left( \frac{1}{2 \pi \alpha'} g_{\mu\nu} \df_\rho 
 + F_{\mu\nu} \df_\phi  + \frac{1}{4\pi} U_{\mu\nu} \right) X^\nu = 0
\label{bosonicBCfields}
\eeq
as the boundary condition for the string world-sheet.  Recalling the conventions from the 
action, $\df_\sigma$ is the derivative normal to the boundary and $\df_\phi$ is the 
derivative tangential 
to the boundary.  We now create a state $| B \rangle$ that obeys the above condition as an 
operator equation.  To do this we use reparameterize the string world sheet in terms of 
holomorphic and antiholomorphic variables $z=\rho e^{i\phi}$ and $\bar z = \rho e^{-i\phi}$
and use the standard mode expansion for $X$ as a function of $z$ \ref{xexpansion}
\beq
X^\mu(z,\bar z) = x^\mu + p^\mu \ln |z^2| + \sum_{m \neq 0} \frac{1}{m} \left(
\frac{\alpha_m^\mu}{z^m} + \frac{ \tilde \alpha_m^\mu}{\bar z^m} \right).
\eeq
we find that in terms of the mode operators the boundary conditions read
\beq
\left( g + 2\pi \alpha' F + \frac{\alpha'}{2} \frac{U}{n} \right)_{\mu\nu} \alpha^\mu_n
+ \left( g - 2\pi \alpha' F - \frac{\alpha'}{2} \frac{U}{n} \right)_{\mu\nu} \tilde
\alpha^\mu_{-n} = 0.
\label{bosonicBC}
\eeq 
The condition for the boundary state to obey (\ref{bosonicBCfields})
can then be restated in terms of (\ref{bosonicBC})
to be 
\beq 
\left[\left( g + 2\pi \alpha' F + \frac{\alpha'}{2} \frac{U}{n} \right)_{\mu\nu}
\alpha^\mu_n
+ \left( g - 2\pi \alpha' F - \frac{\alpha'}{2} \frac{U}{n} \right)_{\mu\nu} \tilde
\alpha^\mu_{-n} \right] |B\rangle = 0,
\nonumber \\
\label{conditiononB}
\eeq
\beq
\left[ g_{\mu\nu} p^\mu - i \frac{\alpha'}{2} U_{\mu\nu} x^\mu \right] |B\rangle
 = 0.
\label{conditiononP}
\eeq
To satisfy this it is clear that $|B \rangle$ must be a coherent state, and it is given by
\cite{Akhmedov:2001yh}
\beqn
|B\rangle &=& {\cal N} \prod_{n\geq1} \exp  \left( -
\left( \frac{g - 2 \pi \alpha' F - \frac{\alpha'}{2} \frac{U}{n} }{
 g + 2 \pi \alpha' F + \frac{\alpha'}{2} \frac{U}{n} } \right)_{\mu\nu}
\frac{\alpha^\mu_{-n} \tilde \alpha^\nu_{-n} }{n} \right)
\nonumber \\ &~& ~~~~~~~~~~~~~~~~~~
\exp \left(- \frac{\alpha'}{4} x^\mu U_{\mu\nu} x^\nu \right) | 0 \rangle
\nonumber \\
&=&  {\cal N} \prod_{n\geq1} 
\exp  \left( - \Lambda^n_{\mu\nu} \alpha^\mu_{-n} \tilde \alpha^\nu_{-n} \right) 
\exp \left(- \frac{\alpha'}{4} x^\mu U_{\mu\nu} x^\nu \right) | 0 \rangle
\label{eq:nmbs1}
\eeqn
where ${\cal N}$ is a normalization constant which must be determined, and we define
\beq
\label{eq:nmbs2}
\Lambda^{n}_{\mu\nu} = \frac{1}{n} \left(
\frac{g - 2 \pi \alpha' F - \frac{\alpha'}{2} \frac{U}{n} }{
 g + 2 \pi \alpha' F + \frac{\alpha'}{2} \frac{U}{n} } \right)_{\mu\nu}
\eeq
for future convenience.

\subsection{Conformal Transformation of Bosonic Boundary State}

Clearly this boundary state is not conformally invariant due to the addition of the interaction 
with the tachyon field.  The two cases where we expect conformal invariance are at the two 
fixed points of renormalization group flow, namely $U=0$ and $U=\infty$,
which correspond respectively to the case of Neumann or Dirichlet boundary conditions on the 
boundary of the string world sheet \cite{Kutasov:2000qp}.  Note that in the case of Dirichlet boundary conditions the 
interaction with the background electromagnetic field is eliminated, as would be expected from 
the sigma model point of view.  
Because of this it is interesting to examine how the boundary state transforms under the 
PSL(2,R) symmetry that is broken by the presence of the $U$ term in the boundary state.  In the 
two conformally invariant cases this leaves the action invariant.
The action of PSL(2,R) on the complex coordinates of the
disk is to perform the mapping
\beq
z \rightarrow w(z) = \frac{a z + b}{\bar b z + \bar a}
\label{conftranfandnorm}
\eeq
where $a$ and $b$ satisfy the relation
\beq
|a^2| - |b^2| = 1.
\eeq
This transformation maps the interior of the unit disk to itself, the exterior
to the exterior and the boundary to the boundary.
Moreover, this transformation of the coordinates induces a
mapping which intermixes the oscillator modes.
To see this consider the definition of the oscillator modes
\beq
\alpha_m^\mu = \sqrt{ \frac{2}{\alpha'} }
\oint \frac{dz}{2\pi } z^m \df X^\mu(z)
\eeq
where the contour is the boundary of the unit disk, and the mode expansion of 
$X$ is
\beq
\df X^\mu(z) = -i \sqrt{ \frac{\alpha'}{2} } \sum_m \frac{\alpha_m^\mu}{z^{m+1} }.
\label{modeexp}
\eeq
Now, using the fact that $X$ is a scalar, or equivalently the fact that
$\df X$ is a (1,0) tensor, we see that
\beq
\alpha_m^\mu = \oint \frac{dz}{2\pi i} z^m \df_w X^\mu(w) \frac{dw}{dz}
.
\eeq
Now, using the fact that a mode expansion for $X$ exists in terms of $w$ with 
coefficients $\alpha'_m$ in exactly the same way as (\ref{modeexp}), we see that
\beq
\alpha_m^\mu = M^{(a,b)}_{mn} \alpha_n^{'\mu}
\eeq
where
\beq
M^{(a,b)}_{mn} = \oint \frac{dz}{2\pi i} z^m \frac{ (\bar b z + \bar a)^{n-1} }{
(a z + b)^{n+1} }.
\label{defofM}
\eeq  

The properties of the matrix $M$ are interesting and facilitate further study.  Some
of the properties of $M^{(a,b)}_{mn}$ are examined in Appendix \ref{sec:bosonic}.  The matrix has 
a block diagonal form so that creation and annihilation operators are not mixed by the 
conformal transformation, and with appropriate normalization of the oscillator modes it can be 
seen to be Hermitian, or equivalently that it preserves the inner product on the space of 
operators.  The exact form $M$ as a function of its indices can be easily obtained, 
but for the purposes of this discussion it is easier to to simply note that with the rescaling
${\cal M}_{mp} = \sqrt{\frac{p}{m} } M_{mp}$ for either $m,p > 0$ or $m,p < 0$ then
${\cal M}^{-1}_{mp} = {\cal M}_{mp}^\dagger$.  (The purpose of the rescaling is to normalize the
creation and annihilation operators to have the standard simple harmonic oscillator commutation
relations.)

Using this information we obtain that the modification of the boundary state 
associated with a particular conformal transformation is 
\beqn
|B_{a,b} \rangle &=& {\cal N} \exp \left( \sum_{n=1, j,k=-\infty}^\infty
\alpha^{\mu}_{-k} M^{(a,b)}_{-n-k} \Lambda^n_{\mu\nu} {\bar M^{(a,b)}_{-n-j} } 
\tilde \alpha^{\nu}_{-j}  \right) 
\nonumber \\ &~& ~~~~~~~~~~
\exp \left(- \frac{\alpha'}{4} x^\mu U_{\mu\nu} x^\nu \right)
 | 0 \rangle.
\label{transfboundstate}
\eeqn
In this equation and all following equations we drop the $'$ associated with the transformed 
oscillators for notational simplicity.  Due to the intuition gained
from the conformally invariant 
cases we propose a boundary state
\beq
|B\rangle = \int d^2 a d^2 b \delta(|a^2| - |b^2| -1) |B_{a,b} \rangle
\label{intedboundarystate}
\eeq
which we will show is the state that reproduces the sigma model amplitudes.
This is just the boundary state
(\ref{transfboundstate}) integrated over the Haar measure of  PSL(2,R).

\subsection{Boundary State Single Particle Emission}
\begin{figure}[tp]
  \begin{center}
    \includegraphics[width=0.7\textwidth]{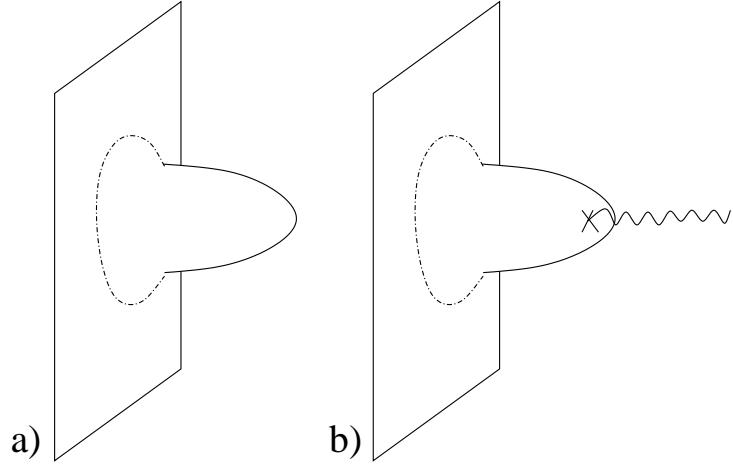}
    \caption{A schematic of the disk tadpole (a) and the emission of one particle by the boundary 
state (b).  \label{fig:bsdisk}}
  \end{center}
\end{figure}

Since we wish to show that the boundary state is an algebraic version of the action 
(\ref{action}) 
we must calculate the emission probability for various particles from the boundary state above.
This has been done in more detail in {\cite{Laidlaw:2001jt}}, (see also \cite{DiVecchia:1999uf,
DiVecchia:1997pr}) 
but we recapitulate the results here 
for 
completeness.

The case of the tachyon is straightforward.  To calculate the emission 
probability for this or any particle from the d-brane described by the 
boundary state we must evaluate the overlap of the Fock space ground 
state with the  transformed boundary state (\ref{transfboundstate}).  
Here, and in subsequent formulae we omit the momentum conserving $\delta$-functions, and the 
integration over the transformation parameters for the boundary state.
For a tachyon with momentum $p^\mu$ we find that the probability for emission from the boundary 
state is
\beqn
\langle 0, p^\mu | B_{a,b} \rangle &=& {\cal N} \exp \left( - p^\mu p^\nu \frac{\alpha'}{2}
\sum_{n=1}^\infty M^{(a,b)}_{-n 0} \Lambda^n_{\mu\nu} \bar M^{(a,b)}_{-n 0} \right)
\nonumber \\
&=& 
 {\cal N}
 \exp \left( - p^\mu p^\nu \frac{\alpha'}{2}
\sum_{n=1}^\infty \frac{1}{n} \left( \frac{ g - 
2\pi \alpha' F - \frac{\alpha'}{2} \frac{U}{n} }
{ g+2\pi \alpha' F+\frac{\alpha'}{2} \frac{U}{n} } 
\right)_{\mu\nu} \frac{ |b|^{2n} }{ |a|^{2n} } 
\right). \nonumber \\
\label{1tachyonemissioncalc}\eeqn
In the above expression we have used the previously defined form for $\Lambda^n_{\mu\nu}$,
the fact that $ M^{(a,b)}_{-n 0} = \left( \frac{ -\bar b}{\bar a} \right)^n$, and the 
conventional normalization 
$\alpha^\mu_o = \sqrt{\frac{\alpha'}{2} } p^\mu$.

Similarly, for an arbitrary massless state  with polarization tensor $P_{\mu\nu}$
and momentum $p^\mu$
\beqn
| P_{\mu\nu} \rangle &=&  P_{\mu\nu} \alpha^\mu_{-1}
\tilde \alpha^\nu_{-1} | 0, p^\mu  \rangle
\label{masslessdefn}
\eeqn
 the overlap
to be calculated is
\beqn
\langle   P_{\mu\nu} | B_{a,b} \rangle
&=&
{\cal N}
\exp
\left( - p_\mu p_\nu \frac{\alpha'}{2} \sum_{n=1}^\infty M^{(a,b)}_{-n0} 
\Lambda^n_{\mu\nu}
\bar M^{(a,b)}_{-n0} \right)
\nonumber \\
&~& P^{\mu\nu}  \left[ - \sum_{n=1}^\infty   M^{(a,b)}_{-n-1} \Lambda^n_{\mu\nu}
\bar M^{(a,b)}_{-n-1}   \right.
\nonumber \\
&~& +p^\alpha p^\beta \frac{\alpha' }{2} \sum_{n=1}^\infty   M^{(a,b)}_{-n-1}
\Lambda^n_{\mu\alpha} 
\bar M^{(a,b)}_{-n0} 
\left. \sum_{m=1}^\infty   M^{(a,b)}_{-m0}
\Lambda^m_{\beta\nu} 
\bar M^{(a,b)}_{-m-1}
\right]
\nonumber \\
&=&
 {\cal N}   
 \exp \left( - p^\mu p^\nu \frac{\alpha'}{2}
\sum_{n=1}^\infty \frac{1}{n} \left( \frac{ g - 2\pi \alpha' F - \frac{\alpha'}{2} \frac{U}{n} }
{ g+2\pi \alpha' F+\frac{\alpha'}{2} \frac{U}{n} } \right)_{\mu\nu} \frac{ |b|^{2n} }{ |a|^{2n} }
\right)
\nonumber \\
&~& 
 P^{\mu\nu}  \left[ - \sum_{n=1}^\infty n
\left( \frac{g - 2 \pi \alpha' F - \frac{\alpha'}{2} \frac{U}{n}   
 }{ g + 2 \pi \alpha' F + \frac{\alpha}{2} \frac{U}{n} } \right)_{\mu\nu}
 \frac{ |b|^{2(n-1)} }{ |a|^{2(n-1) } } \frac{1}{|a^2|^2}   \right.
\nonumber \\
&~& +p^\alpha p^\beta \frac{\alpha' }{2} \sum_{n=1}^\infty
\left( \frac{g - 2 \pi \alpha' F - \frac{\alpha'}{2} \frac{U}{n} }{
 g + 2 \pi \alpha' F + \frac{\alpha}{2} \frac{U}{n} } \right)_{\mu\alpha}
\frac{ |b|^{2(n-1)} }{ |a|^{2(n-1) } } \frac{ -\bar b }{ |a^2| \bar a }
\nonumber \\
&~&
\times \left. \sum_{m=1}^\infty
 \left( \frac{g - 2 \pi \alpha' F - \frac{\alpha'}{2} \frac{U}{m} }{
 g + 2 \pi \alpha' F + \frac{\alpha}{2} \frac{U}{m} } \right)_{\beta\nu}
\frac{ |b|^{2(m-1)} }{ |a|^{2(m-1) } } \frac{ -b}{ |a^2| a}
\right]
\label{gravitonBScalc}
\eeqn
where again the explicit form of the matrices $M$ has been used in the last equality.

This kind of argument can be repeated indefinitely on a state by state 
basis to determine the emission probability for that particular state, but 
we present here another more general
calculation which will prove useful to consider.  In particular the 
state $A$ with momentum $p^\mu$ defined by
\beqn
|A_{\mu\nu\delta} \rangle &=&  
 A_{\mu\nu\delta} \alpha_{-a}^\mu \tilde \alpha_{-b}
^\nu \tilde \alpha_{-c}^\delta |0,p^\mu \rangle,
\eeqn
has its overlap with the boundary state is given by
\beqn  
\langle  A_{\mu\nu\delta}
| B \rangle
&=&
{\cal N}
\exp
\left( - p_\mu p_\nu \frac{\alpha'}{2} \sum_{n=1}^\infty M^{(a,b)}_{-n0}
\Lambda^n_{\mu\nu}
\bar M^{(a,b)}_{-n0} \right) \times
\nonumber \\
&~&
A^{\mu\nu\delta} \sqrt{\frac{\alpha'}{2}}
 \left[ \sum_n ab M^{(a,b)}_{-n -a}    
\Lambda_{\mu\nu}^n
\bar M^{(a,b)}_{-n-b}
p^\alpha \sum_m c M^{(a,b)}_{-m 0} \Lambda_{\alpha \delta}^m \bar M^{(a,b)}_{-m-c}  \right.
\nonumber \\
&~&  + p^\alpha \sum_n ac M^{(a,b)}_{-n -a} \Lambda_{\mu\delta}^n \bar M^{(a,b)}_{-n-c}
\sum_m b M^{(a,b)}_{-m 0} \Lambda_{\alpha \nu}^m \bar M^{(a,b)}_{-m-b}
\nonumber \\ &~&
 - p^\alpha p^\beta p^\gamma 
\frac{\alpha'}{2} \sum_n aM^{(a,b)}_{-n -a} \Lambda_{\mu\alpha}^n 
\bar M^{(a,b)}_{-n0}
\sum_m b M^{(a,b)}_{-m 0} \Lambda_{\beta \nu}^m \bar M^{(a,b)}_{-m-b}
\nonumber \\ &~&
\left. \times \sum_l c M^{(a,b)}_{-l 0} 
\Lambda_{\gamma \delta}^l \bar M^{(a,b)}_{-l-c} 
\right].
\label{arbstateBScalc}
\eeqn
The summation looks formidable, but we note that the contractions of the various matrices 
look suspiciously like those of Green's functions, which it will transpire that they are, but to see 
this requires a simple calculation.  A special case of a more general formula proven in the next 
section shows that for $y = \frac{ az + b}{ \bar b z + \bar a}$ subject to $|a|^2 -|b|^2 =1$ we have that
\beq
\frac{1}{(k-1)!} \df^k z^d(y) |_{y=0} = k \bar M^{(a,b)}_{-d-k}.
\label{specialcase}
\eeq
Note that since the transformation from $z$ to $y$ is one-to-one the above equation makes sense
and is appropriate for the mapping of a point to the origin.
This completes the analysis for the emission of one particle from the boundary state 
$|B_{a,b}\rangle$, however the question becomes more interesting for the emission of more than one 
particle.

\subsection{Boundary State Multiple Particle Emission}

As in the case of emission of one particle by the boundary state it is perhaps the most instructive 
to consider the case of the emission of two tachyons first, and then specialize to more complicated 
correlators.
Ordering the operators appropriately 
for radial (as opposed to anti-radial) quantization and noting that the PSL(2,R) transformation is 
not sufficient to fix the location of both closed string vertex operators.  
Therefore it is necessary to integrate over the position of the second vertex operator.
The quantity that we will wish to compare with in the sigma model is the integration over
insertion points of an arbitrary number of vertex operators, and in this language one, the 
`bra' or `ket' appearing in the overlap equations is singled out as being moved to the origin.
We 
proceed to 
calculate, 
using the previous definitions and mode expansion
\beqn
\langle B_{a,b} | : e^{\left( i k^\mu X_\mu \right)} :
\big|_{\omega} | 0,p^\mu \rangle
&=& {\cal N} \langle 0 | \exp \left( - \sum \alpha_i^\mu M^{(a,b)}_{n i} \Lambda^n_{\mu\nu} 
\bar M^{(a,b)}_{n j} \tilde \alpha ^\nu_j \right)
\nonumber \\ &~&
\exp\left(
  k_\mu \sqrt{\frac{\alpha'}{2} } \sum_{l>0} \frac{1}{l} \left( \alpha^{\mu}_{ -l} 
\omega^l 
+ \tilde \alpha^{\mu}_{ -l}  \bar \omega^l \right)  \right)
\nonumber \\
&~& \exp \left(i k^\mu x_\nu + \sqrt{\frac{\alpha'}{2} } k_\mu \alpha^\mu_0 \ln |\omega|^2 \right)
| 0,p^\mu \rangle
\nonumber \\
&=&
{\cal N}
\exp \left( k^\mu p_\mu \frac{ \alpha'}{2} \ln |\omega|^2 \right)
\nonumber \\
&~& \exp \left( - p^\mu p^\nu \frac{ 
\alpha'}{2} \sum_{n=1}^\infty M^{(a,b)}_{n 0} \Lambda^n_{\mu\nu} 
\bar M^{(a,b)}_{n 0} 
\right) 
\nonumber \\
&~&
\exp \left( - p^\mu k^\nu \frac{ \alpha'}{2} \sum_{n=1,j=0}^\infty M^{(a,b)}_{n 0} 
\Lambda^n_{\mu\nu} \bar M^{(a,b)}_{n j}
\bar \omega^j
\right)
\nonumber \\
&~&
\exp \left( - k^\mu p^\nu \frac{ \alpha'}{2} \sum_{n=1,i=0}^\infty \omega^i M^{(a,b)}_{n i} 
\Lambda^n_{\mu\nu} 
\bar M^{(a,b)}_{n 0}
\right)
\nonumber \\
&~&
\exp \left( - p^\mu k^\nu \frac{ \alpha'}{2} 
\sum_{n=1,i,j=0}^\infty \omega^i M^{(a,b)}_{n i} 
\Lambda^n_{\mu\nu} 
\bar M^{(a,b)}_{n j}
\bar \omega^j
\right).
\nonumber \\
\label{twotachyonamp}
\eeqn 
Just as mentioned following (\ref{arbstateBScalc}), this result is 
reminiscent of a pair of 
exponentiated Green's functions.  

The next natural quantity to calculate is the emission of a more general state in place of 
either, or both tachyons in the previous calculation.  It is of course possible to demonstrate 
the overlap of an arbitrary string state explicitly, but the combinatorial nature of the 
result quickly renders the resulting expressions obscure.  With this in mind we examine the 
slightly more general state that corresponds to the calculation done in the case 
of one particle emission (\ref{arbstateBScalc}).  
\beqn
\langle B_{a,b} |  : e^{\left( i k^\mu X_\mu \right)} :
\big|_{\omega} A_{\mu\nu\delta} \alpha_n^\mu 
\tilde \alpha_p^\nu \tilde \alpha_q^\delta| 0,p^\mu \rangle
= {\cal A}_{{\cal T} 2} \times A^{\mu\nu\delta} 
\nonumber \\
\left[ \sqrt{ \frac{\alpha'}{2} }^3 
\left( - \sum n M_{rn}^{(ab) } \Lambda_{\mu \gamma}^r 
\bar M_{rj}^{(ab)} \bar \omega^j k^\gamma - k_\mu \frac{1}{\omega^n} - \sum 
n M_{rn}^{(ab) } \Lambda_{\mu \gamma}^r
\bar M_{r0}^{(ab)} p^\gamma \right) \right.
\nonumber \\
\left( - \sum k^\gamma \omega^i M_{ri}^{(ab)} \Lambda^r_{\gamma \nu} \bar M^{(ab)}_{rp} p
- k_\nu \frac{1}{\bar \omega^p} - \sum p^\gamma M_{r0}^{(ab)}  \Lambda^r_{\gamma \nu} 
\bar M^{(ab)}_{rp} p
\right)
\nonumber \\
\left( - \sum k^\gamma \omega^i M_{ri}^{(ab)} \Lambda^r_{\gamma \delta} \bar M^{(ab)}_{rq} q
- k_\nu \frac{1}{\bar \omega^q} - \sum p^\gamma M_{r0}^{(ab)}  \Lambda^r_{\gamma \delta}
\bar M^{(ab)}_{rq} q
\right)
\nonumber \\
 + 
\left( - \sum k^\gamma \omega^i M_{ri}^{(ab)} \Lambda^r_{\gamma \delta} \bar M^{(ab)}_{rq} q
- k_\nu \frac{1}{\bar \omega^q} - \sum p^\gamma M_{r0}^{(ab)}  \Lambda^r_{\gamma \delta}
\bar M^{(ab)}_{rq} q
\right)
\nonumber \\
\left. \left( - \sum n M_{rn}^{(ab) } \Lambda_{\mu \nu}^r
\bar  M^{(ab)}_{rp} p
\right) \sqrt{ \frac{\alpha'}{2} }
 \right] + \left( p\leftrightarrow q, \nu \leftrightarrow \delta \right).~~~~~
\label{twovertexbigmess2}
\eeqn
In the above $ {\cal A}_{{\cal T} 2}$ is the result for the boundary state to emit two 
tachyons, which appears as a multiplicative factor and is calculated explicitly above 
(\ref{twotachyonamp}).

Similarly it is possible to calculate the analogous expression for the vertex which emits the 
complicated state at the point $\omega $ on the disk, and using the standard commutation 
relationships as outlined previously we find
\beqn
\langle B_{a,b} | : A_{\mu\nu\delta} \frac{\df^n}{(n-1)!} X^\mu 
\frac{\ddf^p}{(p-1)!} X^\nu \frac{\ddf^q}{(q-1)!}  X^\delta 
e^{\left( i k^\mu 
X_\mu \right)} :
\big|_{\omega} | 0,p^\mu \rangle
=
{\cal A}_{{\cal T} 2} A^{\mu\nu\delta}
\nonumber \\
\Bigg[ - \left( \sum \frac{1}{(n-1)!} \frac{1}{(p-1)!} \frac{m!}{(m-n)!} 
\omega^{m-n}
M_{rm}^{(ab)} \Lambda_{\mu\nu}^r \bar M_{rj}^{(ab)} \frac{j!}{(j-p)!} 
\bar 
\omega^{j-p} \right) 
\nonumber \\ 
+ \bigg\{ \frac{\alpha'}{2} 
\left( - \sum \frac{1}{(n-1)!}  \frac{m!}{(m-n)!} \omega^{m-n}
M_{rm}^{(ab)} \Lambda_{\mu \gamma}^r \bar M^{(ab)}_{rj} \bar \omega^j k^\gamma  \right.~~~  
\nonumber \\
\left.
- \sum \frac{1}{(n-1)!}  \frac{m!}{(m-n)!} \omega^{m-n}
M_{rm}^{(ab)} \Lambda_{\mu \gamma}^r \bar M^{(ab)}_{r0} p^\gamma + p_\mu (-1)^n \omega^{-n} 
\right) ~~
\nonumber \\
\left( - \sum p^\gamma M^{(ab)}_{r0} \Lambda^r_{\gamma \nu} \bar M^{(ab)}_{rj} \frac{1}{(p-1)!}
\frac{j!}{(j-p)!}\bar \omega^{j-p} ~~~~~
\right. \nonumber \\
\left. -  \sum k^\gamma \omega^m M^{(ab)}_{rm} \Lambda^r_{\gamma \nu} \bar M^{(ab)}_{rj} 
\frac{1}{(p-1)!}
\frac{j!}{(j-p)!}\bar \omega^{j-p} +  p_\mu (-1)^p \bar \omega^{-p} \right) \bigg\} \Bigg] ~~
\nonumber \\
\times \left( - \sum p^\gamma M^{(ab)}_{r0} \Lambda^r_{\gamma \delta} \bar M^{(ab)}_{rj} 
\frac{1}{(q-1)!}
\frac{j!}{(j-q)!}\bar \omega^{j-q}        
\right.~~~~~ \nonumber \\
\left. -  \sum k^\gamma \omega^m M^{(ab)}_{rm} \Lambda^r_{\gamma \delta} \bar M^{(ab)}_{rj}
\frac{1}{(q-1)!}
\frac{j!}{(j-q)!}\bar \omega^{j-q} +  p_\mu (-1)^q \bar \omega^{-q} \right)
\nonumber \\
 + \bigg( p\leftrightarrow q, \nu \leftrightarrow \delta \bigg)~~~~~~.
\label{twovertexbigmess}
\eeqn
The above expression can be seen to be the same as that of the emission with the complicated 
vertex at the center, as the case of two tachyon emission would suggest.

\subsection{Bosonic Sigma Model}
\label{bosonicsigmamodel}
Having performed an the calculations from the point of view of the 
raising and lowering 
operators it is now instructive to compare with what should be analogous results from sigma 
model calculations.
We fix our convention that the functional integral is in all cases the
average over the action given in (\ref{action}),
\beq
\langle {\cal O}(X) \rangle = \int {\cal D} X e^{-S(X)}  {\cal O}(X).
\eeq
In addition, the Green's function on the unit disk with Neumann boundary conditions
is determined to be \cite{Hsue:1970ra} 
\beq
G^{\mu\nu}(z,z') = - \alpha' g^{\mu\nu} \left( -  \ln \left| z-z' \right| - \ln
\left| 1- z \bar z' \right| \right),
\label{diskprop}  
\eeq
and it will be useful also to know the bulk to boundary propagator
which is
\beq
G^{\mu\nu}(\rho e^{i \phi}, e^{i \phi'} ) = 2 \alpha' g^{\mu\nu}
\sum_{m=1}^\infty \frac{\rho^m}{m}
\cos[ m(\phi - \phi') ].
\label{diskbbprop}
\eeq
The boundary to boundary propagator can be read off from (\ref{diskbbprop}) as
the limit in which $\rho \rightarrow 1$.
We use $z=\rho e^{i \phi}$ as a parameterization of the points within
the unit disk, so $0 \leq \rho \leq 1$ and $0 \leq \phi <  2\pi$.
Using the bulk to boundary propagator it is possible to integrate out the quadratic
interactions on the boundary {\cite{Fradkin:1985qd}} and to obtain an exact propagator, 
which is given by
\beqn
G^{\mu\nu} (z,z') &=&
-\alpha' g^{\mu\nu} \ln \left| z - z' \right|
\nonumber \\
&~& + \frac{\alpha'}{2}
\sum_{n=1}^\infty \left( \frac{g - 2 \pi \alpha' F - \frac{ \alpha'}{2}
\frac{U}{n} }{ g + 2 \pi \alpha' F + \frac{ \alpha'}{2} \frac{U}{n} } \right)^{ 
\left\{ \mu\nu\right\} }
\frac{(z \bar z')^n + (\bar z z')^n}{n}
\nonumber \\
&~& +
\alpha' \sum_{n=1}^\infty \left( \frac{  2 \pi \alpha' F + 
\frac{ \alpha'}{2}
\frac{U}{n} }{ g + 2 \pi \alpha' F + \frac{ \alpha'}{2} \frac{U}{n} } \right)^{
\left[\mu\nu \right]}
\frac{(z \bar z')^n - (\bar z z')^n}{in}
\nonumber \\
&=&
-\alpha' g^{\mu\nu} \ln \left| z - z' \right|
\nonumber \\ &~&+ \frac{\alpha'}{2} 
\sum_{n=1}^\infty \left( \frac{g - 2 \pi \alpha' F - \frac{ \alpha'}{2}
\frac{U}{n} }{ g + 2 \pi \alpha' F + \frac{ \alpha'}{2} \frac{U}{n} } \right)^{ 
\left\{ \mu\nu \right\} }
\frac{(z \bar z')^n + (\bar z z')^n}{n}
\nonumber \\
&~& +
\frac{\alpha'}{2} \sum_{n=1}^\infty \left( \frac{ g - 2 \pi \alpha' F -
\frac{ \alpha'}{2}
\frac{U}{n} }{ g + 2 \pi \alpha' F + \frac{ \alpha'}{2} \frac{U}{n} } \right)^{
\left[\mu\nu \right] }
\frac{(z \bar z')^n - (\bar z z')^n}{i n}.
\label{bosonicgf}
\eeqn
Note that this expression is appropriately symmetric because the antisymmetry of 
Lorentz indices in the final term is compensated by the antisymmetry of the coordinate 
term.

The first calculation that must be done to determine the normalization of the 
sigma model amplitudes is the partition function.  In this approach the 
oscillator modes of $X$ must be integrated out with the contributions from $F$ and $U$
treated as perturbations.  Since both perturbations are quadratic, all the Feynman 
graphs
that contribute to the free energy can be written and evaluated, and explicitly the 
free energy is given by 
\beq
{\cal F} =
 - \sum_{m=1}^\infty Tr \ln{ \left( g + 2\pi \alpha' F + \frac{\alpha'}{2} \frac{U}{m}  
\right)  }.
\label{messyeqn}
\eeq
See \cite{Fradkin:1985qd,Laidlaw:2000kb} for further calculations done in this 
spirit.
From
(\ref{messyeqn})
we immediately obtain the partition function    
\beqn
Z &=& e^{-T_0} \prod_{m=1}^\infty \frac{1}{\det \left(
g + 2\pi \alpha' F + \frac{\alpha'}{2} \frac{U}{m} \right) }
\int dx_0 e^{- \frac{ U_{\mu\nu} }{4} x_0^\mu x_0^\nu }
\nonumber \\   
&=& \frac{1 }{\det \left( \frac{U}{2}  \right) }  e^{-T_0} \prod_{m=1}^\infty 
\frac{1}{\det \left(
g + 2\pi \alpha' F + \frac{\alpha'}{2} \frac{U}{m} \right) }.
\label{diskpf}
\eeqn
This expression is divergent, but using $\zeta$-function regularization   
{\cite{Kraus:2000nj} } it
can be reduced to
\beq
Z = e^{-T_0} \sqrt{ \det\left( \frac{g + 2\pi \alpha' F}{U/2 } \right) }
\det \Gamma \left( 1 + \frac{\alpha' U /2}{g + 2\pi \alpha' F} \right),
\label{partitionfunction}
\eeq
where $\Gamma(g)$ is the $\Gamma$ function and the dependence of all transcendental 
functions
on the matrices $U$ and $F$ is defined through a  Taylor expansion.

\subsection{Conformal Transformation in the Sigma Model}

We now wish to calculate the expectation value for vertex operators that correspond to 
different
closed string states, however this is a process that must be done with some care.
To  calculate the emission of a closed string in the world-sheet picture
one generally  considers
 a disk emitting an asymptotic closed string state.  This is really a closed string
cylinder
diagram. The standard method is to use
 conformal invariance to map the closed string state to a point
on
the disk, namely the origin, where a corresponding vertex operator is inserted.
On the other hand it has been cogently
argued that it is necessary to have an integrated vertex operator for closed string 
states to
properly couple
{\cite{Craps:2001jp}}, in particular that the graviton must be
produced by an integrated vertex operator to
couple correctly to the energy momentum tensor.
There is no  distinction between a fixed vertex operator and an integrated vertex
operator  in the conformally invariant case because the integration
will only produce a trivial volume factor, however in the case we consider
more care must be taken.  We wish to consider arbitrary locations of the
vertex operators on the string world sheet, and the natural measure to
impose is that of the conformal transformations which map the origin to
a point within the unit disk on the complex plane.

In other words we propose to allow the vertex
operator corresponding to the closed string state to be moved from the origin by a 
conformal
transformation that preserves the area of the unit disk, namely a PSL(2,R) 
transformation.
The method to accomplish this is to go to a new coordinate system
\beq
y = \frac{ a z + b }{\bar b z + \bar a},~~|a^2| - |b^2| = 1,
\eeq
where a vertex operator at the origin $y=0$ would correspond to an insertion of a vertex 
operator
at the point $z = \frac{-b}{a}$.  It is worth noting that in the case of conformal 
invariance,
that is when $U \rightarrow 0$ or $U \rightarrow \infty$ the Green's function remains 
unchanged
in form, the $y$
dependence coming  from  the replacement $z \rightarrow z(y)$. Even in the case of finite 
$U$ the only change to the Green's function is the addition of a term that is harmonic 
within the
unit disk.  The parameter of the integration over the position of the vertex operator
would be to the measure on PSL(2,R), giving an infinite factor in the conformally 
invariant
case {\cite{Shatashvili:1993kk,Craps:2001jp,Liu:1988nz}}.
From this argument we have a definite prescription
for the calculation of vertex operator expectation values, which is to use the conformal
transformation to modify the Green's function, and calculate the expectation values of 
operators
at the origin with this modified Green's function.

\subsection{Sigma Model Single Particle Emission}

Now we will use this prescription to calculate the sigma model 
expectation values of some
operators, and we will start with the simplest, that of the closed string tachyon.
The vertex operator for the tachyon is $: e^{i p_\mu X^\mu(z(y))
} :$, and it is inserted at the point $y=0$.
The normal ordering prescription for all such operators is
that any divergent
pieces will be subtracted, but finite pieces will remain and by inspection we see that
the appropriate subtraction from the Green's function is
\beqn
: {\cal G}^{\mu\nu}( z, z'): &=& G^{\mu\nu}(z,z') - g^{\mu\nu} \alpha' \ln \left| z - z'
\right|
\label{subtractedgf}
\eeqn
Using (\ref{subtractedgf}) we see that
\beqn
\langle: e^{i p_\mu X^\mu(y=0) } : \rangle &=&
 \left. Z e^{- \frac{1}{2} p_\mu p_\nu : {\cal G}^{\mu\nu}
(z(y), z'(y)) : } \right|_{y=0}
\nonumber \\
&=&
Z \exp \left( - \frac{\alpha'}{2} p_\mu p_\nu
\sum_{n=1}^\infty \left( \frac{g - 2 \pi \alpha' F - \frac{ \alpha'}{2}
\frac{U}{n} }{ g + 2 \pi \alpha' F + \frac{ \alpha'}{2} \frac{U}{n} } \right)^{\mu\nu}
\frac{ 1}{n} \frac{ |b^{2n}|}{|a^{2n}|} \right).
\nonumber \\
 \label{tachyonEV}
\eeqn
We recall that our procedure will necessitate an integration over the the parameters of 
the  PSL(2,R) transformation, but comparison with (\ref{1tachyonemissioncalc}) reveals 
that the normalization is fixed by 
\beq
{\cal N} = Z.
\eeq

Having obtained this result fixing the normalization it is natural to 
check the expectation value for other vertex operators to see if 
the relation persists.  We perform a similar analysis 
 for the massless closed string excitations.
In particular the graviton insertion at $y=0$ is given by
\beqn
\langle {\cal V}_h \rangle &=& \langle : - \frac{ 2}{\alpha'}
h_{\mu\nu} \df X^\mu \bar \df X^\nu e^{i p_\mu X^\mu(y=0)} : \rangle
\eeqn
where $h$ is a symmetric traceless tensor and the normalization follows the conventions
of { \cite{Polchinski:1998rq}. } 
This can be analyzed by the same techniques as for the tachyon, noting that there
will be cross contractions between the exponential and the $X$-field prefactors.
Explicitly we obtain
\beqn
\langle {\cal V}_h \rangle &=&
- \frac{ 2}{\alpha'} Z h_{\mu\nu} \left( \df \bar \df' : {\cal 
G}^{\mu\nu}\left(z(y),z'(y)\right) :
+ \df :{\cal G}^{\mu\alpha} \left(z(y), z'(y) \right):
\right.
\nonumber \\
&~& \left. \times \bar \df :{\cal G}^{\mu\beta} \left(z(y), z'(y) \right): (i
p_\alpha) (i p_\beta)
\right)
e^{- \frac{1}{2} p_\mu p_\nu : {\cal G}^{\mu\nu} (z(y), z'(y) ) : }|_{y=0}
\nonumber \\
&=&  
Z h_{\mu\nu}  \left( -
\sum_{n=1}^\infty \left( \frac{g - 2 \pi \alpha' F - \frac{ \alpha'}{2}
\frac{U}{n} }{ g + 2 \pi \alpha' F + \frac{ \alpha'}{2} \frac{U}{n} } \right)^{\mu\nu}
n \frac{|b^{2 (n-1)}| }{|a^{2 (n-1)}| } \frac{1}{|a^2|^2}
\right.
\nonumber \\
&~&  +
\frac{\alpha'}{2}
\sum_{n=1}^\infty \left( \frac{g - 2 \pi \alpha' F - \frac{ \alpha'}{2}
\frac{U}{n} }{ g + 2 \pi \alpha' F + \frac{ \alpha'}{2} \frac{U}{n} } \right)^{\mu\alpha}
\frac{ |b^{2(n-1)} |}{ |a^{2(n-1)} |} \frac{-b}{|a^2|a}
\nonumber \\
&~& \times \left.
\sum_{m=1}^\infty \left( \frac{g - 2 \pi \alpha' F - \frac{ \alpha'}{2}
\frac{U}{m} }{ g + 2 \pi \alpha' F + \frac{ \alpha'}{2} \frac{U}{m} } \right)^{\nu\beta}
\frac{ |b^{2(m-1)} |}{ |a^{2(m-1)} |} \frac{-\bar b}{|a^2|{\bar a}} p_\alpha p_\beta
\right)
\nonumber \\
&~&
\exp \left( - \frac{\alpha'}{2} p_\mu p_\nu
\sum_{n=1}^\infty \left( \frac{g - 2 \pi \alpha' F - \frac{ \alpha'}{2}
\frac{U}{n} }{ g + 2 \pi \alpha' F + \frac{ \alpha'}{2} \frac{U}{n} } \right)^{\mu\nu}
\frac{ 1}{n} \frac{ |b^{2n}|}{|a^{2n}|}  \right).
\label{gravitonEV}
\eeqn
and  by comparing (\ref{gravitonBScalc}) and 
(\ref{gravitonEV}) we see that the relation ${\cal N} = Z$ holds and that the form of the 
two expectation values is identical in detail.

Finally, we can perform the same kind of calculation for a more general closed string   
state, like the one considered in (\ref{arbstateBScalc}).
We consider a state which may be off shell in the sense that it not annihilated by
the positive modes of the $\sigma$-model energy momentum tensor (the Virasoro
generators), may not satisfy the
mass shell condition, and may
not be level matched.
Our explicit choice is to consider the operator
\beqn
\langle {\cal V}_A \rangle &=&
\langle : -i
\left( \frac{2}{\alpha'} \right)^{3/2}
A_{\mu\nu\delta} \frac{\df^a}{(a-1)!}  X^\mu \frac{\bar
\df^b}{(b-1)! }
X^\nu \frac{\bar \df^c}{(c-1)!} X^\gamma e^{i p_\mu X^\mu } : \rangle
\nonumber \\
\eeqn
which is an arbitrary state involving three creation operators.
We find that
\beqn
\langle {\cal V}_A \rangle &=&
 Z A_{\mu\nu\delta}
\left( \frac{2}{\alpha'} \right)^{3/2}
\left( \frac{\df^a}{(a-1)!} \frac{\bar \df'^b}{(b-1)!}
: G^{\mu\nu}(z,z') : \right.
\nonumber \\ &~& ~~~
~~~~~~~~~~~~~~\frac{\bar \df^c}{(c-1)!} :G^{\delta\alpha}(z,z'): p_\alpha
\nonumber \\
&~& +  \frac{\df^a}{(a-1)!} \frac{\bar \df'^c}{(c-1)!}
: G^{\mu\delta}(z,z') : \frac{\bar \df^b}{(b-1)!} :G^{\nu\alpha}(z,z'): p_\alpha
 \nonumber \\ &~&
 -  \frac{\df^a}{(a-1)!} : G^{\mu\alpha}(z,z') :  \frac{\bar \df^b}{(b-1)!}
:G^{\nu\beta}(z,z'): \nonumber \\ &~& ~~~~~~~~~~~~
~~~~~~\left. \frac{\bar \df^c}{(c-1)!} :G^{\delta\gamma}(z,z'):
p_\alpha p_\beta p_\gamma \right)
\nonumber \\ &~&
\left. \times e^{- \frac{1}{2} p_\mu p_\nu : {\cal G}^{\mu\nu} (z, z' ) : }
\right|_{y=0}.
\label{arbstateEV}
\eeqn
Comparing (\ref{arbstateEV}) with (\ref{arbstateBScalc}) and (\ref{specialcase}) 
we observe that the two coincide.

\subsection{Sigma Model Multiple Particle Emission}

In the previous section we demonstrated that the sigma model calculation of the
particle emission coincides with that calculated using the boundary state, so we now 
look at the emission of two particles.  
We expect that 
the two particle emission 
amplitude will depend upon the relative position of the two vertex operators 
since, even in the conformally  invariant case, there are not enough free parameters to fix two 
closed string vertex operators on the disk world sheet.
We first calculate the expectation value of the emission of two tachyons, with momenta $p$ and $k$.
\beqn
\langle  : e^{i k_\mu X^\mu} : \Big|_{\omega} 
: e^{i p_\nu X^\nu } : \Big|_0 \rangle
&=& Z \exp \left( - \frac{ k_\mu k_\nu}{2} G^{\mu\nu}\left(z(\omega), z(\omega) \right) \right)
\nonumber \\
&~& \times
  \exp \left( - \frac{ p_\mu p_\nu}{2} G^{\mu\nu} \left( z(0), z(0) \right) \right)
\nonumber \\ &~& \times
\exp \left( - \frac{ k_\mu  p_\nu}{2} G^{\mu\nu}\left( z(\omega), z(0) \right) \right)
\nonumber \\
&=&
{{\cal A}}_{{\cal T} 2 \sigma }
\eeqn

This is the necessary first step in determining a more arbitrary amplitude.
To make contact with the more complicate amplitudes calculated in (\ref{twovertexbigmess2}) and
(\ref{twovertexbigmess}) we consider
the expression
\beqn
\langle
A_{\mu\nu \delta}  : \frac{ \df^n}{(n-1)!} X^\mu \frac{ \bar \df^p }{(p-1)!} X^\nu
\frac{ \bar \df^q }{(q-1)! } X^\delta e^{i k_\mu X^\mu} : \Big|_\omega 
 : e^{ i p_\mu X^\mu} : \Big|_0 \rangle
\nonumber \\
=
{{\cal A}}_{{\cal T} 2 \sigma }
A_{\mu\nu\delta}
\left[ \left( \frac{ i k_\alpha}{(n-1)!} \df^n G^{\mu\alpha} (z(\omega), z'(\omega) )
+ \frac{i p_\alpha}{(n-1)!} \df^n G^{\mu\alpha} (z(\omega),z'(0) ) \right) \right.
\nonumber \\
\times
\left( \frac{ i k_\beta}{(p-1)!} \bar \df^p G^{\nu\beta}(z(\omega),z'(\omega) )
+ \frac{ i p_\beta}{ (p-1)!} \bar \df^p G^{\nu\beta} ( z(\omega),z'(0) ) \right)
\nonumber \\
\times
\left( \frac{i k_\gamma}{ (q-1)!} \bar \df^q G^{\delta\gamma} (z(\omega),z'(\omega) )
+ \frac{ i p_\gamma}{ (q-1)!} \bar \df^q G^{\delta\gamma}(z(\omega),z'(0) ) \right)
\nonumber \\
+
\frac{ \df^n}{(n-1)!} \frac{\bar \df^{'p}}{(p-1)!}  G^{\mu\nu}(z(\omega),z'(\omega) )
\nonumber \\
\times \left( \frac{i k_\gamma}{ (q-1)!} \bar \df^q G^{\delta\gamma}(z(\omega),z'(\omega) )
+ \frac{ i p_\gamma}{ (q-1)!}\bar \df^q G^{\delta\gamma}(z(\omega),z'(0) ) \right)
\nonumber \\
+
\frac{ \df^n}{(n-1)!}  \frac{\bar \df^{'q}}{(q-1)!} G^{\mu\delta}(z(\omega),z'(\omega) )
\nonumber \\
\times
\left. \left( \frac{ i k_\beta}{(p-1)!} \bar \df^p G^{\nu\beta}(z(\omega),z'(\omega) )
+ \frac{ i p_\beta}{ (p-1)!} \bar \df^p G^{\nu\beta} ( z(\omega),z'(0) ) \right)
\right]
\nonumber \\
\label{arbstatesigmamodel}
\eeqn
Also note that if we consider 
\beqn
\langle
 : 
e^{i k_\mu X^\mu}  : \Big|_\omega
A_{\mu\nu \delta}  : 
\frac{ \df^n}{(n-1)!} X^\mu \frac{ \bar \df^p }{(p-1)!} X^\nu
\frac{ \bar \df^q }{(q-1)! } e^{ i p_\mu X^\mu} : \Big|_0 \rangle
\nonumber 
\eeqn
we see that it gives the above expression (\ref{arbstatesigmamodel})
with $\omega \leftrightarrow 0$.

To demonstrate the general equivalence of the boundary state approach with that 
of the sigma model the sums that appear in the general expressions of boundary state
matrix elements must be shown to coincide with the expressions that appear above.
To this end consider first the sum that appears in (\ref{twotachyonamp}),
\beqn
\sum_{m=0}^\infty \omega^m M^{(a,b)}_{n m}
&=& \sum_{m=0}^\infty \oint \frac{dz}{2\pi i} \omega^m z^n \frac{ (\bar b z + \bar a)^{m-1} }{
(a z + b)^{m+1} } 
\nonumber \\
&=&
\oint \frac{dz}{2\pi i}  z^n \frac{1}{ (\bar b z + \bar a) (a-\omega  \bar b) } 
\left( z - \frac{\bar a \omega - b}{-\bar b \omega +a } \right)^{-1}
\nonumber \\
&=&
\left( \frac{ \bar a \omega - b}{-\bar b \omega + a} \right)^c.
\eeqn
This derivation uses the normalization condition on $a$ and $b$, and can be seen to be equal
to $z^n (y)$ which is the inverse transform of (\ref{conftranfandnorm}).

The other sum that appears generally in this analysis is
\beqn
\sum_{m=0}^{\infty} \frac{m!}{(m-n)!} \omega^{m-n}
M_{rm}^{(ab)}
\nonumber
\eeqn
as seen in (\ref{twovertexbigmess}).  In the case $n>m$ we have used the shorthand 
\beqn 
\frac{m!}{(m-n)!} = m(m-1) \ldots (m-n+1).
\nonumber 
\eeqn
Now consider
\beqn
\sum_{m=0}^{\infty} \frac{m!}{(m-n)!} \omega^{m-n}
M_{rm}^{(ab)}
&=& \sum_{m=0}^{\infty} \oint \frac{dz}{2\pi i} \frac{m!}{(m-n)!} \omega^{m-n}
z^r \frac{ (\bar b z + \bar a)^{m-1} }{ (az + b)^{m+1} }
\nonumber \\
&=& n! \oint \frac{dz}{2\pi i} z^r \frac{ (\bar b z + \bar a )^{n-1} }{ (a-\bar b \omega)^{n+1} }
\left( z - \frac{ \bar a \omega - b }{- \bar b \omega + a} \right)^{-n-1}
\nonumber \\
&=& \left. \df^n z^r (\bar b z + \bar a)^{n-1} (a-\bar b \omega)^{-(n+1)} \right|_{z = 
\frac{ \bar a \omega - b }{- \bar b \omega + a} }
\nonumber \\
&=& \left. \df^n \left( \frac{ \bar a z - b}{ -\bar b z + a } \right)^r \right|_{z=\omega}.
\label{inductionneeded}
\eeqn
The last equality in this can be shown by induction, the case for $n=1$ is trivial, and 
so we demonstrate the induction.
First note that use of the Leibnitz rule gives 
\beqn
\left. \df^k z^n (\bar b z + \bar a)^{k-1} (a-\bar b \omega)^{-(k+1)} \right|_{z =
\frac{ \bar a \omega - b }{- \bar b \omega + a} }
= ~~~~~~~~~
\nonumber \\
\frac{ (\bar a \omega - b)^{n-k} }{ (- \bar b \omega + a )^{n+k} } 
 \sum_{j=0}^k \left( \matrix{ k \cr j } \right) 
\frac{n!}{(n-(k-j))!} \frac{(k-1)!}{(k-j-1)!} \bar b^{j} (\bar a \omega - b)^j.
\nonumber \\
\eeqn
Now consider for $k=k_0 +1$ the expression can be manipulated
\beqn
\df^k 
\left. \left( \frac{ \bar a z - b}{ -\bar b z + a } \right)^n \right|_{z=\omega}
&=& \df \left( \df^{k_0} \left( \frac{ \bar a z - b}{ -\bar b z + a } \right)^n \right)
\nonumber \\
&=& \df \frac{ (\bar a z- b)^{n-k_0} }{ (- \bar b z+ a )^{n+k_0} }
 \sum_{j=0}^{k_0} \left( \matrix{ k_0 \cr j } \right) \frac{n!}{(n-(k_0-j))!} 
\nonumber \\ &~& ~~~~~~~~ \frac{(k_0-1)!}{(k_0-j-1)!} 
\bar b^{j} (\bar a z - b)^j
\nonumber \\
&=& \frac{ (\bar a z- b)^{n-(k_0+1)} }{ (- \bar b z+ a )^{n+(k_0+1) } }
\sum_{j=0}^{k_0} \left( \matrix{ k_0 \cr j } \right) \frac{n!}{(n-(k_0-j))!}
\nonumber \\ &~&  ~~~~~\times \frac{(k_0-1)!}{(k_0-j-1)!} \bar b^{j} (\bar a z - b)^j
\nonumber \\ &~& ~~~~~ \times 
\left( (n+j-k_0) + (2 k_0 -j) \bar b (\bar a z - b) \right)
\nonumber \\
&=&
\frac{ (\bar a z- b)^{n-(k_0+1)} }{ (- \bar b z+ a )^{n+(k_0+1) } }
\sum_{j=0}^{k_0+1} \left( \matrix{ k_0+1\cr j } \right) 
\nonumber \\ &~& ~~\frac{n!}{(n - (k_0+1-j))!}
\frac{ k_0!}{(k_0-j)!} 
\bar b^{j} (\bar a z - b)^j
\eeqn
as desired.
This demonstrates the induction step, and the validity of ({\ref{inductionneeded}}).
Note that similar results can be obtained for expressions with negative indices on $M_{nk}^{(a,b)}$
and negative powers of $\omega$.  These are obtained from considering the boundary state on the 
right of the matrix elements.  This have to be interpreted as a dual description of the boundary
states presented.  This is because radial quantization and the operator state correspondence imply
that in this case the domain of interest is the complex plane with the unit disk excluded.  This is
equally a fundamental region of the plane, and the conformal transformation between the two is
$\omega \rightarrow \frac{1}{\bar \omega}$, a fact which is intimated at by the fact that (for $n,k>0$)
$M_{-n-k}^{(a,b)} = \bar M_{nk}^{(a,b)}$.

Now we have demonstrated that the results obtained from the boundary state calculations
exactly match those of the sigma model after the propagator including the boundary perturbations
has been obtained, and the resulting expression has been transformed into a new coordinate system.
This shows that the boundary state renders all matrix elements that would otherwise be
calculated in the sigma model obtainable by
algebraic manipulations.  This observation will be important as we generalize these results
to higher genus surfaces.  We also remark that the result explicitly presented for 
the emission of two closed string states clearly generalizes to the emission of an arbitrary
number of such particles.  Mechanically this can be seen because the commutation of two such
vertex operators to produce a normal-ordered expression produces the familiar logarithmic
term, and the boundary state gives the $F$ and $U$ dependence within the inner product.

\subsection{Bosonic Boundary State Summary}

In the preceding sections we have developed the bosonic boundary state.
It is a coherent state involving the holomorphic and antiholomorphic 
creation operators which satisfies the boundary conditions associated with the 
boundary conditions (\ref{conditiononB}) and (\ref{conditiononP}), and is given by
(\ref{intedboundarystate}), the
content of which we repeat for convenience.
\beqn
|B\rangle = \int d^2 a d^2 b \delta(|a^2| - |b^2| -1) |B_{a,b} \rangle
\nonumber
\eeqn
with
\beqn
|B_{a,b} \rangle &=& Z \exp \left( \sum_{n=1, j,k=-\infty}^\infty
\alpha^{\mu}_{-k} M^{(a,b)}_{-n-k} \Lambda^n_{\mu\nu} {\bar M^{(a,b)}_{-n-j} }
\tilde \alpha^{\nu}_{-j}  \right)
\nonumber \\ &~&
\exp \left(- \frac{\alpha'}{4} x^\mu U_{\mu\nu} x^\nu \right)
 | 0 \rangle.
\nonumber
\eeqn
We have shown that this state gives the  overlap with 
an arbitrary number of closed string states in the sense that it reproduces the
string sigma model calculation of those same amplitudes.  The reason
for the integration over the  PSL(2,R) group is that in the operator state 
correspondence (a pedagogical overview of which is given in \cite{Polchinski:1998rq}) 
the external state $\langle a |$ at the end of the overlap $\langle a | B \rangle$
is defined by a limiting process which takes it to infinite world-sheet time, thereby
fixing it at the origin.  
Since the object to which this must be compared is an amplitude with
integration over  the positions of the 
inserted vertex operators it is necessary to mimic this with an integration over
PSL(2,R).  

It is also useful to note that the construction  parallels that of \cite{Ishibashi:1989kg}, 
which has as a 
boundary condition that the two dimensional conformal symmetry is not broken, and
this can be stated as 
\beqn
\left( L_n - \tilde L_{-n} \right) | B \rangle = 0
\label{conformalcond}
\eeqn
where the $L$s are the Virasoro generators.  For oscillator number $n=0$ this is nothing but the
level matching condition.  
Our results show that the level matching condition is not satisfied without the 
integration over the conformal group.  This can be seen in (\ref{arbstateBScalc}) because 
there was no condition on the indices $a,b,c$. The properties of the conformal transformation
matrices $M^{(a,b)}_{mn}$ are such that upon integration over PSL(2,R) the level matching condition
is enforced.  This is because the matrices depend upon the phases of $b$ and $a$, and
if the numbers $b$s and $\bar b$s are not matched in a particular overlap an integral of the form
$\int d\phi e^{i n \phi}$ results and vanishes.  
Upon integration at each level (\ref{conformalcond}) is satisfied, as can be seen using 
the properties derived in Appendix \ref{app:conftrans} for the matrices $M^{(a,b)}_{mn}$.

\section{Bosonic Amplitudes in the Euler Number Expansion}
\label{sec:expansion}

Since the overlap of the boundary state with either single or multiple particle
states
has been shown to coincide with that calculated in the sigma model,  we have the tools that are 
needed to proceed and determine higher order contributions 
in the string loop sense to the vacuum energy of the object described by the boundary
state.  We will proceed by utilizing a sewing construction to
relate higher order amplitudes to products of lower order amplitudes.  
The procedure outlined is envisioned to produce an arbitrary number of interactions 
with the boundary state at the oriented tree level, and an arbitrary number of handles
and interactions with the boundary state in the unoriented sector.  As is well known, the 
description of higher genus orientable surfaces is a more difficult subject and the 
construction will produce results that are implicit rather than explicit.
\begin{figure}[tp]
\begin{center}
    \includegraphics[width=0.4\textwidth]{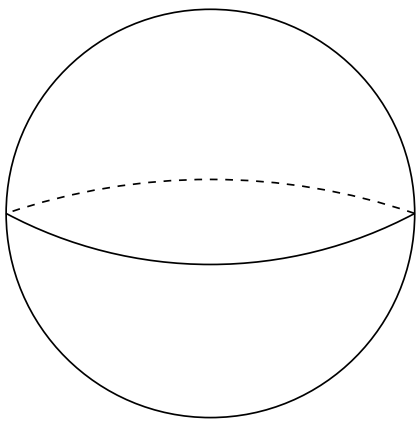}
    \caption[The Sphere]{The sphere presented schematically.  The sphere's 
contribution to the partition function is not included because it has no boundary.
 \label{fig:sphere}}
  \end{center}
\end{figure}
The final result, excluding terms without
boundary, will be several terms in the Euler number expansion so that
\beqn
{\cal Z} = Z_{disk} + Z_{ann} + Z_{MS} + \ldots
\eeqn
where $Z_{ann}$ refers to the annulus partition function, $Z_{MS}$ refers to that
of the Mobius strip, and each term 
carries the appropriate power of the open string coupling constant.  
The results in this section will be organized by Euler number, and where appropriate
compared with other similar results in the literature.

\subsection{$\chi=1$}

There are two surfaces with $\chi=1$, the disk and $RP^2$.  
\begin{figure}[tp]
\begin{center}
    \includegraphics[width=0.8\textwidth]{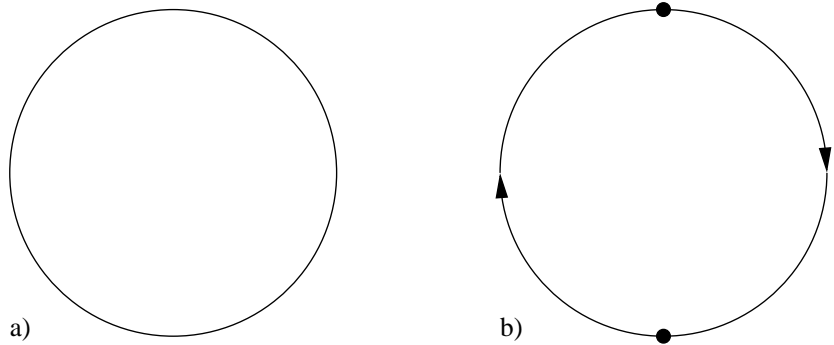}
    \caption[Surfaces with $\chi=1$]{A drawing of the two 
surfaces with $\chi=1$.  Both the disk (a) and the surface $RP^2$ (b) are shown. 
 \label{fig:chi1}}
  \end{center}
\end{figure}  
The non-orientable surface 
$RP^2$, see {\cite{Itoyama:2002dd}} for details in a similar context, has no interaction with 
the fields $F$ and $U$ and so is not of interest for this analysis.  The disk by contrast 
has been analyzed previously in this work and the contribution to the partition function
for the boundary state is given by its overlap with the unit operator (equivalently the 
tachyon with zero momentum), as given in (\ref{diskpf}).  Both the disk and
$RP^2$ are illustrated in
figure \ref{fig:chi1}.

\subsection{$\chi=0$}

There are several surfaces that have an Euler number of 0.  
\begin{figure}[tp]
\begin{center}
    \includegraphics[width=0.8\textwidth]{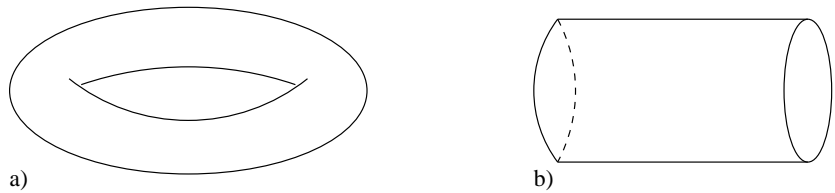}
    \caption[Orientable surfaces with $\chi=0$]{The two orientable surfaces with
$\chi=0$:  the annulus (a) and the annulus (b). The annulus is shown in a manner that
reminds its role as a closed string propagator.
 \label{fig:chi0o}}
  \end{center}
\end{figure}
The easiest to discuss in this
is the torus, which is immaterial for the same reason that $RP^2$ was among the surfaces
with $\chi=1$, namely that it has no interactions with $F$ or $U$.
Similarly the Klein bottle, the unoriented equivalent of the torus, will not contribute to
the partition function.
\begin{figure}[tp]
\begin{center}
    \includegraphics[width=0.8\textwidth]{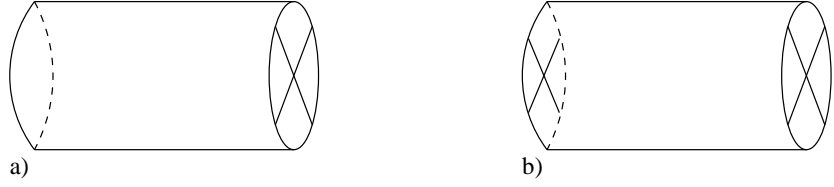}
    \caption[Non-orientable surfaces with $\chi=0$]{
The two non-orientable surfaces with $\chi=0$.  The Klein bottle (a) has no boundary, while
the Mobius strip (b) has both a cross-cap and a boundary.
 \label{fig:chi0no}}
  \end{center}
\end{figure}
We are left with the annulus and with the 
Mobius strip as the 
nontrivial 
contributions at
this level.  The annulus can be thought of as the tree level closed string exchange
channel.  The Mobius strip is the non-orientable analogue of the disk.

We consider first the annulus that was analyzed in detail in {\cite{Laidlaw:2001jt}}, we 
recapitulate some of the salient results.
Suppressing for brevity the integrations over the parameters of the conformal
transformations we have that
\beqn
 Z_{ann} = \langle B_{a,b} | \frac{1}{\Delta} | B_{a',b'} \rangle.
\eeqn
Using the integral representation of the closed string propagator
\beqn
\frac{1}{\Delta} = \frac{1}{4\pi} \int \frac{d^2 z}{|z|^2} z^{L_0 -1} \bar z^{L_0 -1}
\nonumber 
\eeqn
and suppressing the $z$ integrals we obtain
\beqn
 Z_{ann}
&=& Z_{disk}^2 \langle 0 | \exp{  \left(-\alpha^\mu_i M^{(a,b)}_{ni} \Lambda^n_{\mu\nu}
\bar M^{(a,b)}_{nj} \tilde \alpha^\nu_j \right) } 
z^{\sum \alpha_{-n} \alpha_n} \bar z^{\sum \tilde \alpha_{-n} \tilde \alpha_n}
\nonumber \\
&~&
\exp{ \left( - \alpha^\gamma_{-k} M^{(a',b')}_{-m-k} \Lambda^m_{\gamma\delta}
\bar M^{(a',b')}_{-m-l} \tilde \alpha^\delta_{-l} \right)  } | 0 \rangle
\nonumber \\
&=& Z_{disk}^2 \langle 0 | \exp{ \left( -\alpha^\mu_i M^{(a,b)}_{ni} \Lambda^n_{\mu\nu}
\bar M^{(a,b)}_{nj} \tilde \alpha^\nu_j \right) }
\nonumber \\
&~&
\exp{\left( -z^k \alpha^\gamma_{-k} M^{(a',b')}_{-m-k} \Lambda^m_{\gamma\delta}
\bar M^{(a',b')}_{-m-l} \tilde \alpha^\delta_{-l} \bar z^l \right)  } | 0 \rangle
\nonumber \\
&=&  Z_{disk}^2 F(p) \exp \left( \sum_{k=1}^\infty \frac{1}{k} g^{\mu\nu} 
\delta_{rs}
\left\{ \left[ r M^{(a,b)}_{nr} \Lambda^n_{\mu\alpha} \bar M^{(a,b)}_{nj} j \bar z^j
g^{\alpha \delta} 
\right. \right. \right.
\nonumber \\
&~& ~~~~~~~~~~~
~~~ \left. \left. \left. \bar M^{(a',b')}_{-m-j} \Lambda^m_{\nu\delta} M^{(a',b')}_{-m-s} z^s
\right]^k \right\}^{rs}_{\mu\nu} \right)
\label{annulusbyBS}
\eeqn
Verifying the first equality requires the use of the Baker-Hausdorff formula for 
commutators of exponentials, and the second equality is an application of Wick's theorem.
The term in the last exponential is understood to have its powers defined with contraction
of both the Lorentz and oscillator indices.  The number $F(p)$ is a Gaussian factor 
dependent on the (otherwise implicit) momentum of each boundary state, which can be read 
off from the boundary conditions (\ref{conditiononP}).
Explicitly the form of $F(p)$ is given by
\beqn
 F(p) &=& \exp \left\{ p^\mu p^\nu \left[ \left( 
\delta_{0j} g_{\mu\delta}- M^{(a,b)}_{n0} 
\Lambda^n_{\mu\delta}
\bar M^{(a,b)}_{nj} \right) \right. \right.
\nonumber \\
&~&
\left( \frac{1}{ \delta_{jk} g_{\delta\gamma} - j \bar z^j \bar M^{(a',b')}_{-m-j} \Lambda^m_{\eta\delta}
M^{(a',b')}_{-m-l} g^{\eta\zeta} z^l l M^{(a,b)}_{rl} \Lambda^r_{\zeta\gamma} \bar M^{(a,b)}_{rk} }
\right)^{\delta\gamma}_{jk} 
\nonumber \\
&~&
\left. \left. \left( \delta_{k0} g_{\gamma\nu} - k \bar z^k \bar M^{(a',b')}_{sk} \Lambda^s_{\nu \gamma}
M^{(a',b')}_{s0} \right) - g_{\mu\nu} \right] \right\}.
\eeqn
In addition this is multiplied by terms coming from the zero mode part of the propagator.  
In the preceding 
equations the oscillator index has been chosen as positive or zero to
make the negative signs meaningful.  In all cases, repeated indices indicate summation.

Equation \ref{annulusbyBS} is a concrete realization of the proposal of \cite{Craps:2001jp} for the 
calculation of loop corrections to the tachyon action.  This proposal calculates the tree level
couplings to closed strings for off-shell boundary interactions and shows that the correct
procedure is to use  integrated vertex operators to calculate these couplings.  It further
argues that  otherwise
the vertex operator does not couple correctly to the background fields, for instance in the case of the
graviton a non-integrated vertex operator does not couple to the standard energy momentum tensor.
By demanding closed string factorization of the one loop amplitudes \cite{Craps:2001jp} determine that
the partition function for the string amplitude with two boundaries is
\beqn
Z(S_{bdy}) = \sum_I \int dp Z(V_I(p),S_{bdy}) \frac{ f(p^2 + m_I^2)^2 }{p^2 + m_I^2 } 
Z(V_I(-p),S_{bdy}).
\label{klresult}
\eeqn
In (\ref{klresult}) $V_I(p)$ is a vertex operator for particle $I$ with momentum $p$ and 
$f(p^2 + m_I^2)$ is a function which goes to $1$ when the exchanged particle is on mass-shell, 
producing the expected poles in the particle exchange, and 
\beqn
Z(V_I(p),S_{bdy}) = \int dX e^{-S} V_I(p).
\eeqn
For the quadratic tachyon background we have shown that $\langle V_I | B \rangle$ gives $ Z(V_I(p),S_{bdy})$
so the result (\ref{annulusbyBS}) completes the summation over $I$ in (\ref{klresult}).

Considering (\ref{annulusbyBS}) we note that the cases of $U \rightarrow 0$ and $U \rightarrow \infty$ 
give a particularly
simple form for the matrices $M \Lambda \bar M$.  
We have
\beqn
\left. M^{(a,b)}_{k m} \Lambda^k_{\mu\nu} \bar M^{(a,b)}_{kn} 
\right|_{U \rightarrow 0} &=& M^{(a,b)}_{k m} \frac{1}{k} \left( \frac{ g - 2\pi \alpha' F}
{g+ 2\pi \alpha' F} \right)_{\mu\nu} \bar M^{(a,b)}_{kn}
\nonumber \\
&=& \left( \frac{ g - 2\pi \alpha' F}
{g+ 2\pi \alpha' F} \right)_{\mu\nu} \frac{1}{m} \delta_{mn},
\eeqn
and similarly
\beqn
\left. M^{(a,b)}_{k m} \Lambda^k_{\mu\nu} \bar M^{(a,b)}_{kn}
\right|_{U \rightarrow \infty}  &=&  -g_{\mu\nu} \frac{1}{m} \delta_{mn}.
\eeqn
These results can be obtained by explicit contour integration using the definition of
$M$ and are derived in Appendix \ref{app:conftrans}.  We can see that the $U=0$ case gives the 
boundary state of a background gauge field
{\cite{Lee:2001ey}} and when $U=\infty$ a localized object appears.  In fact this parameter
$U$ interpolates between Neumann and Dirichlet boundary conditions {\cite{Kraus:2000nj}}.

It is worthwhile to check the result obtained in (\ref{annulusbyBS}) in the known case where only 
the field $F$ is present.  Then the boundary conditions enforce that $p=0$, and with the above
simplification we find
\beqn
Z_{ann}(F) &=& 
\exp \left( \sum_{k=1}^\infty \frac{1}{k} g^{\mu\nu} \delta_{rs}
\left\{ \left[ r \delta_{rj} \left( \frac{g - 2\pi\alpha'F}{g + 2\pi\alpha' F}\right)_{\mu
\alpha} \frac{1}{j} g^{\alpha\delta}
\right. \right. \right.
\nonumber \\
&~& \left. \left. \left. ~~~~~~~~~~~ j \bar z^j \delta_{js} \left( \frac{g - 2\pi\alpha'F}{g + 2\pi\alpha' 
F}\right)_{\delta\nu}
 z^s
\right]^k \right\}^{rs}_{\mu\nu} \right)
 Z_{disk}^2 F(0)
\nonumber \\
&=& \exp \left( - \sum_{r=0}^\infty Tr \ln \left( g - \frac{1 + |z|^{2r} }{1- |z|^{2r} }
4 \pi \alpha' F + 4 \pi^2 \alpha'^2 F^2 \right) 
\right. 
\nonumber \\
&~&
~~~~~~ - \sum_{r=0}^\infty Tr \ln \left( g (1 - |z|^{2r})
\right)
\nonumber \\
&~& ~~~~~
\left. - \frac{1}{2} Tr \ln \left( g + 4 \pi \alpha' F + 4 \pi^2 \alpha'^2 F^2 \right) \right)
Z_{disk}^2
\nonumber \\
&=&
\prod_{r=1}^\infty (1 - |z|^{2r})^{-D} \prod_{r=1}^\infty \det \left( 
g - \frac{1 + |z|^{2r} }{1- |z|^{2r} }
4 \pi \alpha' F + 4 \pi^2 \alpha'^2 F^2 \right)^{-1}. 
\nonumber \\ 
\label{eq:gotFbs}
\eeqn
This result agrees upon the inclusion of the ghost contribution with that obtained 
in {\cite{Fradkin:1985qd}}.  Note that the partition function for the disk 
is cancelled by the term constant in $r$ which is then summed using $\zeta$ function
regularization, mimicking the calculation
of \cite{Fradkin:1985qd}
that produced the Born-Infeld action at disk level.

Now, considering the fact that, as mentioned, the field $U$ governs the 
interpolation between Neumann and Dirichlet boundary conditions and 
that we expect the space filling branes to be unstable, it 
is also interesting to examine how these expressions for
$Z_{ann}$ vary with $U$ around the two fixed points.
In particular, ignoring the linear terms in $U$ in the normalization, which
can be seen (\ref{partitionfunction}) to be divergent, the expression
for $Z_{ann}$ near $U=0$ is
\beqn
Z_{ann} &=& Z_{ann}(U=0) +
Tr\left( U \frac{\partial}{\partial U} Z_{ann}(U=0) \right)
 + \ldots
\eeqn
Immediately upon differentiation we see that the linear term will be given by
\beqn
Tr\left( U \frac{\partial}{\partial U} Z_{ann}(U=0) \right)
 &=& \int d^2a d^2b \delta( |a^2| - |b^2| -1) {\cal V}_{PSL(2,R)}
{\cal N}^2
 e^{2t}
\nonumber \\ &~&
\frac{1}{\det \left( 1-
e^{-2tb} \left( \frac{ g - 2 \pi \alpha' F}{g+2 \pi \alpha' F} \right)^2
\right)  }
\nonumber \\
&~& \times Tr\left( \frac{ (g- 2\pi \alpha' F )/(g + 2\pi \alpha' F)}{
(g + 2\pi \alpha' F)^2 - (g- 2\pi \alpha' F)^2 } U \right)
\nonumber \\ &~& \times \frac{-4 b e^{-2b} }{1 - e^{-2 b} }
\bar M^{(a,b)}_{-n-b} \frac{1}{n^2} M^{(a,b)}_{-n-b}.
\eeqn
The factor of $\frac{e^{-2b}}{ 1- e^{-2b} }$ comes from the fact that all the other
$\Lambda$ terms become trivial because we have evaluated them at $U=0$
which was noted to be conformally invariant, and from summing the terms $e^{-b}$ which
stand between these.  Likewise note that the factor $1/n^2$ instead of $1/n$ between $\bar M$ and $M$ comes
from the fact that $U$ enters always as $U/n$.
Also, one of the integrals over the PSL(2,R) groups becomes trivial, and relabeling gives the
factor ${\cal V}_{PSL(2,R)}$ and only one integral.
Now, we evaluate
\beqn
\sum_{n\geq 1} \bar M_{-n-b} \frac{1}{n^2} M_{-n-b}
&=&
\sum_{n \geq 1} \oint \frac{dz }{2 \pi i} \frac{ d \bar z}{ -2 \pi i}
\frac{1}{n^2} \frac{1}{ z^n \bar z^n }
\nonumber \\ &~&
\frac{ (\bar a \bar z + \bar b)^{b-1} }{(b \bar z + a )^{b+1} }  
\frac{ (a  z + b)^{b-1} }{(\bar b z + \bar a )^{b+1} }
\nonumber \\
&~&
\frac{1}{n!^2} \partial_z^{n-1} \partial_{\bar z}^{n-1}
\left.\frac{ (\bar a \bar z + \bar b)^{b-1} }{(b \bar z + a )^{b+1} }
\frac{ (a  z + b)^{b-1} }{(\bar b z + \bar a )^{b+1} }
\right|_{z,\bar z = 0}
\eeqn
and we find  that  when we include the
 integration over PSL(2,R) 
the expression becomes
\beqn
 \int d^2a d^2b \delta( |a^2| - |b^2| -1)
\times \sum_{n\geq 1} \bar M_{-n-b} \frac{1}{n^2} M_{-n-b}
= ~~~~~~~~~~~~~~~~~~~~~~~~~
\nonumber \\
 \int d^2a d^2b \delta( |a^2| - |b^2| -1)
\sum_{n\geq 1} \sum_{q=0}^{min( n-1, b-1)} \frac{1}{(nb)^2}
\nonumber \\ 
\left( \frac{ b+n-q-1 !}{q! n-q-1 ! b-q-1!}
\right)^2 \left( \frac{ |b^2 |}{|a^2| } \right)^{b+n-2q-2} \frac{1}{|a^2|^2}
\nonumber \\
\eeqn
and we have used the fact that upon integration over the phase of $a$ and $b$ we will
have  orthogonality
in the sum.
We find that the
contribution is
\beqn
Tr\left( U \frac{\partial}{\partial U} Z_{ann}(U=0) \right)
 &=&
\int d^2a d^2b \delta( |a^2| - |b^2| -1) {\cal V}_{PSL(2,R)}
{\cal N}^2
 e^{2t}
\nonumber \\
&~& \times \frac{1}{\det \left( 1-
e^{-2tb} \left( \frac{ g - 2 \pi \alpha' F}{g+2 \pi \alpha' F} \right)^2
\right)  }
\nonumber \\
&~&  
\times Tr\left( \frac{ (g- 2\pi \alpha' F )/(g + 2\pi \alpha' F)}{
(g + 2\pi \alpha' F)^2 - (g- 2\pi \alpha' F)^2 } U \right)
\nonumber \\ &~&
\sum_{n,m \geq 1}
\frac{-4 m e^{-2m} }{1 - e^{-2 m} }
 \sum_{q=0}^{min( n-1, m-1)} \bigg( \frac{1}{(nm)^2}  \frac{1}{|a^2|^2} 
\nonumber \\ &~&
\left( \frac{ (m+n-q-1) !}{q! (n-q-1) ! (m-q-1)!}
\right)^2
\left( \frac{ |b^2 |}{|a^2| } \right)^{m+n-2q-2}  \bigg).
\nonumber \\
\label{varabout0}
\eeqn

A similar calculation can be done around the condensate ($U \rightarrow 
\infty$ with $1/U$ the natural expansion parameter) and 
it is found
that
\beqn
Tr\left( \frac{1}{U} \frac{\partial}{\partial (1/U)} Z_{ann}\left( \frac{1}{U} = 0 \right) \right)
&=&
\int d^2a d^2b \delta( |a^2| - |b^2| -1) 
\nonumber \\ &~&
{\cal V}_{PSL(2,R)}
 \frac{{\cal N}^2
 e^{2t}}{\det \left( 1-
e^{-2tb}
\right)  }
\nonumber \\ &~&
\times 4 Tr\left(
\frac{1}{ U} \right)
\sum_{n,m \geq 1}
\frac{m e^{-2m} }{1 - e^{-2 m} }
\bar M^{(a,b)}_{-n-m} M^{(a,b)}_{-n-m}.
\nonumber \\
\label{varaboutbig}
\eeqn
Because the natural coefficient for $\frac{1}{U}$ is $n$ the $n$ dependence between the
matrices $M$ is suppressed.
Evaluations show that $
\bar M_{-n-a} M_{-n-b}$ has zero entries on diagonal, so this variation vanishes about the
condensate.
This comparison between
(\ref{varabout0}) and (\ref{varaboutbig})
shows that the case of Neumann boundary conditions, (corresponding
to $U =
0$) is unstable with respect to variations of the tachyon condensate
since the linear variation does not vanish, but that Dirichlet boundary
conditions, obtained as $U \rightarrow \infty$ are stable.  This illustrates the well known
phenomenon of tachyon condensation and gives a mechanism to see explicitly how the
open string tachyon has been removed from the excitations of the condensed state.

In a similar method we can obtain the partition function for the Mobius strip in this 
background as well.  We use the crosscap state elaborated on in {\cite{Itoyama:2002dd} }.
\beqn 
| {\cal C} \rangle = \exp \left( - \sum_{n=1}^\infty \frac{ (-1)^n}{n} \alpha_{-n} \tilde
\alpha_{-n} \right) | 0 \rangle.
\eeqn
Using this in analogy with the development of (\ref{annulusbyBS}) we find that the 
\beqn
 Z_{Mobius}
&=& \langle B_{a,b} | \frac{1}{\Delta} | {\cal C} \rangle
\nonumber \\
&=& Z_{disk} Z_{ RP^2 }
 \exp \left( \sum_{k=1}^\infty \frac{1}{k} g^{\mu\nu} \delta_{rs}
\left\{ \left[ r M^{(a,b)}_{nr} \Lambda^n_{\mu\nu} \bar M^{(a,b)}_{nj} 
\right. \right. \right.
\nonumber \\
&~& ~~~~~~~~~~~~~~~~~~ \left. \left. \left. j \bar z^j
 \frac{ (-1)^s }{s} \delta_{js} z^s
\right]^k \right\}^{rs}_{\mu\nu} \right).
\label{mobiusbyBS}
\eeqn
As in the case of the annulus, we find that the contributions $Z_{disk}$
cancel explicitly when we go to the $U=0$ limit, 
where conformal invariance
is restored.  In that limit we find
\beqn
 Z_{Mobius}(F) 
&=& 
\prod_{m=1}^{\infty} \det \left( g + F \frac{1 + (-1)^m |z|^{2m} }{ 1 - (-1)^m |z|^{2m} }
\right)^{-1}.
\eeqn

Finally it is amusing to check and make sure that an analogous calculation will go 
through and reproduce the known partition function for the Klein bottle.
Instead of the two copies of the boundary state two crosscaps are inserted, and the resulting 
expression 
\beqn
Z_{K^2} &=& Z_{RP^2}^2 \exp \left( -\sum_{r=1}^\infty g^{\mu\nu} \ln \left( g_{\mu\nu}
(1 - |z|^{2r}) \right) \right)
\eeqn
which can be seen to reduce to Dedekind $\eta$-functions \cite{Green:1987sp}, in agreement with the known result
{\cite{Polchinski:1998rq}}.

\subsection{ $\chi = -1$}

To extend it beyond $\chi=0$ the boundary state formalism requires careful contemplation.
We propose the following method which allows the construction of states of arbitrary 
Euler number, and for the non-orientable sector in principle a complete description of the
dynamics.  The procedure proposed is as follows; using the sewing construction for higher genus
amplitudes (described in \cite{Polchinski:1998rq} among others)
 and armed with the result proved 
earlier in this paper that the emission 
of any particle from the bosonic boundary state corresponds with the expectation of a
vertex operator inserted at a definite position on the disk, we propose to add any number of
interactions with the brane described by the boundary state and any number of cross caps.

To recapitulate, the idea motivating the sewing construction is to create a higher genus amplitude
by joining two lower genus amplitudes by inserting a vertex operator on each of the lower genus amplitudes
and summing over the vertex operator.
Explicitly the construction is
\beqn
\langle : A_1 : \ldots :A_n: \rangle_{M} &=& \int_\omega \sum_{V} \omega^{h_V} \bar \omega^{\tilde h_V}
\langle :A_1: 
\ldots :V: \rangle_{M_1}
\langle :V: \ldots :A_n: \rangle_{M_2} \nonumber \\
\eeqn 
with $M= M_1 \# M_2$ and $\ldots$ represents arbitrary vertex insertions.
This construction is tantamount to adding a closed string propagator between the two manifolds with 
vertex operators on them.
Since we have shown that the emission of one particle from the disk with $F$ and $U$ on its
boundary matches the overlap obtained from the boundary state
\beqn
\langle V | B_{a,b} \rangle &=& \langle :V: \rangle_{T_0,U,F}
\eeqn
we can then use this to obtain the contribution of a boundary with the fields
$U$ and $F$ at it.
This sort of construction was considered in {\cite{Callan:1987px}}.

The novel feature presented here is the generalization of the boundary state and cross-cap operators
through the state operator correspondence.  The fact that sphere amplitude for three string scattering
is conformally invariant is used, in combination with the fact that both $|{\cal C} \rangle$ and 
$|B_{a,b} \rangle$ both have a well defined overlap with any closed string state allows us to take
the expression 
\beqn
\frac{1}{\Delta} | B_{a,b} \rangle &=& \int \frac{dz d\bar z}{|z|^4} |z|^{p^2}
\exp{\left( -z^k \alpha^\gamma_{-k} M^{(a',b')}_{-m-k} \Lambda^m_{\gamma\delta}
\bar M^{(a',b')}_{-m-l} \tilde \alpha^\delta_{-l} \bar z^l \right)  } | 0 \rangle
\nonumber \\
\eeqn
and its equivalent using $|{\cal C} \rangle$ to 
(suppressing prefactors)
\beqn
\exp{\left( -z^k \frac{\df^k}{(k-1)!} X^\gamma  M^{(a',b')}_{-m-k} \Lambda^m_{\gamma\delta}
\bar M^{(a',b')}_{-m-l} \frac{\bar \df^l}{(l-1)!} X^\delta \bar z^l \right)  } 
\label{Boperatorosc}
\eeqn
by use of the operator state correspondence.  These states are inserted within 
expectation values to give higher genus contributions.

There are several different states with $\chi = -1$.  The most obvious are the four possible
\begin{figure}[tp]
\begin{center}
    \includegraphics[width=0.8\textwidth]{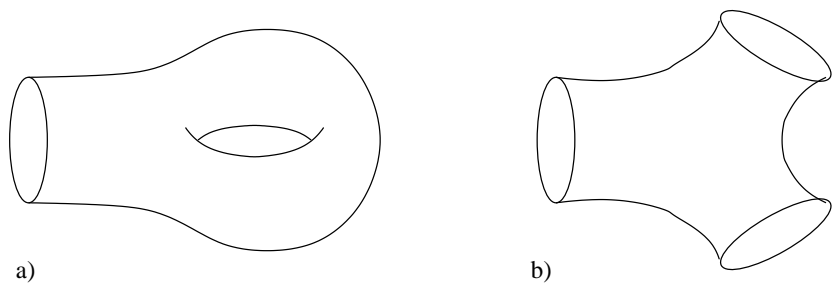}
    \caption[Orientable surfaces with $\chi=-1$]{
The two orientable surfaces with $\chi=-1$.  
They are a one-loop correction to the disk amplitude (a) and a surface with
three interactions (b), topologically equivalent to a pair of pants.
 \label{fig:chim1o}}            
  \end{center}
\end{figure}  
insertions of boundary states and 
cross caps, and the addition of a handle 
to either a boundary
state or cross cap (thereby going from $\chi =1$ to $\chi =-1$ because increasing the genus
by $1$ decreases the Euler number by $2$).  
\begin{figure}[tp]
\begin{center}
    \includegraphics[width=0.8\textwidth]{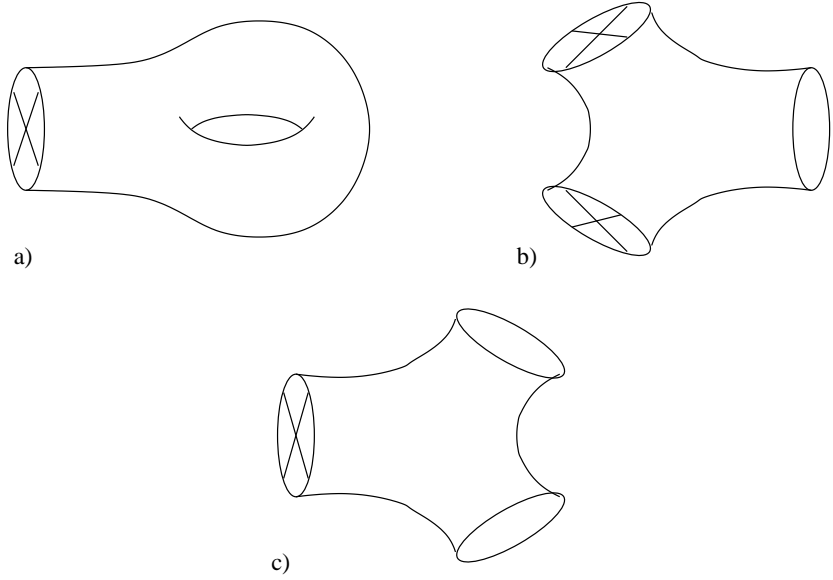}
    \caption[Non-orientable surfaces with $\chi=-1$]{
The three non-orientable surfaces with $\chi=-1$.  
 \label{fig:chim1no}}            
  \end{center}
\end{figure}  
Note that the state with three 
cross caps and the 
state with a cross cap and a handle are topologically equivalent.

To obtain the amplitude for three boundaries we calculate
\beqn
Z_{'pants'} &=& \langle B_{a,b} | \frac{1}{\Delta} :B_{a',b'}: \frac{1}{\Delta} |B_{a'',b''} \rangle
\label{eq:pants}
\eeqn
where $:B_{a',b'}:$ is as given in (\ref{Boperatorosc}). 
Noting that the coefficient of $\alpha_m$ in $\frac{\df^n}{(n-1)! } X$ is
\beqn
\frac{\df^n}{(n-1)! } X &=& \sum_{a=-\infty}^\infty D_{na} \alpha_a,
\label{deqn}
 \\
D_{na} &=& (-1)^{n-1} \frac{ (a+1) \ldots (n+a-1) }{(n-1)! },
\eeqn
we proceed to calculate
\beqn
Z_{'pants'} &=& Z_{disk}^3 
F_0 \left(p \right)
 \exp \left( \sum_k \frac{1}{k} \delta_{na}
\left( n C_{nm}(1) m C_{am}(3)
\right)^k \right)
\nonumber \\
&~&
\exp \left( \sum_k \frac{1}{k} \delta_{na} 
\left( n C_{nm}(1) m D_{n' -a} C_{n'm'}(2) 
\bar D_{m' -m} \right)^k \right)
\nonumber \\
&~& \exp \left( \sum_k \frac{1}{k} \delta_{na} 
\left( n C_{nm}(3) m D_{n' a} C_{n'm'}(2)
\bar D_{m' m} \right)^k \right) 
\nonumber \\
&~& \exp \Bigg( \sum_k \frac{1}{k} \delta_{na}
\left(  n C_{nm}(1) m D_{n' j} C_{n'm'}(2)
\bar D_{m' -m} \right. \nonumber \\ &~& ~~~~~~~~~~~~~~
\left. j C_{jk}(3) k D_{n'' -a} C_{n''m''}(2)
\bar D_{m'' k} \right)^k \Bigg)
\eeqn
Where as in (\ref{annulusbyBS}) $F_0 \left(p \right)$ is a complicated function which is
Gaussian in the momentum of the boundary state, the integrals are implicit, and the expression
$C_{nm}(i)$ is an abbreviation
\beqn
C_{nm}(i) &=& z_i^n M^{(a,b)}_{kn} \Lambda^k_{\mu\nu} \bar M^{(a,b)}_{km} \bar z_i^m
\label{ceqn}
\eeqn
with $i$ an index indicating the integration from 
which the closed string propagator $z_i$ came from.

From this we see immediately that 
the contributions for the genus expansion become increasingly 
complicated as 
$\chi$ increases.  In the particularly simple case of a vanishing 
tachyon we
obtain a product of exponentials of hypergeometric functions.  
In particular for the case of the constant $F$ field we obtain
\beqn
Z_{'pants'}(F) &=& Z_{disk}^3
\exp \left( - \sum_n Tr \ln \left( 1 - |z_1 z_3|^{2n} \left( \frac{g-2\pi\alpha' F}{g+2\pi\alpha' F}
\right)^2 \right) \right)
\nonumber \\
&~& 
\exp \left( - \sum_{na} Tr \ln \Big( 1 - n |z_1|^{2n} |z_2|^2 \right.
\nonumber \\
&~& ~~~~~~~ \left. F(-n+1,-a+1;2;|z_2|^2) 
 \left( \frac{g-2\pi\alpha' F}{g+2\pi\alpha' F}
\right)^2 \Big) \right)
\nonumber \\
&~&
\exp \left( - \sum_{na} Tr \ln \Big( 1 - n |z_3|^{2n} |z_2|^2 \right. 
\nonumber \\ 
&~& ~~~~~~~ \left. F(n+1,a+1;2;|z_2|^2)
 \left( \frac{g-2\pi\alpha' F}{g+2\pi\alpha' F}
\right)^2 \Big) \right)
\nonumber \\
&~&
\exp \left( - \sum_{nma} Tr \ln \left( 1 - n |z_1|^{2n} |z_2|^2 F(-n+1,m+1;2;|z_2|^2)
\right. \right. \nonumber \\
&~& \left. \left. m |z_3|^{2m} |z_2|^2 F(m+1,-a+1;2;|z_2|^2)
 \left( \frac{g-2\pi\alpha' F}{g+2\pi\alpha' F}
\right)^2 \right) \right).
\nonumber \\
\eeqn
In the above $F(a,b;c;x)$ is the hypergeometric function defined by its series expansion
\beqn
F(a,b;c;x) &=& \sum_{n=0}^\infty \frac{ (a+n-1)! (b+n-1)! (c-1)! }{ n! (a-1)! (b-1)! (c+n-1)! }
x^n,
\eeqn
and the logarithm is interpreted as its series expansion, and both Lorentz and oscillator 
indices are summed over.
Note that this expression has many of the properties that we expect for the partition function
on a twice punctured disk.  In particular this depends on three parameters (the $z_i$ terms arising
from the integration over the propagators to the various boundary states) which can be 
identified as the Teichmuller parameters for this surface \cite{Schiffer:1954}.  
In the limit of any of these parameters going to zero the dominant contribution is from the 
annulus amplitude.  The 
analogous amplitude with any number of cross-caps gives a similar 
expression with the following modifications, for each cross-cap the argument in the 
hypergeometric expression acquires a negative sign, and the corresponding matrix of 
Lorentz indices undergoes the substitution $\frac{ g- 2\pi \alpha' F}{g+2\pi\alpha' F}
\rightarrow g$.

The other two diagrams that must be calculated are the corrections to the disk and to $RP^2$ which
come from the addition of a handle.  This addition is achieved by taking the trace, weighted by 
a factor exponentiated to the level number (coming from the propagator within the handle), 
which is an 
identical operation to taking the expectation 
value of this operator on the torus.
For this calculation it is necessary to take the trace of an operator that 
generically has the normal ordered form
\beqn
:\exp \left( - \alpha_n {\cal M}_{nm} \tilde \alpha_m \right) :
\nonumber 
\eeqn
where the indices on ${\cal M}$ can be either positive or negative, with ${\cal M}$ defined
by
\beqn
{\cal M}_{mn} &=& D_{n'm} C_{n'm'} \bar D_{m'n}.
\eeqn
After a considerable amount of algebra we find by summing over all states in 
the Fock space that
\beqn
Tr \left(
\omega^h \tilde \omega^{\tilde h} :\exp \left( - \alpha_n {\cal M}_{nm} \tilde \alpha_m 
\right) : 
\right)
&=& \prod_{n=1}^\infty \frac{1}{\left| 1-\omega^{h_i} \right|^2 } 
\times 
\nonumber \\
&~&
\prod_{n=1}^\infty \frac{ 1}{ 1 - \left( 
\frac{ |a| \omega^{h_a} }{ 1- \omega^{h_a} }
{\cal M}_{ab} \frac{ |b| \tilde \omega^{\tilde h_b} }{ 1 - \tilde \omega^{\tilde h_b} }  
{\cal M}_{-c-b}  \right)^n }. 
\nonumber \\
\label{trace}
\eeqn
This expression uses the convention that the sums within the denominator run over positive
and negative indices.  This suppresses the contribution from the momentum of the loop which is
given by a Gaussian,
\beqn
F(p) &=&  \exp \Bigg\{ p p \Bigg[ \left(
\delta_{0j} - {\cal M}_{0j} \frac{ |j| \tilde \omega^{h_j} }{ 1- \tilde \omega^{h_j} } 
\right) 
\left( \frac{1}{ \delta_{kj} - 
{\cal M}_{kj} \frac{ |k| \omega^{h_k} }{ 1- \omega^{h_k} }
{\cal M}_{kl} } \right)
\nonumber \\
&~&
~~~~~~~~~~~ \left( \delta_{0l}  - {\cal M}_{0l} 
\frac{ |l| \tilde \omega^{h_l} }{ 1- \tilde \omega^{h_l} } \right) -1  \Bigg] \Bigg\}.
\eeqn
The specialization to the case of only interactions with a background $F$ field is given by
substituting $|z|^2 F(a+1,b+1;2;|z|^2)$ for ${\cal M}_{ab}$.

It is interesting at this point to compare the results for this procedure with those
obtained by the standard method of constructing the Green's function on an arbitrary surface
\cite{Schiffer:1954}, and then integrating out the boundary interaction as described previously
(\ref{bosonicgf}).  The Green's function of a unit disk with Neumann boundary conditions with 
a puncture of radius $\epsilon$ at $z=0$ and a puncture of radius $\delta$ at $z = r e^{i \psi}$
is given by 
\beqn
G'(z,z') &=& G(z,z') + \left( \ln \epsilon \right)^{-1} G(z, 0) G(z',0)  
\nonumber \\
&~& + \left( \ln \delta\right)^{-1} G(z, r e^{i \psi}) G(z', r e^{i \psi})
\nonumber \\
&~&
- Re  \left( 4 \delta^2 \left( \frac{1}{ z - r e^{i \psi} } + \frac{\bar z}{1- \bar z r e^{i \psi} }
\right) 
\right. \nonumber \\
&~& ~~~~~~~~~~~~
~~~~~\times
 \left. \left(  \frac{1}{ \bar z' - r e^{-i \psi} } + \frac{z'}{1-z' r e^{-i \psi} } \right) \right)
\nonumber \\
&~& - Re \left( 4 \epsilon^2 \left( z^{-1} + \bar z \right) \left( \bar z'^{-1} + z' \right)
\right) + O(\epsilon^2) + O(\epsilon^2 \delta^2) + O(\delta^2).
\nonumber \\
\eeqn
In the above the explicit form of the Green's function for the disk (\ref{diskprop}) has been 
substituted into the last two lines.
Integrating out the background field $F$ can be done by recasting this as a one dimensional 
$3\times3$ matrix model.  When this is done the interaction with a field on the boundary
can be integrated out, much as was done for the $2 \times 2$ case in \cite{Fradkin:1985qd}, and
the resulting expression contains the lowest order terms (in the Teichmuller parameter) 
of the hypergeometric  functions obtained previously.  Similarly there is  a procedure for obtaining
the Green's function for the disk with a handle added between balls of radius 
$\epsilon$ centered at $z=0$ and $z=r e^{i \psi}$.  This
gives
\beqn
G'(z,z') &=& G(z,z') + \left( \ln \epsilon \right)^{-1} \left( G(z,0) - G(z,r e^{i \psi}) \right)
\left( G(z',0) - G(z',r e^{i \psi}) \right)
\nonumber \\
&~&  - Re \left[ 4 \epsilon^2 \left( z^{-1} + \bar z \right) 
\left( \frac{1}{ z' - r e^{i \psi} } + \frac{\bar z'}{1- \bar z' r e^{i \psi} }
\right)
\right.
\nonumber \\
&~& \left.
+  \left( z'^{-1} + \bar z' \right) 
\left( \frac{1}{ z - r e^{i \psi} } + \frac{\bar z}{1- \bar z r e^{i \psi} }
\right) \right] + O(\epsilon^2) .
\eeqn
As in the case of the disk with holes removed, this Green's function can be then used to integrate 
out the quadratic perturbation, obtaining results that are consistent with those presented 
in (\ref{trace}).

\begin{table}
\begin{center}
\begin{tabular}{ | c | c | }
\hline
(b,h) & Z \\
\hline
(3,0) & $
 \int \exp \Bigg[ - \sum_n Tr \ln \left( 1 - |z_1 z_3|^{2n} \left( \frac{g-2\pi\alpha' 
F}{g+2\pi\alpha' F}
\right)^2 \right)  $
 \\
 & $
 - \sum_{na} Tr \ln \left( 1 - n |z_1|^{2n} |z_2|^2 F(-n+1,-a+1;2;|z_2|^2)
 \left( \frac{g-2\pi\alpha' F}{g+2\pi\alpha' F}
\right)^2 \right)  $
 \\
 & $
- \sum_{na} Tr \ln \left( 1 - n |z_3|^{2n} |z_2|^2 F(n+1,a+1;2;|z_2|^2)
 \left( \frac{g-2\pi\alpha' F}{g+2\pi\alpha' F}
\right)^2 \right)  $
 \\
 & $
- \sum_{nma} Tr \ln \left( 1 - n |z_1|^{2n} |z_2|^2 F(-n+1,m+1;2;|z_2|^2)
\right.  $
\\
 & $
\left. m |z_3|^{2m} |z_2|^2 F(m+1,-a+1;2;|z_2|^2)
 \left( \frac{g-2\pi\alpha' F}{g+2\pi\alpha' F}
\right)^2 \right) \Bigg]  $
\\
\hline
(1,1)& $ \int \prod_{n=1}^\infty \frac{1}{\left| 1-\omega^{h_i} \right|^2 } $
 \\
 & $ 
\prod_{n=1}^\infty \frac{ 1}{ 1 - \left(
\frac{ |a| \omega^{h_a} }{ 1- \omega^{h_a} }
|z|^2 F(a+1,b+1;2;|z|^2)
 \frac{ |b| \tilde \omega^{\tilde h_b} }{ 1 - \tilde \omega^{\tilde h_b} }
|z|^2 F(-c+1,-b+1;2;|z|^2) \left( \frac{g - 2\pi \alpha' F}{g+2 \pi \alpha'F} \right)^2
  \right)^n 
} $ 
 \\
 & $
\exp \Bigg[ p p \left[ \left(
\delta_{0j} - |z|^2 F(1,j+1;2;|z|^2)  \left( \frac{g - 2\pi \alpha' F}{ 
g+2 \pi \alpha'F} \right) 
\frac{ |j| \tilde \omega^{h_j} }{ 1- \tilde \omega^{h_j} }
\right)
\right.  $ 
 \\
 & $
\left( \frac{1}{ \delta_{kj} -
|z|^2 F(k+1,j+1;2;|z|^2)  \left( \frac{g - 2\pi \alpha' F}{g+2 \pi \alpha'F} \right)^2
 \frac{ |k| 
\omega^{h_k} }{ 1- \omega^{h_k} }
|z|^2 F(k+1,l+1;2;|z|^2) } \right) $
 \\
 & $  \left. \left( \delta_{0l}  - |z|^2 F(1,l+1;2;|z|^2) 
\left( \frac{g - 2\pi \alpha' F}{g+2 \pi \alpha'F} 
\right)
\frac{ |l| \tilde \omega^{h_l} }{ 1- \tilde \omega^{h_l} } \right) -1  \right] \Bigg] $ 
\label{table:mooren}
\\ \hline
\end{tabular} 
\end{center}
\caption[Orientable Surfaces $\chi=-1$]{The partition functions for the 
orientable surfaces with $\chi=-1$ in the case of $U=0$.  The number of 
boundaries and handles (b,h) is listed.}
\end{table}
\begin{table}
\begin{center}
\begin{tabular}{ | c | c | }
\hline
(b,c)& Z \\
\hline
(2,1) & $
 \int \exp \Bigg[ - \sum_n Tr \ln \left( 1 - |z_1 z_3|^{2n} \left( \frac{g-2\pi\alpha' 
F}{g+2\pi\alpha' 
F}
\right)^2 \right) 
$ \\
 & $
- \sum_{na} Tr \ln \left( 1 + n |z_1|^{2n} |z_2|^2 F(-n+1,-a+1;2;-|z_2|^2)
 \left( \frac{g-2\pi\alpha' F}{g+2\pi\alpha' F}
\right) \right) 
$ \\
 & $
 - \sum_{na} Tr \ln \left( 1 - n |z_3|^{2n} |z_2|^2 F(n+1,a+1;2;-|z_2|^2)
 \left( \frac{g-2\pi\alpha' F}{g+2\pi\alpha' F}
\right) \right)
$ \\
 & $
 - \sum_{nma} Tr \ln \left( 1 - n |z_1|^{2n} |z_2|^2 F(-n+1,m+1;2;-|z_2|^2)
 \right. $ \\
 & $ \left. m |z_3|^{2m} |z_2|^2 F(m+1,-a+1;2;-|z_2|^2)
 \left( \frac{g-2\pi\alpha' F}{g+2\pi\alpha' F}
\right)^2 \right) \Bigg]
$ \\
(1,2)& $ \int \exp \Bigg[ - \sum_n Tr \ln \left( 1 - (-1)^n 
|z_1 z_3|^{2n} \left( \frac{g-2\pi\alpha' 
F}{g+2\pi\alpha'
F}
\right) \right) 
$ \\
 &
$ - \sum_{na} Tr \ln \left( 1 + n |z_1|^{2n} |z_2|^2 F(-n+1,-a+1;2;-|z_2|^2)
 \left( \frac{g-2\pi\alpha' F}{g+2\pi\alpha' F}
\right) \right) 
$ \\
 &
$- \sum_{na} Tr \ln \left( 1 - n (-1)^n |z_3|^{2n} |z_2|^2 F(n+1,a+1;2;-|z_2|^2)
 \right) 
$ \\
 &
$  - \sum_{nma} Tr \ln \left( 1 - n |z_1|^{2n} |z_2|^2 F(-n+1,m+1;2;-|z_2|^2)
\right.  $ 
\\
 & $ \left. m (-1)^m |z_3|^{2m} |z_2|^2 F(m+1,-a+1;2;-|z_2|^2)
 \left( \frac{g-2\pi\alpha' F}{g+2\pi\alpha' F}
\right) \right) \Bigg] $
\label{table:monoren} 
\\ \hline
\end{tabular}
\end{center}
\caption[Non-Orientable Surfaces $\chi=-1$]{The partition functions for the
non-orientable surfaces with $\chi=-1$ in the case of $U=0$.  The number of
boundaries and crosscaps (b,c) is listed.}
\end{table}

\subsection{ $\chi = -2$}

\begin{figure}[tp]
\begin{center}
    \includegraphics[width=0.8\textwidth]{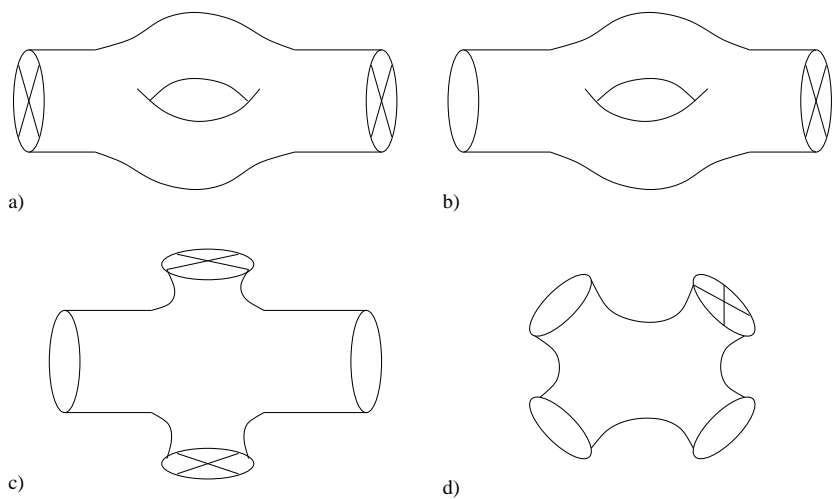}
    \caption[Non-orientable surfaces with $\chi=-2$]{
The four non-orientable surfaces with $\chi=-2$.  
 \label{fig:chim2no}}            
  \end{center}
\end{figure}

As the surfaces increase in complexity there are an increasing number of orientable and non-orientable
surfaces with boundary at each Euler number, and consequently a larger number of amplitudes to calculate.
\begin{figure}[tp]
\begin{center}
    \includegraphics[width=0.8\textwidth]{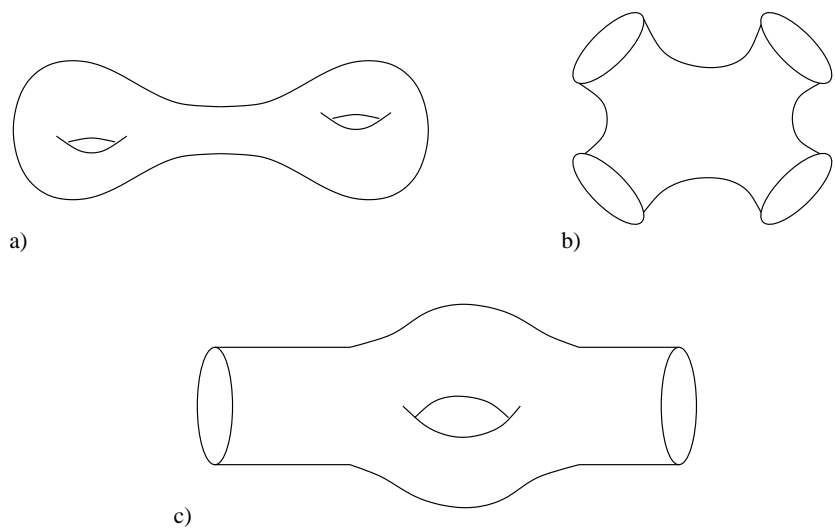}
    \caption[Orientable surfaces with $\chi=-2$]{
The three orientable surfaces with $\chi=-2$.  The double torus (a) has no boundary, the surface with
two boundaries (c) can be thought of as a one loop correction to the annulus, and the surface with four
boundaries (b) is topologically a shirt (in the same spirit that \ref{fig:chim1o}b is a pair of pants)  
 \label{fig:chim2o}}            
  \end{center}
\end{figure}  
The surfaces in question are 
illustrated in Figures \ref{fig:chim2o} and \ref{fig:chim2no}
 and can be described as follows.  
Among 
orientable surfaces there are three, one with two loops and no boundaries, one with one loop
and two boundaries, corresponding to a `one-loop' modification of a string propagator, and a surface
with four boundaries and no loops, which by analogy with the exposition in previous sections can be thought
of as a tree level interaction between 4 separate D-branes.  Similarly in the non-orientable sector
there are a number of different possibilities, and they are most simply classified keeping in 
mind the fact that the insertion of two cross-caps can be exchanged for a loop on a non-orientable
surface.  There is a surface with four cross-caps, the non-orientable analog of the two loop orientable
graph, a surface with three cross-caps and one boundary, which will be the higher loop generalization of
the interaction with the Mobius strip, the surface with two boundaries and two cross-caps which is the 
non-orientable contribution to the one-loop modification of the closed string propagator, and a surface with
three boundaries and one cross-cap.  We will examine each of these surfaces in turn, as in the previous
sections.

Once again the contributions of first listed surfaces, both in the orientable and nonorientable sectors
can be ignored in our investigation of the partition function for the tachyon field.  As before this
is because these surfaces do not interact with this because they have no boundaries.

Next we consider the surfaces with no handles and at least one boundary.  (This is actually
all of the surfaces except for Figure \ref{fig:chim2o}c because surfaces with two crosscaps are 
equivalent to a non-orientable surface with a handle.)  In analogy with (\ref{eq:pants}) we calculate
for Figure \ref{fig:chim2o}b with the partition function given by 
\beq
Z_{'shirt'} = \int \langle B_{a_1,b_1} | \frac{1}{\Delta}
: B_{a_2,b_2} : \frac{1}{\Delta}
: B_{a_3,b_3} : \frac{1}{\Delta}
| B_{a_4,b_4} \rangle.
\eeq
Using the convention from the $\chi=-1$ case we implicitly assume that the internal $B$s have
a propagator inside of them, and can be
then written using \ref{Boperatorosc} which gives
\beq
: B_{a_i,b_i} : = 
\exp{\left( - \alpha_p D_{np} C_{nm}(i) \bar D_{mq} \tilde \alpha_q \right)},
\eeq
with 
\beqn
C_{nm}(i) &=& z_i^n M^{(a_i,b_i)}_{kn} \Lambda^k_{\mu\nu} \bar M^{(a_i,b_i)}_{km} \bar z_i^m
\eeqn
as in (\ref{ceqn}), and $D$ is as described in (\ref{deqn}).
Using this input, and noting that there is one factor of $\frac{1}{\Delta}$ that
is not accounted for and whose parameter $z$ we give the
subscript $5$ to, we can recast the expression for this amplitude as
\beqn
Z_{'shirt'} &=& \langle \tilde B(1) | :\tilde B(2): 
:\exp \left( - \alpha_k z_5^{-k} D_{-nk} C_{nm}(3) \bar D_{-mj} \bar z_5^{-j} \alpha_j \right)
: |\tilde B(5\times 4) \rangle
\nonumber \\
\eeqn
where the term $\tilde B$ denotes the inclusion of the propagator, and $5\times4$ in this case 
denotes the multiplication of $z_5$ and $z_4$ which only appear in the last term in the 
combination $z_5 z_4$, so we rescale, absorbing $z_4$ into the normalization of $z_5$.
Upon performing the calculations we find 
\beqn
Z_{'shirt'} &=& Z_{disk}^4 F'_0(p)
\exp \left(  \sum_k \delta_{na} \left( n C_{nm}(1) m C_{am}(5) \right)^k \right)
\nonumber \\
&~& \exp \left(  \sum_k \delta_{na} \left( n C_{nm}(1) m D_{n'-a}C_{n'm'}(3) \bar D_{m' -m}
\right)^k \right)
\nonumber \\
&~& \exp \left(  \sum_k \delta_{na} \left( n C_{nm}(1) m D_{n'-a}C_{n'm'}(2) \bar D_{m' -m}   
\right)^k \right)
\nonumber \\
&~& \exp \left(  \sum_k \delta_{na} \left( n D_{n''n} C_{nm}(2)\bar 
D_{m''m} m D_{n'-a}C_{n'm'}(3) \bar D_{m' -m}
\right)^k \right)
\nonumber \\
&~& \exp \left(  \sum_k \delta_{na} \left( n D_{n''n} C_{nm}(2)\bar 
D_{m''m} m C_{am}(5)
\right)^k \right)
\nonumber \\
&~& \exp \left(  \sum_k \delta_{na} \left( n D_{n''n} C_{nm}(3)\bar 
D_{m''m} m C_{am}(5)               
\right)^k \right)
\nonumber \\
&~& \exp \bigg( \sum_k \frac{1}{k} \delta_{na}
\left(  n D_{qn} C_{qp}(2) \bar D_{pm} m D_{n' j} C_{n'm'}(3)
\bar D_{m' -m} \right. 
\nonumber \\ &~& ~~~~~
\left. j C_{jk}(4) k D_{n'' -a} C_{n''m''}(3)
\bar D_{m'' k} \right)^k \bigg)
\nonumber \\
&~& \exp \bigg( \sum_k \frac{1}{k} \delta_{na}
\left(  n C_{nm}(1) m D_{n' j} C_{n'm'}(2)
\bar D_{m' -m} \right.
\nonumber \\
&~& ~~~~~ \left. j D_{j'j} C_{j'k'}(3)\bar D_{k'k} k D_{n'' -a} C_{n''m''}(2)
\bar D_{m'' k} \right)^k \bigg)
\nonumber \\
&~& \exp \bigg( \sum_k \frac{1}{k} \delta_{na}
\Big(  n C_{nm}(1) m 
\nonumber \\ &~& ~~~~~~
\left( D_{n' j} C_{n'm'}(2)
\bar D_{m' -m} + D_{n' j} C_{n'm'}(3)
\bar D_{m' -m} \right)  
j C_{jk}(4) k 
\nonumber \\ &~& ~~~~\left(
D_{n'' -a} C_{n''m''}(2)
\bar D_{m'' k} + D_{n'' -a} C_{n''m''}(3)
\bar D_{m'' k} \right) \Big)^k \bigg)
\nonumber \\
&~& \exp \bigg( \sum_k \frac{1}{k} \delta_{na}
\left(  n C_{nm}(1) m D_{n' j} C_{n'm'}(2)
\bar D_{m' -m} D_{n' j} C_{n'm'}(3)
\bar D_{m' -m} 
\right. \nonumber \\ &~& ~~~ \left. 
j C_{jk}(4) k
D_{n'' -a} C_{n''m''}(3)
\bar D_{m'' k} D_{n'' -a} C_{n''m''}(2)
\bar D_{m'' k} \right)^k \bigg)
\eeqn
This is clearly a lot more complicated than the corresponding result for the three boundary case.
This can be generalized to the case of any number of crosscaps by substituting into the expression
for $C$ the term $\sum_n \frac{(-1)^n}{n} \delta^{\mu\nu}$ in the place of
$\sum_n \frac{1}{n} \left( \frac{g - 2\pi \alpha' F - \frac{\alpha'}{2} \frac{U}{n} }{g +
2\pi \alpha' F +\frac{\alpha'}{2} \frac{U}{n} }\right)^{\mu\nu}$.

There remains only one general type of diagram to be concerned with, and that is the torus amplitude
with two boundaries.  This can be also constructed in the following manner, which we choose to emphasize the
factorization properties \cite{Callan:1987px,Craps:2001jp}, 
because the torus can be thought of as the exchange of 
two 
closed string propagators.  With this in mind we propose the following construction, each of the
two arms of the torus is thought of as a closed string propagator occupying its own Fock space, and the
operators that make up the two boundary states that form the two ends of this surface are allowed to 
be in either of the Fock spaces, and we average over all possible choices.  This can be thought of as allowing
the excitations from the boundary state to propagate along either of the two closed string propagators, and 
averaging over all possible choices in in analogy with the general spirit of path integrals.
Then the partition function will be
\beqn
Z &=& \int Z_{disk}^2 \langle 0,p | \prod_{i,j=1,2} 
 \exp{  \left(-\alpha^{\mu i}_{n'} z_1^{n'} M^{(a,b)}_{nn'} \Lambda^n_{\mu\nu}
\bar M^{(a,b)}_{nm'} \bar z_1^{m'} \tilde \alpha^{\nu j}_{m'} \right) }
\nonumber \\
&~& ~~~~~y_1^{\sum \alpha^1_{-n} \alpha^1_n} \bar y_1^{\sum \tilde \alpha^1_{-n} \tilde \alpha^1_n}
y_2^{\sum \alpha^2_{-n} \alpha^2_n} \bar y_2^{\sum \tilde \alpha^2_{-n} \tilde \alpha^2_n}
\nonumber \\
&~& 
~~\prod_{i',j'=1,2}
\exp{ \left( - \alpha^{\gamma i'}_{-k} z_2^k M^{(a',b')}_{-m-k} \Lambda^m_{\gamma\delta}
\bar M^{(a',b')}_{-m-l} z_2^l \tilde \alpha^{\delta j'}_{-l} \right)  } | 0,p \rangle
\label{twobdyann}
\eeqn
where the superscript on the $\alpha$ operators in addition to the Lorentz index indicates the
Fock space to which it belongs.  Now, (\ref{twobdyann}) can be evaluated giving
\beqn
Z &=& \int Z_{disk}^2 \langle 0,p | \prod_{i,j=1,2}
 \exp{  \left(-\alpha^{\mu i}_{n'} z_1^{n'} M^{(a,b)}_{nn'} \Lambda^n_{\mu\nu}
\bar M^{(a,b)}_{nm'} \bar z_1^{m'} \tilde \alpha^{\nu j}_{m'} \right) }
\nonumber \\
&~&
\prod_{i',j'=1,2}
\exp{ \left( - \alpha^{\gamma i'}_{-k} y_i^k z_2^k M^{(a',b')}_{-m-k} \Lambda^m_{\gamma\delta}
\bar M^{(a',b')}_{-m-l} z_2^l y_j^l \tilde \alpha^{\delta j'}_{-l} \right)  } | 0,p \rangle
\eeqn

\subsection{$\chi = -3$}

As for $\chi =-2$ there are a number of different surfaces of this genus that can be obtained with
the insertion of handles, cross-caps, and boundaries.  The method presented above provides a concrete
proposal for the construction of these higher genus amplitudes for all $\chi \le -1$.  The construction
is particularly appropriate for what can be interpreted as tree level scattering amplitudes for
an arbitrary number of closed strings emitted from the brane described by the boundary state.

\subsection{Beyond the Born-Infeld Action}

In the preceding we have further explored the bosonic boundary state formalism \cite{Callan:1987px}
 and discussed its 
extension to the off-shell case including interaction with a tachyon field of quadratic
profile.  The boundary state has been shown to reproduce the $\sigma$ model calculations 
for emission of any number of closed string states, as detailed in the correspondence
\beqn
\langle V_1 | :V_2: \ldots |B_{a,b} \rangle &=& \langle :V_1: :V_2: \ldots \rangle_{T_0,U,F}.
\eeqn
This can be restated as the fact that the boundary state  
encodes the bosonic string 
propagator in an algebraic manner.

It has been shown that the inner product of two of the boundary states also reproduces the
$\sigma$ model calculations for a world-sheet of the appropriate genus.
We also present a generalization of this to higher genus, the results of which become progressively
more complicated.
In the case of vanishing tachyon field we obtain the following  expansion in the 
open string coupling constant $g_o$
\beqn
Z_{F} &=& \sum_\chi g_o^\chi Z_\chi \nonumber \\
&=& g_o^{-1} \sqrt{ \det \left( g + 2 \pi \alpha' F \right) }
\nonumber \\
&~& + \int \prod_r \left( 1 - |z^2|^r \right)^{-D} \prod_r \det \left(
g- \frac{1+ |z^2|^r}{1-|z^2|^r} 4 \pi\alpha' F + 4 \pi^2 \alpha'^2 F^2 \right)^{-1}
\nonumber \\
&~& + \int \prod_r \left( 1 - (-1)^r |z^2|^r \right)^{-D} \prod_r \det \left(
g- \frac{1+ (-1)^r |z^2|^r}{1-(-1)^r|z^2|^r} 2\pi \alpha' F \right)^{-1}
\nonumber \\
&~& + g_o^1 (Z_{o30} + Z_{o11} + Z_{n21} + Z_{n12} ) \nonumber \\
&~& + O(g_o^2)
\eeqn
where $Z_{xij}$ is the partition function given in Table \ref{table:mooren}
or \ref{table:monoren} for orientable $x=o$ and non-orientable $x=n$
surfaces with $i,j$ boundaries and handles (or boundaries and cross-caps if appropriate).
This is a generalization of the Born Infeld action taking into account higher loop stringy corrections,
specifically including contributions from Euler number $\chi=-1$ in addition
to the $\chi=1$ and $\chi=0$ terms previously in the literature, and including the contributions from
non-orientable surfaces such as the Mobius strip.
The construction presented can be generalized to higher 
genus with particular success in the case of the sphere with a number of boundaries and cross-caps added.	
It quickly becomes apparent that the simplifications obtained by the method of encoding the Green's
function in the boundary state are  overwhelmed by the increase in the parameters associated with the
various boundary states.

\section{Fermionic Boundary State}

Despite the details shown in the previous sections the fermionic contribution to the 
boundary states is in fact more involved than that for the $X$ fields.  This stems in part
from the fact that the fermions have a more involved world-sheet action, involving the Ramond and 
Neveu-Schwarz sectors corresponding to different boundary conditions for the fermions
\cite{Polchinski:1998rr}.  Another complication
that will appear briefly is that there are 
branch cuts in the integrals that define the matrices relating the the oscillators before and after
a conformal transformation, and this introduces some subtlety of treatment, however, that is for
the Ramond sector fermions, whose zero modes make them inappropriate for the study of
tachyon condensation \cite{Kutasov:2000aq}, 
especially considering that the lowest lying states in that sector are
bosonic.

We start as in the bosonic case with the consideration of the world-sheet 
action  \cite{Kutasov:2000qp,Arutyunov:2001nz,Alishahiha:2001tg} 
\beqn
S_{ferm} = \int_M \left( \psi^\mu_+ \df_- \psi^\nu_+ + \psi^\mu_- \df_+ 
\psi^\nu_- \right)
+  \oint_{\df M} F_{\mu\nu} 
\left( \psi^\mu_+ \psi^\nu_+ - \psi^\mu_-
\psi^\nu_- \right) + \nonumber \\ 
U_{\mu\nu} 
\left( \psi^\mu_+ \frac{1}{\df_\phi} \psi^\nu_+ - \psi^\mu_- 
\frac{1}{\df_\phi}
\psi^\nu_- \right) 
\label{fermionicaction}
\eeqn
where as in the case of the $X$ fields there is a boundary interaction 
with a constant gauge field, and the term involving the tachyon profile
$U$ is a simple generalization of the result in \cite{Marino:2001qc}, and is 
appropriate to the NS sector since that is the sector with the tachyon, as 
well as that the fermions not having zero modes renders the inverse
integral well defined
\beqn
\frac{1}{\df_\phi} \psi(\phi) = \frac{1}{2} \int d\phi' 
\epsilon(\phi-\phi') \psi(\phi')
\eeqn
where $\epsilon$ is a step function: $\epsilon(x) =1$ for $x>0$ and  
$\epsilon(x)=-1$ for $x<0$.
Combining the previously mentioned expansion for $\psi_+$ and $\psi_-$ 
with and expanding as in {\cite{DiVecchia:1999fx,DiVecchia:1999rh}} we obtain the boundary conditions
which must be satisfied
\beqn
\left(
g+ 2\pi\alpha' F + \frac{\alpha'}{2} \frac{U}{n} \right)_{\mu\nu} 
\psi_n^\nu + i \eta
\left(
g - 2\pi\alpha' F - \frac{\alpha'}{2} \frac{U}{n} \right)_{\mu\nu} 
\tilde \psi_{-n}^\nu
=0
\eeqn
where the factor of $i$ comes from the conformal rotation of the 
world-sheet coordinates, and $\eta = \pm 1$ will accomplish the 
GSO projection with the selections {\cite{DiVecchia:1999fx,DiVecchia:1999rh}}
\beqn
2 | B_\psi \rangle = | B_+ \rangle - | B_- \rangle.
\label{eq:fermionicbs}
\eeqn
where $|B_\pm \rangle$ are the coherent states that satisfy the 
boundary conditions with the corresponding positive or negative value for 
$\eta$, explicitly
\beqn
|B_\pm \rangle = {\cal N}_{f} \exp \left[  \pm i \sum \psi^\mu_{-n} \chi^n_{\mu\nu}
\tilde \psi^\nu_{-n} \right] | 0 \rangle
\eeqn
with
\beqn
\chi^n_{\mu\nu} = \left( 
\frac{ 
g - 2\pi\alpha' F - \frac{\alpha'}{2}
\frac{U}{n} 
}{
g + 2\pi\alpha' F + \frac{\alpha'}{2}
\frac{U}{n} } \right)_{\mu\nu}
\eeqn
Note that in the case of the Ramond sector the above difference becomes a 
sum and there is no contribution from the tachyon profile $U$, 
and also that $\chi$ is closely related to the bosonic term 
$\Lambda$.
We specialize this discussion to the Neveu-Schwarz sector, as that is the 
case which can draw the parallel with the discussion in the bosonic 
sector.

Now we examine how the conformal transformation which redefines world-sheet 
coordinates  (\ref{conftranfandnorm})  acts on the ($\frac{1}{2},0$) degrees of 
freedom.
Using the standard mode expansion \cite{Green:1987sp,Polchinski:1998rr} the relationship between
modes before and after transformation is
\beqn
\psi_m &=& N_{mn} \psi'_n \nonumber \\
N_{mn} &=& \oint \frac{dz}{2\pi i} z^{m-1/2} \frac{ (\bar b z + \bar 
a)^{n-1/2} }{ (az+b )^{n+1/2} } 
\label{eq:defofN}
\eeqn
The expression for $N$ also contains an arbitrary
 phase that comes from the choice 
of branch for the square root of the Jacobean for the transformation, 
which can be ignored because in all cases we deal with bilinears in 
this, and also a relative sign can be absorbed in the definition of 
$\eta$.  An examination of the properties of $N$, as well as its bosonic 
partner are found in Appendix \ref{app:conftrans}. 

\subsection{Particle Emission from Fermionic Boundary State}

Now, in analogy with the development we can calculate the emission amplitude for 
a state in the NS-NS sector by taking the overlap with the appropriate element of the
Fock space.  As the development here is very similar to that in the bosonic case 
we only present a representative sample of the possible overlaps.  First the 
massless state, corresponding to among other things the graviton, which is given by
\beqn
|P_{\mu\nu} \rangle = P_{\mu\nu} \psi^\mu_{-1/2} \tilde \psi^\nu_{-1/2} | 0 \rangle \nonumber 
\eeqn
within the Fock space.
The overlap of this with the boundary state given above is then
\beqn
\langle P_{\mu\nu} | B \rangle &=& \int d^2a d^2b \delta(|a^2| - |b^2| -1) 
i {\cal N}_f  P^{\mu\nu} N^{(a,b)}_{m 1/2} \chi^m_{\mu\nu} \bar N^{(a,b)}_{m 1/2}
\nonumber \\
&=& \int d^2a d^2b \delta(|a^2| - |b^2| -1)
 i {\cal N}_f  P^{\mu\nu} \chi^m_{\mu\nu} \frac{ |b|^{2m-1} }{ |a|^{2m+1} }
\eeqn
with the bosonic part implicit and calculated previously  
(\ref{1tachyonemissioncalc}).  By contrast
the overlap with a state with higher number of excitations is somewhat
longer.
An example is to consider the state 
\beqn
|{\cal P}_{\mu\nu\alpha\beta} \rangle = {\cal P}_{\mu\nu\alpha\beta}
\psi^\alpha_{-1/2} \psi^\mu_{-1/2} \tilde \psi^\beta_{-1/2} 
 \tilde \psi^\nu_{-1/2} | 0 \rangle \nonumber
\eeqn
where there are the obvious symmetry and antisymmetry relations
between the indices.  The overlap is then
\beqn
\langle {\cal P}_{\mu\nu\alpha\beta}| B \rangle &=&
- \int d^2a d^2b \delta(|a^2| - |b^2| -1)
{\cal P}^{\mu\nu\alpha\beta} \nonumber \\
&& ~~~\bigg( N^{(a,b)}_{m 1/2} \chi^m_{\mu\alpha} \bar N^{(a,b)}_{m 1/2}
N^{(a,b)}_{n 1/2} \chi^n_{\nu\beta} \bar N^{(a,b)}_{n 1/2}
\nonumber \\ && ~~~- N^{(a,b)}_{m 1/2} \chi^m_{\nu\alpha} \bar N^{(a,b)}_{m 1/2}
N^{(a,b)}_{n 1/2} \chi^n_{\mu\beta} \bar N^{(a,b)}_{n 1/2}
\bigg)
\eeqn
The expressions for the matrices $N^{(a,b)}_{m 1/2}$ can be found in Appendix \ref{sec:fermionic}.

\subsection{Particle Emission in the Superstring Sigma Model}

Now, we pursue the analogy with the bosonic case further by calculating the 
disk amplitude for emission of the corresponding 
particle.  We start by mentioning the two point functions
for the NS fermions on the disk in the free case, which are
respectively
(with $G_\psi(z,w) = \langle \psi(z) \psi(w) \rangle$ and the obvious notation
for the conjugate fields
\beqn
G_\psi(z,w) = \frac{\alpha'}{i} \left( \frac{ \sqrt{ z w} }{z-w} 
- \frac{ \sqrt{ z \bar w} }{1 - z \bar w} \right)
\\
\tilde G_\psi(z,w) = \frac{\alpha'}{i} \left( - 
\frac{ \sqrt{ \bar z \bar w } }{\bar z - \bar w} 
+ \frac{ \sqrt{ \bar z w} }{1 - \bar z  w} \right)
\eeqn
Note in passing that these reproduce the well known expression
\cite{Kutasov:2000aq} for the correlators of fermions on
the boundary of the world-sheet, and when we parameterize $z=e^{i \phi}$ and $w = e^{i \phi'}$ we
obtain as the sum of the holomorphic and antiholomorphic propagators
\beqn
\langle \psi(\phi) \psi(\phi') \rangle + \langle \tilde \psi(\phi) \tilde \psi(\phi') \rangle
= \frac{ -2}{ \sin \left( \frac{ \phi - \phi'}{2} \right) }
\eeqn
in agreement with \cite{Kutasov:2000aq}. 
Just as in the case of the bosonic fields it is possible to integrate out the
boundary interactions and obtain the modified propagator which satisfies 
the boundary conditions, obtaining, now including the Lorentz indices,
\beqn
\frac{i}{\alpha'} G_\psi(z,w)^{\mu\nu}  &=& \frac{ \sqrt{ z w} }{z-w} 
g^{\mu\nu} 
\nonumber \\
&&  - \sum_{r \in \mathbb{Z}+1/2 >0} 
\left( \frac{g - 2 \pi \alpha' F - \frac{\alpha'}{2}\frac{U}{r}  }{
g +2 \pi \alpha' F +\frac{\alpha'}{2}\frac{U}{r}  } 
\right)^{\left\{\mu\nu\right\}} \mathrm{Im} \left( z \bar w \right)^r
\nonumber \\
&&  - \sum_{r \in \mathbb{Z}+1/2 >0}
\left( \frac{g - 2 \pi \alpha' F - 
\frac{\alpha'}{2}\frac{U}{r}  }{
g +2 \pi \alpha' F +\frac{\alpha'}{2}\frac{U}{r}  }
\right)^{\left[\mu\nu\right]} \mathrm{Re} \left( z \bar w \right)^r              
\label{eq:fermholo}
\eeqn
and the corresponding expression for the conjugate fields
\beqn
\frac{-i}{\alpha'} G_\psi(z,w)^{\mu\nu}  &=& 
\frac{ - \sqrt{ \bar z \bar w} }{\bar z- \bar w}
g^{\mu\nu} 
\nonumber \\
&&
+ \sum_{r \in \mathbb{Z}+1/2 >0}
\left( \frac{g - 2 \pi \alpha' F - \frac{\alpha'}{2}\frac{U}{r}  }{
g +2 \pi \alpha' F +\frac{\alpha'}{2}\frac{U}{r}  }
\right)^{\left\{\mu\nu\right\}} \mathrm{Im} 
\left( \bar z w \right)^r                      
\nonumber \\
&&
+ \sum_{r \in \mathbb{Z}+1/2 >0}  
\left( \frac{g - 2 \pi \alpha' F - \frac{\alpha'}{2}\frac{U}{r}  }{
g +2 \pi \alpha' F +\frac{\alpha'}{2}\frac{U}{r}  } 
\right)^{\left[\mu\nu\right]} 
\mathrm{Re} \left( \bar z w \right)^r
\label{eq:fermantiholo}
\eeqn
which reproduce the results from {\cite{Arutyunov:2001nz,Viswanathan:2001cs}}.  
In a similar way the partition 
function
from the fermions can be evaluated to obtain
\beqn
Z_\psi = \prod_{r \in \mathbb{Z}+1/2 >0} \det \left( g + 2\pi \alpha' F + \frac{\alpha'}{2} \frac{U}{r} \right)
\eeqn
We note that in the case of vanishing tachyon profile $U$ the disk level partition functions 
of the bosons and
the fermions on the world-sheet the two partition functions cancel each other in agreement with 
\cite{Bachas:1992bh}.

Preliminaries aside we may now calculate the expectation value of the vertex operator 
corresponding to the state discussed previously.  From the point of view of a purely
sigma model calculation, the vertex operator from the $(-1,-1)$ picture 
\beqn
| P_{\mu\nu} \rangle \rightarrow : P_{\mu\nu} \psi^\mu \tilde \psi^\nu e^{i k X} :
\eeqn
will vanish under path integral averaging.
In contrast to this however, the fact that this system is annihilated by 
the BRST charge operator $Q$ suggests that the $(0,0)$ picture is more appropriate
in any case, and the corresponding vertex operator is
\beqn
| P_{\mu\nu} \rangle \rightarrow : P_{\mu\nu}
\left( \df X^\mu + i k_\alpha \psi^\alpha \psi^\mu \right)
\left( \ddf X^\nu + i k_\beta \tilde \psi^\beta \tilde \psi^\nu \right) e^{i k X} :
\eeqn
and so averaging we find
\beqn
\langle : P_{\mu\nu}
\left( \df X^\mu + i k_\alpha \psi^\alpha \psi^\mu \right)
\left( \ddf X^\nu + i k_\beta \tilde \psi^\beta \tilde \psi^\nu \right) e^{i k X} :
\rangle = ~~~~~~~~~~~~~~~~~
\nonumber \\
 P_{\mu\nu} \left( \df \ddf G_X^{\mu\nu} -  k_\alpha k_\beta
\df G_X^{\mu\alpha} G_X^{\beta \nu} + i k_\alpha \df G_X^{\mu\alpha}  
i k_\beta \tilde G_\psi^{\beta\nu} + i k_\beta \ddf G_X^{\mu\beta}  i 
k_\alpha G_\psi^{\alpha \nu} \right. \nonumber \\
 \left.- k_\beta \tilde G_\psi^{\beta\nu}
k_\alpha G_\psi^{\alpha \nu} \right) \exp \left( - \frac{1}{2} k_\mu k_\nu G_X^{\mu\nu} \right)
\eeqn
where the Green's functions can be evaluated from expressions (\ref{eq:fermholo}) 
and (\ref{eq:fermantiholo}).
For the bosonic parts of this expression it has already been shown that the boundary state
encodes the interaction content of the sigma model, and for the fermionic degrees of freedom there are relevant
calculations that
can be found in
Appendix \ref{app:gf}.
The coincidence of this with the calculation from the boundary state is another 
independent check of the boundary state giving the correct overlap with closed
string states, which is now for the perturbative superstring.

\subsection{Euler Number Expansion for Fermions}

To follow the same analysis for the fermions in Euler number expansion as for the bosons,
the same steps are necessary.  First the observation is repeated that the sphere and 
$RP^2$ do not have any interactions with these boundary fields and so are not of interest
in constructing a stringy action for these fields on the brane, and as before the
disk case has been calculated explicitly.  Paralleling the development before we 
can interpret the annulus amplitude as either a tree level self interaction
diagram for the brane fields, or in the case of distinct branes as a single particle exchange.

So, for the case $\chi=0$ we take
 the fermion boundary state for the NS sector and 
the overlap given by 
\beqn
\langle B_\psi | \frac{1}{\Delta} | B_\psi \rangle &=&
\exp \left( \sum_k \frac{1}{k} Tr \left( \left[ N^{(a,b)}_{nr} \chi^n_\mu\alpha \bar N^{(a,b)}_{nj} \bar z^j 
g^{\alpha\gamma} \right. \right. \right.
\nonumber \\ 
&~& \left. \left. \left. ~~~~~~~~~~~~~~~~
\bar M^{(a',b')}_{-m-j} \chi^m_{\nu\delta} M_{-m-s}^{(a',b')} z^s \right]^k \right)_{\mu\nu}^{rs}
\right) \eeqn
which in the case of vanishing tachyon gives the opposite contribution to (\ref{eq:gotFbs}) 
as seen explicitly
in the calculations \cite{Bachas:1992bh}.

\section{Ghosts and Antighosts}

For completeness, we now mention the ghost and antighost systems, but since they
do not couple to the boundary interactions, the discussion will be brief. (see 
\cite{DiVecchia:1999fx,DiVecchia:1999rh} 
for a more detailed discussion)  
It has been mentioned before that the boundary state is annihilated by the
operator $Q_{BRST} + \tilde Q_{BRST}$.   Expanding $Q$ in terms of ghosts
$b$ and $c$, this condition together with the 
known form of the boundary state for the bosonic coordinates leads to the
conditions
\beqn
\left( c_n + \tilde c_{-n} \right) |B_{bc} \rangle &=& 0 \\
\left( b_n - \tilde b_{-n} \right) |B_{bc} \rangle &=& 0.
\eeqn
Due to the anticommutation relationships between $b$ and $c$,
\beqn
\left\{ b_n , c_m \right\} &=& \delta_{m+n,0}
\\
\left\{ b_n , b_m \right\} &=& \left\{ c_n , c_m \right\} =0
\eeqn
it is immediately possible to see that this coherent state is given by
\beqn
 |B_{bc} \rangle &=& \exp \left( \sum_n c_{-n} \tilde b_{-n} + \tilde c_{-n} b_{-n} \right)
\frac{1}{2} (c_0 + \tilde c_0) |0\rangle
\label{eq:bcbs}
\eeqn
where $ |0\rangle$ is a state which is annihilated by $c_n$ for $n \geq 1$ and by 
$b_n$ for $n \geq 0$.

Similarly the antighosts arise for the case of the superstring, and we mention here those
appropriate for the NS sector, as that was where the tachyon field caused interest.  
The superghosts contribute to the energy momentum tensor of the string as do all the other
fields, and by decomposing the $Q_{BRST}$ 
into its components and defining $\eta$ as in the fermionic case 
the $\beta \gamma$ modes relate according to 
\beqn
\left( \gamma_n + i \eta \tilde \gamma_{-n} \right) |B_{\beta \gamma} \rangle &=& 0 \\
\left( \beta_n + i \eta \tilde \beta_{-n} \right) |B_{\beta \gamma} \rangle &=& 0
\eeqn
and this gives a superghost boundary state as
\beqn
|B_{\beta \gamma} \pm \rangle &=& \exp \left( \pm i  \sum_{n \in \mathbb{Z} + 1/2 >0}
 \gamma_{-n} \tilde \beta_{-n}
- \beta_{-n} \tilde \gamma_{-n} \right)
\label{eq:sgbs}
\eeqn
because the commutators for the $\beta \gamma$ 
system are 
\beqn
\left[ \beta_n , \gamma_m \right] &=& \delta_{m+n,0}
\\
\left[ \beta_n , \beta_m \right] &=& \left[ \gamma_n , \gamma_m \right] =0.
\eeqn

\section{Summary}

While the boundary states for the ghosts and antighosts are well known 
\cite{DiVecchia:1999fx,DiVecchia:1999uf}, we have developed in this chapter
the boundary states for both bosons and fermions corresponding to the 
boundary string field theory actions
\beqn
S\left( g, F, T_0, U \right) &=&
\frac{1}{ 4 \pi \alpha'} \int_M  ~
\df^a X^\mu \df_a X_\mu
+  \psi^\mu_+ \df_- \psi^\nu_+ + \psi^\mu_- \df_+
\psi^\nu_-
\nonumber \\
&~& + \oint_{\df M} \left(\frac{1}{2}
F_{\mu\nu} X^\nu \df_\phi X^\mu + \frac{1}{2\pi}
T_0 + \frac{1}{8\pi} U_{\mu\nu} X^\mu X^\nu \right),
\nonumber \\
&~&
+  \oint_{\df M} F_{\mu\nu}
\left( \psi^\mu_+ \psi^\nu_+ - \psi^\mu_-
\psi^\nu_- \right) +  U_{\mu\nu}
\left( \psi^\mu_+ \frac{1}{\df_\phi} \psi^\nu_+ - \psi^\mu_-
\frac{1}{\df_\phi}
\psi^\nu_- \right)
\nonumber \\
\eeqn
which is given by 
\beqn
|B\rangle = Z |B_X \rangle |B_\psi \rangle |B_{bc}\rangle |B_{\beta\gamma} \rangle   
\eeqn
with the normalization determined by the comparison 
of the overlap with closed string states to the analogous calculation in
the world-sheet sigma model, and the integration over PSL(2,R) implicit.
$|B_X \rangle$, $|B_\psi\rangle$, $|B_{bc}\rangle$, and $|B_{\beta\gamma} \rangle$
are given respectively by equations \ref{transfboundstate}, \ref{eq:fermionicbs}, 
\ref{eq:bcbs}, and \ref{eq:sgbs}.
This boundary state correctly reproduces the emission of particles by the brane 
described by the boundary interaction, and can be thought of as a 
state interpolating between the renormalization group fixed points of tachyon condensation.


\chapter{Generalized Boundary Interactions}

\label{ch:generalize}

In the previous chapter we considered exclusively the states in which the background consisted
of  a tachyon field with a quadratic boundary interaction, and also a constant antisymmetric
gauge field.  While these are interesting and have generated a great deal of investigation and
study  \cite{Tseytlin:2000mt,Bachas:1992bh,Kutasov:2000qp,Akhmedov:2001yh}
 they clearly cannot be the whole story, because
they do not exhaust the possible interactions on the 
boundary of the string world-sheet, including the possibility
of interactions higher than quadratic.  The programme in string theory is to
regard these as coupling constants that generate higher order interactions on the string world-sheet 
and on the boundary; a famous example of which is the spacetime metric tensor which appears in the
string action, when it is expanded around Minkowski spacetime it gives a massless two dimensional
theory
plus interaction terms,
\beqn
\int \df X^\mu \ddf X^\nu G_{\mu\nu}\left( X^\alpha \right) 
\rightarrow 
\int \df X^\mu \ddf X^\nu \left( \eta_{\mu\nu} + \left(\df_X^{\alpha} G_{\mu\nu}
\left( X^\beta_0 \right) \right) X^\alpha + \ldots
\right). \nonumber
\eeqn 
A  general 
theory which has such 
arbitrary interactions 
is difficult to solve analytically without a great deal of symmetry 
\cite{Green:1987sp}.  While the study of general
world-sheet actions is beyond the scope of this thesis, a much more general class of boundary
interactions is available for investigation.  

There are two directions, not mutually exclusive,
that this can take, the first is to investigate additional constant fields and couplings in the 
context of the boundary state.  These will add to the spectrum of possible fields and charges
on the hyper-surface that spans the edge of the string world sheet.  The second is to 
have non-quadratic interactions on the boundary, which will, in general, make it
difficult to write out a nice and compact expression for the boundary state,
but none the less, it is possible to recover some kind of dynamics for the strings
from this.  In all of this consideration there are still two general principles that govern 
the analysis, that the boundary states algebrize the world-sheet action, and that 
the effect of conformal transformations is accounted for.

\section{Additional Boundary State Fields}
\label{sec:addlquadfields}

We wish to demonstrate that the boundary state is not applicable only to the case
of a tachyon field and gauge field, but also to 
fields of different world-sheet dimension.  We will first examine the 
addition of different fields into the boundary state, 
which amounts to, in the bosonic case
adding fields and interactions in the form
\beq
S = \int_M \df X \ddf X + \int_{\df M} T(X^\mu) + A_\mu(X) \df X^\mu + B_\mu(X) \df^2 X^\mu
 + C_\mu(X) \df^3 X^\mu 
+\ldots 
\label{eq:additionalfields}
\eeq
For the purely quadratic case, each of those fields is expanded to linear order with the exception
of the tachyon discussed before.
The partition function and world-sheet two point functions 
are straightforward
 generalizations
of the case examined in (\ref{bosonicgf}).
In particular we find that the disk propagator becomes 
\beqn
G^{\mu\nu} (z,z') &=&
-\alpha' g^{\mu\nu} \ln \left| z - z' \right|
\nonumber \\
&&+ \frac{\alpha'}{2}
\sum_{n=1}^\infty 
\frac{(z \bar z')^n + (\bar z z')^n}{n}
\nonumber \\
&~& ~~~~
\left( \frac{g  - \frac{ \alpha'}{2}
\frac{U}{n} - 2 \pi \alpha' F - \alpha' n B - \alpha' n^2 C +\ldots 
}{ g + \frac{ \alpha'}{2} \frac{U}{n} 
+ 2 \pi \alpha' F + \alpha' n B + \alpha' n^2 C + \ldots} \right)^{
\left\{ \mu\nu \right\} }
\nonumber \\
&~& +
\frac{\alpha'}{2} \sum_{n=1}^\infty 
\frac{(z \bar z')^n - (\bar z z')^n}{i n}.
\nonumber \\
&~& ~~~~\left( \frac{g  - \frac{ \alpha'}{2}
\frac{U}{n} - 2 \pi \alpha' F - \alpha' n B - \alpha' n^2 C +\ldots
}{ g + \frac{ \alpha'}{2} \frac{U}{n}
+ 2 \pi \alpha' F + \alpha' n B + \alpha' n^2 C + \ldots} \right)^{
\left\{ \mu\nu \right\} }
\eeqn
Here we have used the expansion (\ref{tachyonprofile}) for the 
tachyon $T(X)$, assumed a prefactor of $\frac{1}{4\pi}$ for 
all the additional fields, and used the convention that $B$, $C$, and the higher terms in this expansion
are the field strength associated with the corresponding field in (\ref{eq:additionalfields}), so
\beqn 
F_{\mu\nu} &=& \df_\mu A_\nu(x) - \df_\nu A_\mu(X) \nonumber \\
B_{\mu\nu} &=& \df_\mu B_\nu(X) + \df_\nu  B_\mu(X) 
\nonumber \\
C_{\mu\nu} &=& \df_\mu C_\nu(X) - \df_\nu  C_\mu(X)
\nonumber \\
\ldots
\label{defoffields}
\eeqn
In this background  the disk partition function becomes
\beqn
Z  &=& \frac{1 }{\det \left( \frac{U}{2}  \right) }  e^{-T_0} \prod_{m=1}^\infty
\frac{1}{\det \left(
g  + \frac{\alpha'}{2} \frac{U}{m} + 2\pi \alpha' F + 
 \alpha' n B + \alpha' n^2 C + \ldots \right) }. \nonumber \\
\label{lotsofieldspf}
\eeqn

Additional fields such as those described above were discussed in \cite{Li:1993za}, with
boundary interaction 
\beqn
S_{bdy} &=& a + \frac{1}{8\pi} \oint d\theta d\theta'
X^\mu(\theta) u_{\mu\nu}(\theta - \theta') X^\nu(\theta')
\label{mostgeneralint}
\eeqn
where $\theta$ parameterizes the boundary.
The boundary coupling $u$ in 
(\ref{mostgeneralint}) preserves locality in the sense that the Taylor expansion
consists of derivatives of a $\delta$-function
\beqn
u^{\mu\nu}(\theta - \theta') &=& \sum t^{\mu\nu} \frac{\df^r}{\df_\theta^r} \delta ( \theta - \theta').
\eeqn
The boundary coupling $u$ can also be Fourier analyzed as
\beqn
u_k^{\mu\nu} = u^{\nu\mu}_{-k} = \oint d\theta u^{\mu\nu} (\theta) e^{-i k \theta}.
\label{eq:fouriertranform4}
\eeqn 
In the language of (\ref{eq:additionalfields}) this corresponds to having
the couplings
\beqn
u_k^{\mu\nu} = U^{\mu\nu} + k F^{\mu\nu} + k^2 B^{\mu\nu} + k^3 C^{\mu\nu} + \ldots
\label{uexplicitly}
\eeqn 
and the partition function is determined to be \cite{Li:1993za}
\beqn
Z = \frac{1}{\det \left( \frac{u_0}{2} \right) } e^{-a} \prod_{k=1}^\infty
\det \left( 1 + \frac{u_k}{k} \right)^{-1}
\eeqn
in agreement with (\ref{lotsofieldspf}) with the identifications $a = T_0$, $u_0 = U$ and
(\ref{uexplicitly}).  It is also shown in
\cite{Li:1993za} that using a point splitting regularization it 
is possible to introduce a short distance cut-off and truncate the expansion (\ref{eq:additionalfields})
and then renormalize with respect to this cut-off.  For our purposes it is sufficient
to note, using the formal relationship
\beqn
X(\theta) X(\theta + \epsilon) = \sum_{n=0}^\infty \frac{\epsilon^n}{n!} X(\theta) \frac{\df^n}{\df_\theta^n} 
X(\theta)
\eeqn
it is possible to use a boundary state to describe a non-local boundary interaction.
The boundary state corresponding to 
the action
(\ref{eq:additionalfields})
is, just as in (\ref{transfboundstate})
\beqn
|B_{a,b} \rangle &=&  Z \exp \left( \sum_{n=1, j,k=-\infty}^\infty
\alpha^{\mu}_{-k} M^{(a,b)}_{-n-k} { \Xi}^n_{\mu\nu} {\bar M^{(a,b)}_{-n-j} }
\tilde \alpha^{\nu}_{-j}  \right)
\nonumber \\ &~&
\exp \left(- \frac{\alpha'}{4} x^\mu U_{\mu\nu} x^\nu \right)
 | 0 \rangle.
\eeqn 
with
\beqn
{\Xi}^n_{\mu\nu} &=& \frac{1}{n} 
\left( \frac{g  - \frac{ \alpha'}{2}
\frac{U}{n} - 2 \pi \alpha' F - \alpha' n B - \alpha' n^2 C +\ldots
}{ g + \frac{ \alpha'}{2} \frac{U}{n}
+ 2 \pi \alpha' F + \alpha' n B + \alpha' n^2 C + \ldots} \right)^{
 \mu\nu  
}.
\eeqn
This will reproduce all the $\sigma$ model amplitudes \cite{Li:1993za,Laidlaw:2001jt}.  We can
also consider the case of interaction terms that are explicitly non-local.  Due to the (anti)symmetry 
requirement (\ref{eq:fouriertranform4}) still holds and so 
using equation \ref{uexplicitly} the non-local can be recast into a set of local boundary interactions,
potentially infinite in number.

This generalization has the following interesting property, it was noted that the 
transition from $U=0$ to $U = \infty$ was characterizing the transition between 
Neumann and Dirichlet boundary conditions.  This could be seen in the boundary state
expression because for large values of $U$ the coefficient in the exponential simplified
to the well known expression for the boundary state for a $D$ brane.  Now, restricting attention 
to the case of the additional field $B$ which appeared as $\oint_{\df M}  B_\mu \df^2 X^\mu$ or equivalently
the first term in the Taylor expansion $\oint_{\df M} B_{\mu\nu} \df X^\mu \df X^\nu$ where $B_{\mu\nu} =
\df_\mu B_\nu + \df_\nu B_\mu$,
as detailed in equation \ref{defoffields}
 we see that the statement is still the same.  In the case that $B 
\rightarrow
\infty$ the strings will satisfy the 
regular Dirichlet boundary conditions, but they will not have
a condition upon their zero  mode. 
We can contrast the effects of the tachyon's $U$ with this $B$.  World-sheet excitations
with large enough mode number will overcome the effect of $U$, since it appears as $U/n 
\rightarrow 0$ as $n \rightarrow \infty$, but by contrast $B$ appears as $nB$ which grows
with increasing $n$ and  
makes the system `more' Dirichlet 
in the UV.  Of course, this is just another way of saying that the coupling $U$ is irrelevant
in world-sheet power counting, and that $B$ is relevant.  
To speculate what the effects of this kind of background field are, consider the case of a region of
space where $B$ is non-zero.  An end of a string in that region will not leave for the same
reason that it would not leave a brane's surface.  A large region like that could model an extended object
which traps strings near its boundaries.

Another similar point is that if there are large $U$ and $B$ couplings on the string boundary, there 
will be a finite number of modes which dominate the partition function, equivalently the action 
for these background fields, eliminating the need to regularize the expression.  We now examine
this thought somewhat more systematically.

\section{Time Dependent Tachyons}

It is also possible to study tachyons of a more general profile, as discussed in
\cite{Recknagel:1998ih,Rey:2003zj,Sen:2002nu,Sen:2002in,Gaberdiel:2001zq}.  
The motivation for this kind of study  is to examine the 
time
dependence of the tachyons described by a boundary state as opposed to simply the 
static solutions that describe either the tachyon vacuum or static D-branes of lower dimension.
This allows a more sophisticated analysis of the dynamics that describe the decay of
a space-filling brane into one of smaller dimension.

A model for the study of this process is the boundary interaction term
\beqn
\delta S = \tilde \lambda \oint_{\df M} \cosh X^0 
\label{eq:hyp}
\eeqn
where the bulk action is the standard bosonic string action and the term 
$\tilde \lambda$ is written to conform with the conventions of \cite{Sen:2002vv,Rey:2003zj}.
It has been previously shown \cite{Recknagel:1998ih} that this type of deformation is amenable
to study.  In fact, a compact expression for the boundary state for this type
of perturbation is known to be
\beqn
| {\cal B} \rangle = {\cal N} \sum_j \sum_m D^j_{m,-m}(R) |j,m,m \rangle \rangle
\eeqn
where $j$ runs over non-negative integer and half integers and can be interpreted
as a spin, $m$ stands in the roll of projection of spin $j$, $R$ is a rotation matrix 
in $SU(2)$ which can be parameterized as $R = \left( \matrix{ a &b \cr -\bar b & \bar a }
\right)$, and $|j,m,m \rangle \rangle$ is a Virasoro Ishibashi state
\cite{Ishibashi:1989kg}
associated with
the primary state $|j,m,m \rangle$ with momentum $2m$ and conformal weight $(j^2, j^2)$.
The matrix elements of $D$ are defined, for the
parameterization of $R$ given, by the formula \cite{Gaberdiel:2001zq}
\beqn
D^j_{m,n}(R) &=& \sum_{k = \mathrm{max} (0,n-m)}^{\mathrm{min}(j-m,j+n)} \frac{
\sqrt{ (j+m)!(j-m)!(j+n)!(j-n)! } }{ (j-m-k)! (j+n-k)! k! (m-n+k)! } 
\nonumber \\
&~& ~~~~~~~~~~a^{j+n-k}
\bar a^{j+n-k} b^k (- \bar b)^{m-n+k}
\eeqn
and also the primary state can be expressed, up to a phase
as
\beqn
|j,m,m \rangle = k {\cal O}_{j,m} \exp \left( 2 i m X \right) |0 \rangle
\eeqn
where ${\cal O}_{j,m}$ are a combination of oscillators with left- and right-moving
level of $j^2 - m^2$.  Since the potential has been specialized to be in the 
$X^0$ direction it is possible to pick out the coefficient of the part of the 
boundary state that has no dependence on the $\alpha^0$ or $\tilde \alpha^0$ 
oscillators.  Note that the other 25 bosonic directions are given by a boundary
state like (\ref{eq:nmbs1}) but with all external fields vanishing giving
\beqn
| B_{25} \rangle = \exp \left( - \sum_{i=1}^{25} \sum_{n \geq 1} \frac{ \alpha_{-n}^i \alpha_{n}^i }{n} \right)
\eeqn
as in \cite{DiVecchia:1999rh}.  
Now, fixing the phases
by comparison with known configurations \cite{Sen:2002nu,Sen:2002vv}
 we are able to find that for the
hyperbolic cosine perturbation (\ref{eq:hyp}) it is 
\beqn
| {\cal B}_0 \rangle = {\cal N} \left[ 1 + 2 \sum_{n \geq 1} \left( - \sin( \tilde \lambda \pi ) \right)^n 
\cosh \left( n X^0 \right) \right] | 0 \rangle
\eeqn
which can be explicitly summed to give a time dependent constant in front of the 
$X^0$ mode independent part of $| {\cal B}_0 \rangle $, and obtain
$| {\cal B}_0 \rangle = {\cal N} f(x^0) | 0 \rangle$ with
\beqn
F(x^0) = \frac{1}{1 + e^{x^0} \sin \left( \tilde \lambda \pi \right) }
+ \frac{1}{1 + e^{-x^0} \sin \left( \tilde \lambda \pi \right) } -1.
\eeqn
In all of this, $\sin \left( \tilde \lambda \pi \right)$
is a parameter from the $SU(2)$ transformation necessary to put the boundary state in this 
form.

Similarly, it is interesting to find the coefficient of the term associated with the purely time
($00$) component
of the graviton, that is the coefficient of the state 
\beqn
\nonumber
\alpha^0_{-1} \tilde \alpha^0_{-1} | k \rangle \in | {\cal B} \rangle
\eeqn
which is found to be 
\beqn
g(x^0) = \cos \left( 2 \tilde \lambda \pi \right) + 1 - f(x^0) 
\eeqn
The sum of $g$ and $f$ is conserved, independent of $x^0$, and can be interpreted as the conserved energy density on
an unstable d-brane by observing that the sum goes, in the 
small $\tilde \lambda$ regime as
\beqn
f(x^0) + g(x^0) \rightarrow 2 \left( 1 - \tilde \lambda^2 \pi^2 \right)
\label{dbraneenergy}
\eeqn
but on the other hand the d-brane tension is given as
$\frac{1}{2 \pi^2 g^2}$ with $g$ the open string coupling constant, and from the point
of view of string field theory the potential energy for the tachyon field deformed by
$\tilde \lambda$ is $-\frac{ \tilde \lambda^2}{ 2 g^2}$ and summing the two, one obtains
the total energy \cite{Sen:2002nu,Sen:2002vv,Rey:2003zj}
\beqn
\frac{1}{2 \pi^2 g^2} \left( 1 - \tilde \lambda^2 \pi^2 \right) \nonumber
\eeqn
which is proportional to (\ref{dbraneenergy}), and this then shows that it is correct
to interpret the sum $f(x^0) + g(x^0)$ as the total energy density of the system of 
branes.
This kind of construction will give the evolution of the normalization of the
boundary states which describe the spatial d-brane.  The explicit form of $f(x^0)$ is
such that for boundary perturbations where $\sin \left( \tilde \lambda \pi \right) >0$ we have
\beqn
f(x^0) \rightarrow 0 ~\mathrm{as}~ x^0 \rightarrow \infty
\eeqn
which can be interpreted as a decay of the states with Neumann boundary conditions in
all spatial directions.

Following \cite{Rey:2003zj,Sen:2002vv} it is possible to generalize this sort of construction to something
that has spatial inhomogeneities rather than just some time dependence.  A natural candidate 
in the spirit of (\ref{eq:hyp}) is the boundary interaction
\beqn
\delta S = \tilde \lambda \oint_{\df M} \cosh \frac{X^0}{\sqrt{2}} \cos \frac{X^1}{\sqrt{2}}.
\eeqn
As is apparent from the form of the interaction, this will be solvable in the same sense that 
(\ref{eq:hyp}) was, and further it can be seen to decompose into boundary states that are 
purely functions of $X^0 \pm i X^1$, and so the analysis above can be repeated.  The boundary
state describing this decouples as
\beqn
| B \rangle = |B_{X^0,X^1} \rangle \otimes |B_{X^\mu, \mu \neq0,1}\rangle \otimes | B_{b,c} \rangle
\eeqn
where 
\beqn
|B_{X^0,X^1} \rangle = | B_+ \rangle \otimes | B_- \rangle
\eeqn
and following \cite{Rey:2003zj}
it is possible to find that
\beqn
|B_\pm \rangle &=& f(x^0 \pm i x^1) |0\rangle
\nonumber \\
&& + \frac{1}{2} g(x^0 \pm i x^1) \left( \alpha_{-1}^0 \pm i \alpha^1_{-1} \right) \left( 
\tilde \alpha_{-1}^0 \pm i \tilde \alpha^1_{-1} \right) |0\rangle
\nonumber \\
&& + \frac{1}{4}  h_1 (x^0 \pm i x^1) 
 \left( \alpha_{-2}^0 \pm i \alpha^1_{-2} \right)
\left(
\tilde \alpha_{-2}^0 \pm i \tilde \alpha^1_{-2} \right) |0\rangle
\nonumber \\
&&
+ \frac{1}{4}  h_2 (x^0 \pm i x^1)
\left( \alpha_{-1}^0 \pm i \alpha^1_{-1} \right)^2 \left(
\tilde \alpha_{-1}^0 \pm i \tilde \alpha^1_{-1} \right)^2 |0\rangle
\nonumber \\
&& + \frac{i}{4}  h_3 (x^0 \pm i x^1)
\left( \alpha_{-1}^0 \pm i \alpha^1_{-1} \right)^2 \left(
\tilde \alpha_{-2}^0 \pm i \tilde \alpha^1_{-2} \right) |0\rangle
\nonumber \\
&& + \frac{i}{4}  h_3(x^0 \pm i x^1)
 \left( \alpha_{-2}^0 \pm i \alpha^1_{-2} \right)
 \left(
\tilde \alpha_{-1}^0 \pm i \tilde \alpha^1_{-1} \right)^2 |0\rangle
+ \ldots
\eeqn
and similarly the implicit coefficient functions are determined to be
\beqn
 f(x^0 \pm i x^1) &=& \frac{1}{1 + \exp \left( \frac{x^0 \pm i x^1}{\sqrt{2}} \right) \sin( \tilde \lambda \pi/2 ) } 
+ 
\nonumber \\
&~& ~~~
\frac{1}{1 + \exp \left( -\frac{x^0 \pm i x^1}{\sqrt{2}} \right)  \sin( \tilde \lambda \pi/2 )} -1 \\
g(x^0 \pm i x^1) &=& 1 + \cos ( \tilde \lambda \pi ) - f(x^0 \pm i x^1)
\\
h_1(x^0 \pm i x^1) &=& \left( 1 + \cos ( \tilde \lambda \pi ) \right) \left( 1 - \sin( \tilde \lambda \pi/2)  \right)
\cosh \left( \frac{ x^0 \pm i x^1 }{\sqrt{2}} \right) 
\nonumber \\
&~& ~~~- f(x^0 \pm i x^1) \\
h_2(x^0 \pm i x^1) &=& 2 \left( 1 + \cos ( \tilde \lambda \pi ) \right)   \sin( \tilde \lambda \pi/2) 
\cosh \left( \frac{ x^0 \pm i x^1 }{\sqrt{2}} \right) 
\nonumber \\
&~& ~~~ + f(x^0 \pm i x^1) \\
h_3(x^0 \pm i x^1) &=& - \left( 1 + \cos ( \tilde \lambda \pi ) \right) \sin( \tilde \lambda \pi/2)
\sinh  \left( \frac{ x^0 \pm i x^1 }{\sqrt{2}} \right)
\eeqn
Now, using these expressions it is possible to expand the boundary states for arbitrary oscillators, and
we see that, for instance, the coefficient of the tachyon mode (that is to say the Fock space vacuum
$|0\rangle$) is the same as
the coefficient of some of the purely spatial components of the graviton (for instance $(\alpha_{-1}^3 
\tilde \alpha_{-1}^4 + \alpha_{-1}^4 \tilde \alpha_{-1}^3 ) |0\rangle$), with similar relationships occurring
among all oscillator combinations with the same holomorphic and antiholomorphic levels in $X^0 \pm i X^1$.
This can be briefly compared to the case of the quadratic tachyon profile considered in section
\ref{sec:bbs}.  
The similarity to the current consideration is that the states appearing in the boundary state 
expansion
do not necessarily have equal numbers of creation operators at the same level as they would in the
case of pure Neumann or Dirichlet boundary conditions 
(compare $\alpha_{-1} \alpha_{-1} \tilde \alpha_{-2} |0\rangle$ with
$\alpha_{-1} \alpha_{-1}\tilde \alpha_{-1} \tilde \alpha_{-1}|0\rangle$)

The interesting point that arises from this analysis is that, as in the purely time dependent case there
is time evolution of the coefficient functions.  This evolution is analyzed in detail in 
\cite{Rey:2003zj} and it is 
found 
that the energy density evolves off a brane and becomes localized, showing the decay of a space filling
brane into something smaller.

\section{Spherically Symmetric Tachyon Condensation}

Another possible generalization within the study of tachyon condensation is
to consider, as in \cite{Grignani:2002rx}, a more symmetric case in which the symmetry
renders the analysis more tractable.  The problem considered was that of 
the condensation of open string tachyon fields  which have an $O(D)$ symmetric profile.
In the context of the quadratic tachyon profile studied in \ref{sec:bbs} this is
simply the problem of condensation from a space-filling brane to a spherical
symmetric state by decay of the radial direction.

This problem is investigated by using the observation of 
\cite{Callan:1989nz} that the
bulk excitations can be integrated out of the partition function to get an
effective non-local field theory which lives on the boundary.  
The problem is then reduced to  a boundary conformal field theory with $D$
scalar fields on a disc perturbed by relevant boundary operators with
$O(D)$ symmetry.  The model is exactly solvable in the large $D$
limit and admits a tractable $1/D$ expansion, which only is consistent
for tachyon fields that are polynomials.  In the case of tachyon fields that
are polynomial the theory is renormalizable by normal ordering, but in
the case of non-polynomial tachyon potentials it is possible to have large 
anomalous dimensions for the operators and that these may require 
non-perturbative
renormalization which could make the $\beta$-function nonlinear.  This nonlinearity
combined with the vanishing of the $\beta$-function as a field equation for the
tachyon profile gives terms that describe  tachyon
scattering \cite{Klebanov:1988wx,Kostelecky:1999mu}.
However when the tachyon profile, and the other fields are adjusted so that the sigma model
that they define is at an infrared fixed point of the renormalization
group, these background fields are a solution of the classical
equation of motion of string theory.   Witten and
Shatashvili \cite{Witten:1992qy,Shatashvili:1993kk} have argued that  
these equations of motion can be derived from an action which is
derived from the disc partition function.

We start with the world-sheet action
\beqn
S = \int_M  \df X \ddf X + \int_{\df M} T(X)
\nonumber
\eeqn
and breaking the field $X$ into classical and quantum parts in the standard way
\beqn
X = X_c + X_q
\nonumber 
\eeqn
and $X_c$ satisfies the wave equation on the entire surface and $X_q \rightarrow 0$ on the
boundary.
Integrating out one obtains the action decouples between the classical and quantum parts 
as
\beqn
S = \int_M \df X_q \ddf X_q + \int_{\df M} \left( \frac{1}{2} X_c |\df| X_c + T(X_c) \right)
\eeqn
where the term $|\df|$ gives a non-local contribution to the kinetic term, defined by
its Fourier transform 
\beqn
|\df| \delta(\phi - \phi') = \sum_{n >0} \frac{n}{\pi} \cos n (\phi - \phi')
\eeqn
The quantum term is nothing but the partition function of the string with Dirichlet
boundary conditions, and in the absence of the tachyon field the integration over the 
classical fields on the boundary will give the terms to convert from the Dirichlet to
Neumann boundary conditions on the partition function.

Investigating the large $D$ limit of this $O(D)$ invariant model we reparameterize
\beqn 
T(X) \rightarrow D T \left( X^2/D \right).
\eeqn
We introduce auxiliary fields and a source to the partition function of the 
boundary field theory, as
\beqn
Z &=& Z_0 \int dX d\chi d\lambda \exp(-S)  
\label{eq:pflast}
\eeqn
with
\beqn
S &=& \int \frac{d\phi}{2\pi} \left( \frac{1}{2} X^i |\df + 2i \lambda| X^i + D T(\chi) - D i \lambda \chi - J^i X^i  \right)
\eeqn
where $\lambda$ is a scalar that enforces the condition $\chi = \frac{X^i X^i}{D}$ and $J^i$ is a source
term.  
As was apparent from the initial form of the action the zero modes of $X$ are special in
that they naively contribute a constant term in the two dimensional partition function
(\ref{eq:pflast}) and also since in the above action all the $X$ terms appear quadratically it
is convenient to integrate out the non-zero modes ($X_0$) which then gives the effective action
\beqn
S_\mathrm{eff} &=& \frac{D}{2} Tr \ln \left( |\df | + 2i \lambda \right) + D \int \frac{d\phi}{2\pi}
\left( T(\chi) - i \lambda \left( \chi - \frac{X^i X^i}{D} \right) - \frac{ X^i J^i}{D} \right.
\nonumber \\
&& \left. - \frac{1}{2D} \int \frac{d\phi}{2\pi}  \left( J(\phi) - 2 i \lambda X_0  \right)
G_X(\phi, \phi', 2 i  \lambda ) 
\left( J(\phi') - 2 i \lambda X_0 \right)
\right]
\label{eq:odeffact}
\eeqn
where both the trace and the boundary greens function for $X$ are only defined on non-zero modes
as those were the ones integrated out.
In the large $D$ limit the integrals over $\chi$ and $\lambda$ can be done using a saddle 
point method obtaining the equations
\beqn
T'(\chi) = i \lambda
\label{eq:lambdaeqn}
\eeqn
and 
\beqn
\chi = \frac{1}{D} \left( X + x(\phi) \right)^i  \left( X + x(\phi) \right)^i + G_X(\phi, \phi, 
2T'(\chi) ).
\label{eq:chieqn}
\eeqn
In this $x(\phi)$ is the induced classical field 
\beqn
x(\phi) = \int \frac{d\phi'}{2\pi} G_X(\phi, \phi', 2 T'(\chi) )
\left( J(\phi') - 2 T'(\chi(\phi'))  X_0 \right),
\eeqn
and the saddle point relation for $\lambda$ has already been used to simplify
the expressions.  
Thus to leading order in the large $D$ limit, the partition function is given by 
\beqn
Z = Z_0 \int dX_0 \exp - S_\mathrm{eff}[\chi_0, \lambda_0, X_0]
\eeqn
where $\chi_0$ and $\lambda_0$ are solutions of (\ref{eq:chieqn}) and (\ref{eq:lambdaeqn}) respectively.
This analysis can be extended to higher orders in $\frac{1}{D}$ as well.

When considering the effective action ({\ref{eq:odeffact}}) we note that there are divergences
that must be renormalized.  It was argued in \cite{Grignani:2002rx} that while the logarithm of the mode
number contributes a divergence that can be regularized by $\zeta$ function regularization
there remains a truly divergent term which multiplies the tachyon, and which can be subtracted
by the renormalization transformation
\beqn
T(\chi) \rightarrow : T \left( \chi - 2 \zeta(1) - 2 c \right) :
\eeqn
where $c$ is an arbitrary constant that should be fixed by a renormalization prescription.
With the substitution of this
into the effective action above, one can obtain 
an expression for an effective action that is finite, up to an arbitrary parameter that
was
discussed in \cite{Tseytlin:2000mt}.  

Interpreting the $\zeta$ function as being involved with the cutoff of the theory at
large world-sheet momentum we can see that taking the logarithmic derivative of
$:T:$ will give a linear $\beta$-function for the tachyon field at this order 
\cite{Klebanov:1988wx,Kostelecky:1999mu}
\beqn
\beta(:T:) = - :T: - 2 :T':
\eeqn
which is the large $D$ limit of the tachyon wave operator.

A transparent way to understand the content of the classical
partition function is to consider the limit where $T(X)$ is a smooth 
function and to expand in derivatives of $T$.  To do this, we set the
source $J$ to zero.  Then, we expect that the condensate $\chi$ is a
constant, independent of $\phi$.  Then, the Green function can easily
be evaluated.  It is most useful to consider an expansion of
(\ref{eq:chieqn}) (after renormalization)
\beqn
 \chi&=& \frac{\hat X^2}{D} -2
c_1+2\sum_{p=1}^\infty\zeta(p+1)\left(-2T'(\chi)\right)^p
\eeqn
and sums of this type appear  in the analysis 
in 
\cite{Grignani:2002rx}
Substituting into (\ref{eq:odeffact}) we find 
\beqn
Z &=& Z_0 \int dX_0 e^{-DT(X_0^2/D) } 
\nonumber \\ &~& ~~~~~~~~ \times \left(1-2 c_1DT'(
\frac{X_0^2}{D}) +2D\zeta(2)\left[T'\left(\frac{ X_0^2}{D}\right)\right]^2
+\ldots\right) \nonumber \\
\eeqn
and the omitted terms are of higher orders in derivatives of $T$ by its argument.
Now calculating the action, as discussed in {\cite{Grignani:2002rx}} and around 
(\ref{bosonicbfaction})
given by 
\beqn
S&=&\left( 1+\int \beta(T)\frac{\delta}{\delta T}\right)Z
\nonumber \\
&=& Z_0 \int d X_0 e^{-DT\left(\frac{X_0^2}{D}\right)} \left\{1+DT\left(
\frac{X_0^2}{D}\right)
+2DT'\left(\frac{X_0^2}{D}\right) \right. \nonumber \\
&~& ~~~~~~\left. \left[1-
c_1 D T\left(\frac{X_0^2}{D}\right)\right]\right\}
\eeqn
Which exactly coincides with the result of \cite{Tseytlin:2000mt}.


\chapter{Conclusions and Future Directions}

\label{ch:conclusions}

In this thesis 
we investigated the interplay between interactions on the string
world-sheet boundary, conformal invariance, and tachyon condensation.
We have reviewed some of the background and developments which motivate
the study of tachyon condensation.  We developed a boundary state 
appropriate for non-conformally invariant boundary interactions 
\cite{Laidlaw:2001jt,Akhmedov:2001yh},
used this boundary state to calculate higher genus string diagrams \cite{Laidlaw:2002qu}.  Where
possible we verified that the amplitudes we
obtained coincide with the known results calculated with other 
methods.  We have commented on the applicability of our boundary state
to other boundary interactions, including ones that violate world-sheet locality, 
and explored other ways to analyze tachyon condensation in Chapter \ref{ch:generalize} 
\cite{Grignani:2002rx}.

The boundary state 
\beqn
|B\rangle = \int d^2 a d^2 b \delta(|a^2| - |b^2| -1) |B_{a,b} \rangle
\nonumber
\eeqn
with
\beqn
|B_{a,b} \rangle &=& Z \exp \left( \sum_{n=1, j,k=-\infty}^\infty   
\alpha^{\mu}_{-k} M^{(a,b)}_{-n-k} \Lambda^n_{\mu\nu} {\bar M^{(a,b)}_{-n-j} }
\tilde \alpha^{\nu}_{-j}  \right)
\nonumber \\ &~&
\exp \left(- \frac{\alpha'}{4} x^\mu U_{\mu\nu} x^\nu \right)
 | 0 \rangle.
\nonumber
\eeqn
has been shown to correctly reproduce sigma model particle emission 
amplitudes, and thus describes a brane in the process of tachyon condensation.
As the parameter $U$ runs under RG flow from $0$ to $\infty$ the string world-sheet 
undergoes a change from Neumann to Dirichlet boundary conditions, and this 
boundary state gives a smooth interpolation between the two.  The 
normalization coefficient $Z$ has been shown elsewhere \cite{Kutasov:2000qp}
to correctly reproduce the expected \cite{Sen:1999mh} ratios between 
brane tensions during tachyon condensation, and this strengthens the 
interpretation of $|B\rangle$ as a brane.  
A similar boundary state was found in the superstring case as well, with
the same properties.

We also use the boundary state to calculate higher genus amplitudes.  
For the case of a conformally invariant boundary we exactly reproduce the
known results at $\chi = 0$ for a constant background gauge field.  
We also provide a concrete realization of the proposal \cite{Craps:2001jp}
for the string loop corrections to tachyon condensation, manifestly
reproducing the closed string factorization properties in the off-shell 
case considered.

Chapter \ref{ch:generalize} examines other boundary interactions, and details several different
methods of probing their structure.  
We review the construction of boundary state for time dependent backgrounds.   It
exhibits many similarities to the conformally integrated boundary state defined above which
suggests that these boundary states are also appropriate for 
the examination of the  time dependent structure of tachyon decay. Also,
we examined the $1/D$ expansion as an additional way of probing the properties
of tachyon condensation.

This work highlights several opportunities for future research and investigation.  
The boundary state constructed in Chapter \ref{ch:boundaries} is well understood in 
the context of boundary string field theory.  As this state represents a tachyon in the
process of condensing, it would be very interesting to study its representation in
cubic string field theory.  
It would similarly be interesting to extend the analysis in section \ref{sec:expansion}
to higher genus, and also to attempt cross-checks on the quantities calculated there.
Also, as alluded to in Chapter
\ref{ch:generalize} there is a natural connection between the Ishibashi states \cite{Ishibashi:1989kg}
used to describe the time dependent tachyon condensation and the boundary state $|B\rangle$, and
it is possible to include time dependent coefficients for the spatial directions in analogy with 
\cite{Rey:2003zj,Sen:2002nu}.

\bibliographystyle{plain}
\bibliography{PhD}

\appendix

\chapter{Properties of the Conformal Transformation Matrices}

\label{app:conftrans} 
In this appendix we examine some of the properties of the matrices that 
perform the conformal transformation which
maps
\beqn
\omega = \frac{az+b}{\bar b z + a}
\nonumber
\eeqn
on the degrees of freedom in the bosonic and
fermionic sectors respectively.  

\newcommand{\intzw}{\oint \frac{dz}{2\pi i} \frac{d\omega}{2\pi i} }
\newcommand{\intz}{\oint \frac{dz}{2\pi i} }
\newcommand{\intw}{\oint \frac{d\omega}{2\pi i} }

\section{Bosonic Matrix $M_{mn}^{(a,b)}$}

\label{sec:bosonic}

As discussed in chapter \ref{ch:boundaries}
 the matrix that maps the bosonic degrees of freedom to
one another under the conformal transformation above is
\beqn
M_{mn}^{(a,b)} = \intz z^m \frac{ (\bar b z + \bar a)^{n-1} }{( az + b)^{n+1} }
\eeqn
with the contour for the integral around the unit circle, as seen in (\ref{defofM}).
This matrix has a simple block structure, and the elements in each block can be
evaluated and are enumerated below.  There are a total of nine cases.

First, $m>0$, $n>0$ has a pole of order $n+1$ at $-\frac{b}{a}$, and can be evaluated as
\beqn
M_{mn}^{(a,b)} = \left. \frac{1}{n!} \df^n \frac{1}{a^{n+1} } z^m (\bar b z + \bar a)^{n-1} \right|_{-b/a}
\eeqn
and some of the elements of this are given explicitly as
\beqn
M_{1n}^{(a,b)} &=& \frac{1}{n} \frac{\bar b^{n-1} }{a^{n+1} } 
\nonumber \\
M_{m1}^{(a,b)} &=& m \frac{ (-b)^{n-1} }{a^{n+1} }
\eeqn

The case $m>0$, $n=0$ is immediately evaluated as 
\beqn
M_{m0}^{(a,b)} &=& \left( \frac{-b}{a} \right)^m
\eeqn

Examining $m>0$, $n<0$ there are no poles within the contour so the matrix vanishes.

The case $m=0$, $n>0$ can be obtained from the residue theorem as
\beqn
M_{0n}^{(a,b)} &=& 
\left. \frac{1}{n!} \df^n \frac{1}{a^{n+1} } (\bar b z + \bar a)^{n-1} \right|_{-b/a}
=0
\eeqn

Similarly we determine $M_{00}^{(a,b)}=1$, and in the case of $m=0$, $n<0$
there are again no poles within the integration contour so the matrix elements vanish.

Now, for the case of $m<0$, $n>0$ we have poles at both zero and $-b/a$.
\beqn
M_{mn}^{(a,b)} &=& \intz z^{-|m|} \frac{ (\bar b z + \bar a)^{n-1} }{( az + b)^{n+1} }
\eeqn
but with the transformation $z \rightarrow \omega=1/z$ we can rewrite the integral
as 
\beqn
M_{mn}^{(a,b)} &=& \intw w^{|m|} \frac{(\bar b + \bar a \omega)^{n-1} }{( a+b \omega)^{n+1} }
\eeqn
and the negative sign from the differential is compensated for by the switch of integration 
directions.
This new expression can be seen, as for the $m>0$, $n<0$ case to have no poles within
 the contour and thus to vanish.

For the case  $m<0$, $n=0$ there are again two poles, a pole of order $m$ at $0$ and a
simple pole at $z=-b/a$.  This can be evaluated by either performing the redefinition above
 $z \rightarrow \omega=1/z$ in which case it is obvious that the expression is just the 
complex conjugate of the $m>0$, $n=0$ case, or it can be evaluated directly which we do for illustrative
purposes here in the case $m=-2$.
\beqn
M_{20}^{(a,b)} &=& \intz z^{-2} \frac{ 1}{(\bar b z + \bar a) ( az + b) }
\nonumber \\
&=& \left( \frac{-b}{a} \right)^{-2} + \left.  \df \frac{ 1}{(\bar b z + \bar a) ( az + b) } \right|_0
\nonumber \\
&=& \frac{a^2}{b^2} - \frac{1}{b^2 \bar a} - \frac{ \bar b}{b \bar a^2 }
\nonumber \\
&=& \frac{ a^2 \bar a^2 - a \bar a - b \bar b }{b^2 \bar a^2} = \frac{ \bar b^2}{ \bar a^2}
\eeqn
exactly as expected from the previous considerations.

Finally for the case $m<0$, $n<0$ the redefinition $z \rightarrow \omega=1/z$
gives the equality immediately
\beqn 
\bar M_{|m|~|n|}^{(a,b)} = M_{-|m|~-|n|}^{(a,b)}
\eeqn

This analysis confirms a kind of block diagonal structure, and ensures that, as advertised, there
is no mixing between creation and annihilation operators.  There is however a flow to the zero mode which
reflects a natural redefinition of the momentum after a conformal transformation.  This was important in
the work on the bosonic degrees of freedom to ensure that the overlap between the boundary state $|B\rangle$
and a particle matched the sigma model expectation value for the corresponding vertex operator.

While perhaps obvious, we now check that the expected composition laws hold for these matrices.
So, calculating we find
\beqn
M_{mn}^{(a,b)} M_{nk}^{(a',b')} &=& \sum_n \intzw z^m 
\frac{ (\bar b z + \bar a)^{n-1} }{( az + b)^{n+1} }
\omega^n \frac{ (\bar b' \omega + \bar a')^{k-1} }{( a' \omega + b')^{k+1} }
\eeqn
Since we know that the positive and negative elements of this matrix are complex conjugates of
each other, and further the structure on $M_{m0}$ and $M_{0n}$ we can restrict the sum to
be from $1$ to $\infty$ and concentrate only on the annihilation operators, knowing that the
sum will also work for the creation operators, so
\beqn
M_{mn}^{(a,b)} M_{nk}^{(a',b')} &=&
\intzw z^m \frac{\omega}{(a z + b) } \frac{1}{ (a-\bar b \omega) z + (b - \bar a \omega) }
\frac{ (\bar b' \omega + \bar a')^{k-1} }{( a' \omega + b')^{k+1} }
\nonumber \\
&=& \intw \frac{ (\bar b' \omega + \bar a')^{k-1} }{( a' \omega + b')^{k+1} }
\left[ \left( \frac{-b}{a} \right)^m + \left( \frac{ -(b - \bar a \omega) }{ (a-\bar b \omega) } \right)^m \right]
\nonumber \\
&=& \intw \omega^m \frac{ (\bar b' (a \omega +b) + \bar a' (\bar b \omega + \bar a ) )^{k-1} }{
( a' (a \omega +b ) + b'( \bar b \omega + \bar a ) )^{k+1} }
\nonumber \\
&=& M_{mk}^{(a'a + b' \bar b, b a' + b' \bar a)}
\eeqn
and in the second to last line the redefinition $\omega \rightarrow \frac{ a \omega + b}{ \bar b \omega + \bar a }$ was
used.  It can immediately be seen that $|a'a + b' \bar b|^2 - |b a' + b' \bar a|^2 =1$ so this is another conformal transformation
of the same type, as expected.  This also shows that the 
expected inverse matrix transformation for$M_{mn}^{(a,b)}$, which would
be $M_{mn}^{(\bar a, -b)}$ is in fact the inverse.

Finally, we check the claimed property that renders these matrices moot in the conformally invariant case,
explicitly that,
\beqn
M_{km}^{(a,b)} \frac{1}{k} \bar M_{kn}^{(a,b)}  &=& \frac{1}{m} \delta_{mn}
\eeqn
As before the stated property that $M_{0m}=0$ helps, and we restrict to positive $k$, finding
\beqn
M_{km}^{(a,b)} \frac{1}{k} \bar M_{kn}^{(a,b)}  &=& \intzw 
z^k   
\frac{ (\bar b z + \bar a)^{m-1} }{( az + b)^{m+1} }
\frac{1}{k} \omega^k 
\frac{ (b \omega + a)^{n-1} }{ (\bar a \omega + \bar b )^{n+1} }
\nonumber \\
&=& - \intzw \ln (1 - \omega/z) 
\frac{ (\bar b z + \bar a)^{m-1} }{( az + b)^{m+1} }
\frac{ (b \omega + a)^{n-1} }{ (\bar a \omega + \bar b )^{n+1} } \nonumber \\
\eeqn
where we have transformed $z \rightarrow 1/z$.  Integrating along the branch cut which runs from $z=0$ to
$z=\omega$, and redefining again $z \rightarrow \frac{ a z - \bar b }{ - b z + \bar a}$ we obtain 
\beqn
M_{km}^{(a,b)} \frac{1}{k} \bar M_{kn}^{(a,b)}  &=&
\intw \frac{1}{m} \frac{ (b \omega + a)^{n-1} }{ (\bar a \omega + \bar b )^{n+1} }
\left[ \left( \frac{ \bar a \omega + \bar b }{ b \omega + a } \right)^m - \left( \frac{ \bar b}{a} \right)^m \right]
\nonumber \\
&=& \frac{1}{m} \delta_{mn}
\label{eq:desiredresult}
\eeqn

We have now verified the salient points claimed within the text, and shown that these transformations
in fact act as a group, and become trivial in the cases where $\Lambda(n)$ is independent of $n$.

\section{ Fermionic Matrix $N^{(a,b)}_{rm}$} 

\label{sec:fermionic}
We have also described previously the matrix in the fermion NS sector, $N^{(a,b)}_{rm}$ that describes
the mappings between the various fermion creation and annihilation operators.
This matrix was derived {\ref{eq:defofN}} to be
\beqn
N^{(a,b)}_{rm} &=& \intz z^{r-1/2} \frac{ (\bar b z + \bar a )^{m-1/2} }{ (a z + b)^{m + 1/2} }
\eeqn
where $r \in \mathbb{Z} + \frac{1}{2}$
Since there are no zero modes for this, the number of possible options is significantly less, but
as in the case of the bosonic matrix we enumerate them.

In the case $r>0$, $m>0$ we have poles of order $m+1/2$ at $-\frac{b}{a}$.
This can be evaluated to give 
\beqn
N^{(a,b)}_{rm} &=& \left. 
\frac{1}{a^{m+1/2} } \frac{1}{ (m-1/2)! } \df^{m-1/2} z^{r-1/2} 
( \bar b z + \bar a )^{m-1/2} \right|_{-b/a}
\eeqn
and some of the cases that are short to write are
\beqn
N^{(a,b)}_{1/2~ m} &=& \frac{ \bar b^{m-1/2} }{ a^{m+1/2} } \nonumber \\
N^{(a,b)}_{3/2~m} &=& \frac{ \bar b^{m-3/2} }{ a^{m+3/2} } \left( 1 - |b|^2 \right) \nonumber
\eeqn

For the case $r>0$, $m<0$ there are no poles within the integration contour, and so these elements vanish.
Similarly, for the case $r<0$, $m>0$ there are apparently poles at both $0$ and $-b/a$ but just as
in the bosonic case it is possible to make the transformation $z \rightarrow \frac{1}{z}$ which results in
a new function to be integrated with the poles outside of the contour, and hence vanishes.  
The same trick can be used for the case $r<0$, $m<0$ and we explicitly exhibit it for completeness
\beqn
N^{(a,b)}_{rm} &=& \intz z^{r-1/2} \frac{ (\bar b z + \bar a )^{m-1/2} }{ (a z + b)^{m + 1/2} }
\nonumber \\
&& \mathrm{with} z \rightarrow \frac{1}{\omega} \nonumber \\
&=& \intw \frac{1}{\omega} \frac{1}{\omega^{r-1/2} } \frac{ (\bar b+ \bar a \omega )^{m-1/2} }{ (a + b
\omega )^{m + 1/2} } \nonumber \\
&=& \intw \omega^{|r| - 1/2}  \frac{ (a + b
\omega )^{|m| -1/2} }{ (\bar b+ \bar a \omega )^{|m| +1/2 } }
\eeqn
Which shows that 
\beqn
\bar N_{|m|~|n|}^{(a,b)} = N_{-|m|~-|n|}^{(a,b)}
\eeqn
 as desired.  This analysis shows as in the bosonic case that the creation and annihilation 
operators do not mix under these transformations.

To complete the parallel with the bosonic case it is necessary to show the composition law, and
that in the case that the matrices are contracted through a $PSL(2,R)$ invariant exponent in
the boundary state that they contract to a unit matrix.
The first problem is to calculate
\beqn
N^{(a,b)}_{rp} N^{(a',b')}_{pq} &=& \sum_p \intzw z^{r-1/2} \frac{ (\bar b z + \bar a )^{p-1/2} }{ (a z + b)^{p + 1/2} }
\omega^{p-1/2} 
\frac{ (\bar b' \omega + \bar a' )^{q-1/2} }{ (a' \omega + b')^{q + 1/2} }
\nonumber \\
&=& \intzw z^{r-1/2} \frac{ 1}{ z (a-\bar b \omega) - (-b + \bar a \omega) }
\frac{ (\bar b' \omega + \bar a' )^{q-1/2} }{ (a' \omega + b')^{q + 1/2} }
\nonumber \\
&=& \intw \frac{ ( -b + \bar a \omega)^{r-1/2} }{ (a - \bar b \omega )^{r + 1/2} } 
\frac{ (\bar b' \omega + \bar a' )^{q-1/2} }{ (a' \omega + b')^{q + 1/2} }
\nonumber \\
&& \mathrm{with} \omega \rightarrow \frac{a \omega + b}{\bar \omega + \bar a}  \nonumber \\
&=& \intw \omega^{r -1/2} \frac{ (\bar b'' \omega + \bar a'' )^{q-1/2} }{ (a'' \omega + b'')^{q + 1/2} }
\eeqn
with $b''=\bar a b' + a' b$ and $a'' = a' a + b' \bar b$ just as in the bosonic case.

Now we calculate in analogy to {\ref{eq:desiredresult}} the quantity
\beqn
N^{(a,b)}_{rp} \bar N^{(a,b)}_{rq} &=& \sum_r \intzw
z^{r-1/2} \frac{ (\bar b z + \bar a )^{p-1/2} }{ (a z + b)^{p + 1/2} }
\omega^{r-1/2}
\frac{ ( b \omega +  a )^{q-1/2} }{ (\bar a \omega + \bar b)^{q + 1/2} }
\nonumber \\
\eeqn
and by transforming $z \rightarrow \frac{1}{z}$ and summing we find
\beqn
N^{(a,b)}_{rp} \bar N^{(a,b)}_{rq}
&=& \intzw \frac{1}{z-\omega} \frac{ (\bar b + \bar a z)^{p-1/2} }{ (a  + b z)^{p + 
1/2} }
\omega^{r-1/2} \frac{ ( b \omega +  a )^{q-1/2} }{ (\bar a \omega + \bar b)^{q + 1/2} }
\nonumber \\
&=& \intz \frac{ (a + b z)^{q-p-1} }{ (\bar b + \bar a z)^{q-p+1} } = \delta_{qp}
\eeqn

These relations 
show that the matrix of transformations for the fermions in the NS sector has the analogous nice
properties as that of the bosonic transformation.

\chapter{Green's Functions}

\label{app:gf}

Here we present in some detail the calculations of the bosonic and fermionic Green's functions
for the quadratic tachyon background under consideration.  The construction 
presented below will make the generalization to the case of different (quadratic) boundary
interactions that are mentioned in section \ref{sec:addlquadfields}
and the case of more complicated interactions, that is to say higher order than quadratic, 
while not presented explicitly because they are not amenable to exact expression
in a compact manner can be dealt with through standard techniques of field theory.

\section{Bosonic Tree level}

The starting point for this calculation is the action
(\ref{action}) which is rewritten here for convenience
\beqn
S\left( g, F, T_0, U \right) &=&
\frac{1}{ 4 \pi \alpha'} \int_\Sigma d\sigma d\phi ~ 
g_{\mu\nu}  \df^a X^\mu \df_a X_\mu
\nonumber \\
&~& + \int_{\df \Sigma} d\phi \left(\frac{1}{2} 
F_{\mu\nu} X^\nu \df_\phi X^\mu + \frac{1}{2\pi}
T_0 + \frac{1}{8\pi} U_{\mu\nu} X^\mu X^\nu \right).
\nonumber \\
\eeqn
Now, for a disk world-sheet the greens function satisfying Neumann boundary conditions
is determined in
\cite{Hsue:1970ra}
and we wrote it as (\ref{diskprop})
\beq
G^{\mu\nu}(z,z') = - \alpha' g^{\mu\nu} \left( -  \ln \left| z-z' \right| - \ln
\left| 1- z \bar z' \right| \right).
\label{app:diskprop}
\eeq
Clearly it is possible to either calculate exactly from the boundary conditions this 
greens function in the background of (\ref{action}), or we can treat the boundary 
terms as perturbations and perform an explicit sum (an equivalent procedure).

For illustrative purposes we choose the second method, and since the interaction terms
are quadratic there is one term at each order in perturbation theory.  The final bulk
to bulk propagator will be the sum of the propagator with no boundary terms and the 
increasing number of boundary interactions.  For the parameterization of the 
world-sheet $z= \rho e^{i \phi}$ we have the  bulk to boundary propagator
(\ref{diskbbprop}) which is 
\beq
G^{\mu\nu}(\rho e^{i \phi}, e^{i \phi'} ) = 2 \alpha' g^{\mu\nu}
\sum_{m=1}^\infty \frac{\rho^m}{m}
\cos[ m(\phi - \phi') ]
\label{eq:o0gf}
\eeq
and also 
the boundary to boundary propagator
\beq
G^{\mu\nu}(\rho e^{i \phi}, e^{i \phi'} ) = 2 \alpha' g^{\mu\nu}
\sum_{m=1}^\infty
\frac{ \cos[ m(\phi - \phi') ]
}{m}
\eeq
using the identities from \cite{gradshteyn:1965}.

Now, to first order in the perturbing terms the contribution to the 
propagator is
\beqn 
G_1^{\mu\nu}(\rho e^{i \phi},\rho' e^{i \phi'}) &=&
\int d\theta G^{\mu\nu'}(\rho e^{i \phi}, e^{i \theta} ) 
\left( F \df_\theta + \frac{1}{4 \pi} U \right)_{\nu' \mu'}
G^{\mu' \nu} ( e^{i \theta},\rho' e^{i \phi'})
\nonumber \\
&=& (2 \alpha')^2 \int d\theta \sum_{m,m'} \frac{\rho^m \rho'^{m'} }{m m'} 
\cos m (\phi - \theta) 
\nonumber \\ &~& ~~~~~~~~
\times \left( F \df_\theta + \frac{1}{4 \pi} U \right)^{\mu\nu}
\cos m'(\theta-\phi') \nonumber \\
&=& 
(2 \alpha')^2 \pi \sum_m \frac{\rho^m \rho'^m}{m^2}  
\nonumber \\ &~& ~~~~~~~~ \times 
\left(  \frac{1}{4 \pi} U \cos m(\phi - \phi')
- m F \sin m (\phi - \phi') \right)^{\mu\nu}.
\nonumber \\
\label{eq:o1gf}
\eeqn
Similarly the second order contribution can be read off from the concatenation
of (\ref{eq:o1gf}) with 
(\ref{eq:o0gf}) to give
\beqn
G_2^{\mu\nu} (\rho e^{i \phi},\rho' e^{i \phi'})
&=&
(-2 \alpha')^3 \pi \int d\theta 
\sum_m \frac{\rho^m}{m^2}  
\nonumber \\ &~& \left(  \frac{1}{4 \pi} U \cos m(\phi - \theta)
- m F \sin m (\phi - \theta) \right)^{\mu\nu'} 
\nonumber \\ &~&
\left( F \df_\theta + \frac{1}{4 \pi} U \right)_{\nu' \mu'} 
\sum_{m'}  g^{\mu' \nu} \frac{ \rho'^{m'} }{m'} \cos m' (\theta - \phi')
\nonumber \\
&=&
(-2 \alpha')^3 \pi^2 \sum_m \frac{\rho^m \rho'^m}{m^3} 
\bigg( \frac{U^2}{(4\pi)^2} \cos m (\phi - \phi') 
\nonumber \\
&&  - m \left( F \frac{U}{4 \pi} + 
\frac{U}{4 \pi} F \right) \sin m (\phi - \phi') 
\nonumber \\
&& 
+ m^2 F^2 \cos m (\phi - \phi') \bigg)^{\mu \nu}
\label{eq:o2gf}
\eeqn
with the obvious generalization to higher orders. 

Now, in the above we note that all terms in this sum will naturally separate into
terms with $\cos m (\phi - \phi')$ and $\sin m (\phi - \phi')$ and by inspection the
dependence on $F$ and $U$ is such that the coefficient of $\sin m (\phi - \phi')$ is 
naturally combinations of $F$ and $U$ that are antisymmetric in Lorentz indices, just
as those for $\cos m (\phi - \phi')$ are symmetric in $F$ and $U$.
Using the facts 
\beqn
\sin m (\phi - \phi') &=& \frac{ e^{i (\phi - \phi')} - e^{-i (\phi - \phi') } }{2i} \nonumber \\
\cos m (\phi - \phi') &=& 
\frac{ e^{i (\phi - \phi')}+ e^{-i (\phi - \phi') } }{2}
\nonumber 
\eeqn
and the identification $z= \rho e^{i \phi}$,
the sum 
\beqn
G_0^{\mu\nu} + G_1^{\mu\nu} + G_2^{\mu\nu} + \ldots \nonumber
\eeqn 
can then be calculated  as
\beqn
G^{\mu\nu} (z,z') &=&
-\alpha' g^{\mu\nu} \ln \left| z - z' \right|
\nonumber \\ &~& 
+ \frac{\alpha'}{2}
\sum_{n=1}^\infty \left( \frac{g - 2 \pi \alpha' F - \frac{ \alpha'}{2}
\frac{U}{n} }{ g + 2 \pi \alpha' F + \frac{ \alpha'}{2} \frac{U}{n} } \right)^{
\left\{ \mu\nu \right\} }
\frac{(z \bar z')^n + (\bar z z')^n}{n}
\nonumber \\
&~& +
\frac{\alpha'}{2} \sum_{n=1}^\infty \left( \frac{ g - 2 \pi \alpha' F -
\frac{ \alpha'}{2}
\frac{U}{n} }{ g + 2 \pi \alpha' F + \frac{ \alpha'}{2} \frac{U}{n} } \right)^{
\left[\mu\nu \right] }
\frac{(z \bar z')^n - (\bar z z')^n}{i n}.
\eeqn
as noted in (\ref{bosonicgf}).
This includes the $\ln | 1- z \bar z' |$ term in the two $F$ and $U$ dependent terms
as can be seen by the limit that as $F, U \rightarrow 0$ we recover the known expression
(\ref{app:diskprop}), and in the case of $U \rightarrow \infty$, Dirichlet boundary 
conditions, we obtain
\beqn
G^{\mu\nu}(z,z') = - \alpha' g^{\mu\nu} \left( -  \ln \left| z-z' \right| + \ln
\left| 1- z \bar z' \right| \right),
\eeqn 
which is the Dirichlet propagator on the disk.

\section{Fermionic Tree level}

As in the bosonic case we start with the fermionic action 
(\ref{fermionicaction}) which is
\beqn
S_{ferm} = \int_M \left( \psi^\mu_+ \df_- \psi^\nu_+ + \psi^\mu_- \df_+
\psi^\nu_- \right)
+  \oint_{\df M} F_{\mu\nu}
\left( \psi^\mu_+ \psi^\nu_+ - \psi^\mu_-
\psi^\nu_- \right) + \nonumber \\
U_{\mu\nu}
\left( \psi^\mu_+ \frac{1}{\df_\phi} \psi^\nu_+ - \psi^\mu_-
\frac{1}{\df_\phi}
\psi^\nu_- \right).
\eeqn  
The appropriate Green's functions for the free case have been
determined to be \cite{Arutyunov:2001nz,Viswanathan:2001cs}
\beqn
G_\psi(z,w) = \frac{\alpha'}{i} \left( \frac{ \sqrt{ z w} }{z-w}
- \frac{ \sqrt{ z \bar w} }{1 - z \bar w} \right)
\\
\tilde G_\psi(z,w) = \frac{\alpha'}{i} \left( -
\frac{ \sqrt{ \bar z \bar w } }{\bar z - \bar w}
+ \frac{ \sqrt{ \bar z w} }{1 - \bar z  w} \right).
\eeqn        
As in the bosonic case we specialize to the bulk to boundary propagator,
which upon imposition of  the antisymmetry requirement 
on it becomes
\beqn
G^{\mu\nu}_\psi\left( \rho e^{i \phi}, e^{i \phi'} \right)
&=& 2 \alpha' g^{\mu\nu} \sum_{r \in \mathbb{Z}+1/2 >0} \rho^r \sin r 
(\phi - \phi').
\eeqn

Upon insertion of the interaction term associated with the
gauge field, the first order modification to the bulk to 
bulk propagator
is
\beqn
G^{\mu\nu}_{1\psi}\left( \rho e^{i \theta}, \rho' e^{i \theta'} \right)   
&=& (2 \alpha')^2 F^{\mu\nu}
\int d\phi \sum_{r,r' \in \mathbb{Z}+1/2 >0}
\rho^r \rho'^{r'} 
\nonumber \\ && ~~~~~~~~~~~~~~~
\times \sin r ( \theta - \phi) \sin r' (\phi - \theta')
\nonumber \\
&=& (2 \alpha')^2 \pi F^{\mu\nu}
\sum_{r \in \mathbb{Z}+1/2 >0} \left( \rho \rho' \right)^r 
\cos r (\theta - \theta').
\eeqn
Similarly we can determine the order $F^2$ modification as
\beqn
G^{\mu\nu}_{2\psi}\left( \rho e^{i \theta}, \rho' e^{i \theta'} \right)
&=& 
(2 \alpha')^3 \pi (F^2)^{\mu\nu} \int d\phi 
\sum_{r,r' \in \mathbb{Z}+1/2 
>0}
\rho^r \rho'^{r'} 
\nonumber \\ && ~~~~~~~~~~~~~~~
\times \cos r ( \theta - \phi) \sin r' (\phi - \theta')
\nonumber \\                                              
&=& (2 \alpha')^3 \pi^2 (F^2)^{\mu\nu}
\sum_{r \in \mathbb{Z}+1/2 >0} \left( \rho \rho' \right)^r 
\sin r  (\theta - \theta')
\nonumber \\
\eeqn     
with higher order terms determined similarly.

For the insertion of the $U$ interaction term associated with the
tachyon field, it is important to remember the definition of 
$\frac{1}{\df_\theta}$, 
\beqn
\frac{1}{\df_\phi} \psi(\phi) = \frac{1}{2} \int d\phi'
\epsilon(\phi-\phi') \psi(\phi')
\eeqn
where $\epsilon$ is a step function: $\epsilon(x) =1$ for $x>0$ and
$\epsilon(x)=-1$ for $x<0$.   
Using this the lowest order correction to the fermionic Green's function 
due to $U^{\mu\nu}$ is
\beqn
G^{\mu\nu}_{1\psi}\left( \rho e^{i \theta}, \rho' e^{i \theta'} \right)
&=& (2 \alpha')^2 U^{\mu\nu}
\int d\phi 
\sum_{r,r' \in \mathbb{Z}+1/2 >0}
\rho^r \rho'^{r'} 
\nonumber \\ && ~~~~~~~~~~~~~~~
\times \sin r ( \theta - \phi) \frac{1}{\df_\phi}
\sin r' (\phi - \theta')
\nonumber \\
&=& (2 \alpha')^2 \pi U^{\mu\nu}
\int d\phi
\sum_{r,r' \in \mathbb{Z}+1/2 >0}
\rho^r \rho'^{r'} 
\nonumber \\ && ~~~~~~~~~~~~     ~~~
\times \sin r ( \theta - \phi) \frac{ \cos r' (\phi - 
\theta')}{ r'}
\nonumber \\
&=& (2 \alpha')^2 \pi U^{\mu\nu} 
\sum_{r \in \mathbb{Z}+1/2 >0} \left( \rho \rho' \right)^r
\frac{\sin r  (\theta - \theta')}{r} .
\label{lowestorderU}
\eeqn
The steps in (\ref{lowestorderU}) can evidently be repeated
indefinitely and so for the $n$th insertion of $U$ into the
bulk to bulk propagator we obtain
\beqn
G^{\mu\nu}_{n\psi}\left( \rho e^{i \theta}, \rho' e^{i \theta'} \right)
&=&
(2 \alpha')^{n+1} \pi^n (U^n)^{\mu\nu}
\sum_{r \in \mathbb{Z}+1/2 >0} \left( \rho \rho' \right)^r
\frac{\sin r  (\theta - \theta')}{r^n} .
\nonumber \\
\eeqn
It is also clear that for interactions with combinations of
$F$ and $U$ the resultant will depend on $\sin r (\theta - \theta')$
in the case of an even number of $F$s, and on $\cos
r (\theta - \theta')$ for an odd number.  
Summing the contributions of interactions with both $U$ and $F$ 
allows the verification of (\ref{eq:fermholo}).
The Green's function for the antiholomorphic coordinates $\tilde \psi$
 can 
be obtained by an identical argument.



\end{document}